\documentclass[12pt]{scrartcl}
\pdfoutput=1
\usepackage{hyphenat}
\usepackage{mathtools}
\usepackage{amsthm}
\usepackage{amsfonts}
\usepackage{bbm}
\usepackage[margin=1in]{geometry}
\usepackage{physics}
\usepackage[dvipsnames]{xcolor}
\usepackage{pifont}
\usepackage{diagbox}
\usepackage{tikz}
\usepackage{tikz-3dplot}
\usepackage{tkz-euclide}
\usepackage{mathrsfs}
\usepackage{subcaption}
\usepackage{cite}
\usepackage{hyperref}
\usepackage{cleveref}
\usepackage[normalem]{ulem}
\newcommand{\stkout}[1]{\ifmmode\text{\sout{\ensuremath{#1}}}\else\sout{#1}\fi}
\usepackage[author=,final]{fixme}
\fxusetheme{color}
\NewDocumentCommand{\corr}{mmO{}}{\fxwarning*[author=]{\sout{#2}#3}{#1}}
\FXRegisterAuthor{erg}{aerg}{Elias}

\newtheorem{theorem}{Theorem}

\DeclareMathOperator{\Ext}{Ext}
\DeclareMathOperator{\Tor}{Tor}
\DeclareMathOperator{\Hom}{Hom}

\DeclareMathOperator{\gal}{Gal}
\DeclareMathOperator{\diag}{diag}
\DeclareMathOperator{\Span}{Span}
\DeclareMathOperator{\im}{im}
\DeclarePairedDelimiter{\floor}{\lfloor}{\rfloor}
\DeclarePairedDelimiter{\ceil}{\lceil}{\rceil}

\def\dblvdot{\vcenter{\hbox{$\vdot$}\vspace{-1.5pt}\hbox{$\vdot$}}}
\newcommand{\cmark}{\textcolor{blue}{\ding{51}}}%
\newcommand{\xmark}{\textcolor{red}{\ding{55}}}%
\newcommand{\dmark}{\textcolor{blue}{$\vdot$}}%
\newcommand{\ymark}{\textcolor{blue}{$\odot$}}%
\newcommand{\sqmark}{\textcolor{blue}{$\dblvdot$}}%
\newcommand{\nmark}{\textcolor{red}{$\otimes$}}%
\newcommand\be{\begin{equation}}
\newcommand\ee{\end{equation}}

\definecolor{yellow2}{rgb}{0.98, 0.80, 0.20} \definecolor{blue2}{RGB}{65,128,255}
\usetikzlibrary{decorations.pathreplacing,decorations.markings,arrows.meta,shapes.misc}

\tikzset{cross/.style={cross out, draw=black, minimum size=2*(#1-\pgflinewidth), inner sep=0pt, outer sep=0pt},
  cross/.default={1pt}}
\tikzset{dotnode/.style={inner sep=0pt,outer sep=0pt,
    circle,fill,minimum size=5pt}}
\tikzset{smalldot/.style={inner sep=0pt,outer sep=0pt,
    circle,fill,minimum size=3pt}}
\tikzset{->-/.style={decoration={
      markings,
      mark=at position 0.5 with {\arrow{>}}},postaction={decorate}}}
\tikzset{-<-/.style={decoration={
      markings,
      mark=at position 0.5 with {\arrowreversed{>}}},postaction={decorate}}}

\numberwithin{equation}{section}

\title{SymTFT, Protected Gaplessness, and Spontaneous Breaking of Non-invertible Symmetries}
\author{Michele Del Zotto$^{\dagger\sharp}$, Azeem Hasan$^{\flat}$ and Elias Riedel Gårding$^{\sharp}$
  \\[1cm]
  \small\slshape$^\sharp$ Mathematics Institute, Uppsala University,  \\[-0.2cm]
  \small\slshape Box 480, SE-75106 Uppsala, Sweden\\
  \small\slshape$^\sharp$ Center For Geometry And Physics, Uppsala University,  \\[-0.2cm]
  \small\slshape Box 480, SE-75106 Uppsala, Sweden\\
  \small\slshape$^{\flat}$ Department of Mathematical Sciences, IBA \\[-0.2cm]
  \small\slshape Karachi, Pakistan \\
  \small\slshape$^\dagger$ Department of Physics and Astronomy, Uppsala University,  \\[-0.2cm]
  \small\slshape Box 516, SE-75120 Uppsala, Sweden\\
}
\date{}

\begin{document}

\maketitle

\phantom{ciao}
\vspace{-1.5cm}

\begin{center}
  \textbf{Abstract}
\end{center}

\begin{quote}
\noindent In recent years we have learned that several four-dimensional field theories can manifest non-invertible zero-form symmetries generalizing the Kramers-Wannier duality defect of the 2d critical Ising model. Several recent works by various groups have observed a deep interplay among such non-invertible symmetries in 3+1 dimensions, their anomalies, and the properties of the ground state(s). The purpose of this work is to present a first coarse classification of all possible classes of non-invertible symmetries of this type that can either enforce gaplessness or be spontaneously broken in the infrared exploiting the topological symmetry theory formalism. Our methods also generalize to non-invertible KW-like duality symmetries graded by non-abelian finite subgroups. As a first applications of our results we present examples in the context of supersymmetric models. Along the way we notice the potential for further global structures that could be realized by non-SUSY versions of Argyres-Douglas type fixed points.

\end{quote}
\vfill

------------

April, 2025

\thispagestyle{empty}

\newpage

\tableofcontents

\section{Introduction}
\label{sec:introduction}

It is typically very hard to establish universal criteria to decide whether in the deep infrared a given quantum field theory is gapped or gapless. Notable examples of such results are given by Lieb--Schultz--Mattis type theorems in various dimensions (see e.g.~\cite{Lieb:1961fr,Hastings:2003zx,Cho:2017fgz,Jian:2017skd,Else:2019lft,Cordova:2019bsd,Cordova:2019jqi,Ye:2021nop,Brennan:2023kpo,Brennan:2023ynm,Brennan:2024tlw}). The common theme of a subclass of these results is that certain symmetry structures cannot be realized by (trivially) gapped quantum fields. In the context of generalized symmetries \cite{Gaiotto:2014kfa}, recent studies have demonstrated that this effect holds for certain classes of non-invertible duality symmetries \cite{Choi:2021kmx,Kaidi:2021xfk,Choi:2022zal,Kaidi:2022uux,Bashmakov:2022jtl,Choi:2022rfe,Antinucci:2022vyk,Bashmakov:2022uek,Antinucci:2022cdi,Sun:2023xxv,Cordova:2023bja,Antinucci:2023ezl} --- for a sample of recent constructions of non-invertible symmetries in higher dimensions see e.g. \cite{Heidenreich:2021xpr,
Roumpedakis:2022aik,
Bhardwaj:2022yxj,
Hayashi:2022fkw,
Cordova:2022ieu,
Damia:2022bcd,
Bhardwaj:2022lsg,
Bartsch:2022mpm,
GarciaEtxebarria:2022vzq,
Apruzzi:2022rei,
Niro:2022ctq,
Choi:2022fgx,
DelZotto:2024tae,
Yokokura:2022alv,
Bhardwaj:2022kot,
Bhardwaj:2022maz,
Bartsch:2022ytj,
Hsin:2022heo,
Kaidi:2023maf,
Carta:2023bqn,
Bhardwaj:2023ayw,
Damia:2023ses,
Argurio:2023lwl,
Lawrie:2023tdz,
Copetti:2023mcq,
Honda:2024xmk,
Bah:2023ymy,
Chen:2023czk,
Choi:2023pdp,
Cordova:2023her,
Bhardwaj:2023bbf,
Honda:2024yte,
Sela:2024okz,
Cordova:2024ypu,
Arbalestrier:2024oqg,
Hasan:2024aow,
DelZotto:2024ngj,
Gagliano:2025gwr,
Choi:2024rjm,
Cordova:2025eim}, as well as \cite{Cordova:2022ruw,McGreevy:2022oyu,Freed:2022iao,Gomes:2023ahz,Schafer-Nameki:2023jdn,Brennan:2023mmt,Bhardwaj:2023kri,Shao:2023gho,Carqueville:2023jhb,Costa:2024wks} for reviews. In particular, non-invertible symmetry preserving RG flows that cannot lead to a trivially gapped theory in the IR have been found by Apte, C\'ordova and Lam \cite{Apte:2022xtu}---in such cases, the infrared quantum field is either gapless or the symmetry is spontaneously broken. A relation between these effects and anomalies of non-invertible symmetries was then pointed out \cite{Sun:2023xxv,Cordova:2023bja,Antinucci:2023ezl} (see also \cite{Antinucci:2024ltv}). The purpose of this paper is to study these effects systematically, thus providing a first coarse classification-type result. Along the way, we clarify certain aspects of the interplay between topological symmetry theories and topological orders, we discuss some first applications of our results, and we also encounter possible global forms of gauge theories that exhibit (non-invertible) 2-form symmetries arising by gauging (non-anomalous) non-invertible duality symmetries.

\medskip

In a nutshell, symmetry enforced gaplessness is an effect that occurs when a given theory has an anomalous global symmetry that cannot be realized by any topological field theory (TQFT) and is preserved along an RG flow and in the deep IR\cite{Cordova:2019bsd,
Cordova:2019jqi}. The resulting symmetry preserving IR phase must be non-trivial, for anomaly matching, and gapless, since the symmetry is not spontaneously broken and there is no TQFT that can match the anomaly \cite{Wang:2014lca,
Wang:2016gqj,
Sodemann:2016mib,
Wang:2017txt}.

\medskip

As a theoretical laboratory we imagine a UV conformal fixed point with a given finite 0-form symmetry and a relevant deformation. Along the resulting RG flow, the symmetry can be preserved or broken in the IR. If the symmetry is explicitly broken by the relevant deformation, this can still lead to interesting constraints building on a spurion-like analysis, but often one can find classes of relevant deformations that are symmetry preserving. If the symmetry is not explicitly broken it can be either preserved in the IR or spontaneously broken. In the former case, we can either have a gapped or a gapless symmetry preserving IR phase. The gapped quantum field can be either trivial (i.e.~an SPT) or non-trivial (i.e.~a topological order). The gapless case is often organized by a CFT (that can also combine with an SPT or a topological order, giving rise to gapless versions of both). Alternatively, the symmetry can be preserved by the RG flow, but spontaneously broken in the IR. In this latter case, the symmetries become domain walls organizing a family of vacua that can be gapped or gapless as well. Exactly the same pattern arises if one starts with a conformal fixed point with a non-invertible KW-like duality symmetry and considers a relevant deformation. The non-invertible symmetry behave very much like an ordinary 0-form symmetry.\footnote{\, Naively, non-invertible symmetries are associated to topological defects. If these are not explicitly broken by the relevant deformation (thus becoming endable or screened), they should not be sensitive to scales -- being topological, we can stretch them at no energy cost.} There are some key differences related to the fact that non-invertible symmetries have a higher structure \cite{Copetti:2023mcq,DelZotto:2024ngj} (see also \cite{Bhardwaj:2024xcx}), and therefore are more constraining with respect to RG flows (see e.g.~\cite{Cordova:2022ieu} and \cite{DelZotto:2024arv} for some examples). In particular, non-invertible KW-like duality symmetries can lead to vacua with different features when spontaneously broken \cite{Damia:2023ses}, which is to be expected, since their action maps twisted sectors into one another \cite{Cordova:2024iti,Choi:2024tri,Copetti:2024onh}. In this work we exploit the topological symmetry theory approach after \cite{Kapustin:2014gua,Ji:2019jhk,
Gaiotto:2020iye,
Apruzzi:2021nmk,
Freed:2022qnc,Kaidi:2022cpf} to study gapped phases resulting out of non-invertible symmetry preserving RG flows.

\medskip

This paper is organized as follows. The first two sections are review of results we need to proceed with our classification. In particular, in Section \ref{sec:symTFTreview} we present a review of the SymTFT construction of non-invertible $\mathbb G$-duality defects, and in Section \ref{sec:5dtopordo} we give a brief review of the result by Johnson-Freyd and Yu on the classification of 5d topological orders \cite{Johnson-Freyd:2020usu,Johnson-Freyd:2021tbq} (building on results by Lan, Kong, and Wen \cite{Lan:2018vjb,Kong:2020jne}). Towards the end of the Section, we comment on the fact that the Witt equivalence classes relevant for the classification of 5d topological orders are too coarse to characterize the natural equivalences of topological symmetry theories, which would correspond to a first step towards the classification of symmetry structures for 3+1d QFTs.\footnote{\, This is the purpose of a future work (currently in preparation).} In Section \ref{sec:core} we outline the main question of our study, after \cite{Apte:2022xtu,Cordova:2023bja}. We start from a 3+1 dimensional QFT with a 1-form symmetry (possibly anomalous) and map it to a bulk boundary system, where the bulk is a 4+1 topological order, i.e. a BF gauge theory with finite gauge group $\mathbb A$. The problem of characterizing topological IR phases for these systems reduces \textit{coarsely}\footnote{\, By this we mean that we are neglecting possible interplay with the spin structure, torsional homology, and we are working up to stacking of SPTs\label{foot:coarse}.} to the problem of classifying invariant Lagrangian subgroups for the lattice $\mathbb A \times \mathbb A^\vee$. We approach this problem systematically, and present our findings in Section \ref{sec:classificat}. The result is extremely technical, and its derivation, which is a (hard!) exercise in Galois theory is detailed in several (long and technical) Appendices. In Section \ref{sec:RGflows} we present some first applications of our techniques in the context of symmetry preserving RG flows, explaining how to exploit our Theorem to classify the IR gapped phases of 3+1 dimensional field theories with non-invertible symmetries. A further application of our techniques is that we can reproduce also results about the SSB of non-invertible symmetries previously outlined in \cite{Damia:2023ses} from a topological symmetry theory perspective, including the possibility of non-invertible symmetry preserving but supersymmetry breaking RG flows, as an interesting further application. We leave a detailed analysis of the resulting systems to future studies. An especially interesting insight is that there are more global forms for field theories with gaugeable non-invertible duality symmetries. Gauging them, one expects to obtain theories that do not have Lagrangian descriptions, as the corresponding multiples are realized by mutually non-local excitations, somewhat similar to what happens at Argyres-Douglas fixed points of $\mathcal N=2$ theories \cite{Argyres:1995jj}. In this study, however, we neglect several subtleties and we only achieve a coarse classification. It would be interesting to develop it further, and perhaps systematically establish or rule out this possibility. It is well-known, for example, that EM dualities in Maxwell theory have anomalies \cite{Hsieh:2019iba} that would obstruct such gaugings.\footnote{\, In fact, there might be a non-anomalous $\mathbb Z_2$ subgroup leading to non-invertible duality defects after \cite{Cordova:2023ent} which one could use to realize a phantomatic non-Lagrangian electromagnetism (whatever that is).}

\section{SymTFT and non-invertible topological defects}\label{sec:symTFTreview}
In this section, to fix the notations and conventions we will use in this paper we begin with a review of the construction of duality defects, exploiting the topological symmetry theory approach after Freed and Teleman \cite{Freed:2018cec} and Kaidi, Ohmori and Zheng \cite{Kaidi:2022cpf} --- for more about the topological symmetry theory see e.g.~\cite{Gaiotto:2014kfa,Ji:2019jhk,Gaiotto:2020iye,Apruzzi:2021nmk,Freed:2022qnc}

\medskip

\begin{figure}
  \centering
  \begin{tikzpicture}
    \shadedraw [shading=axis,top color =red, bottom color=white, white] (0,0) rectangle (5,3);
    \draw [red] (0,3) -- (5,3);
    \draw [very thick] (0,0) -- (0,3);
    \draw [very thick] (5,0) -- (5,3);
    \node[below] at (0,-0.1){$ \mathcal B$};
    \node[below] at (5,-0.1){$\widehat{\mathcal T}$};
    \node[above] at (2.5,1.2){$\mathcal F$};
    \draw [very thick] (10,0) -- (10,3);
    \draw [-stealth, thick] (6,1.5) -- (9,1.5);
    \node[above] at (7.5,1.5){contract};
    \node[below] at (7.5,1.5){$\simeq$};
    \node[below] at (10,-0.1){$\mathcal{T}_{\mathcal B}$};
  \end{tikzpicture}

  \caption{Topological symmetry theory. The $d$-dimensional field theory $\mathcal T_{\mathcal B}$ is realized via an isomorphism with a $(d+1)$-dimensional topological field theory $\mathcal F$ on a $d$-strip. On the left side of the $d$-strip, we have a $d$-dimensional topological boundary $\mathcal B$ condition of $\mathcal F$, on the right side of the $d$-strip we have a $d$-dimensional $\mathcal F$-relative theory $\widehat{\mathcal T} $. Since the whole $d$-strip construction is topological, we can stack the two boundaries contracting the interval, thus obtaining the desired isomorphism leading to $\mathcal T_{\mathcal B}$.} \label{fig:sandwich}
  \end{figure}
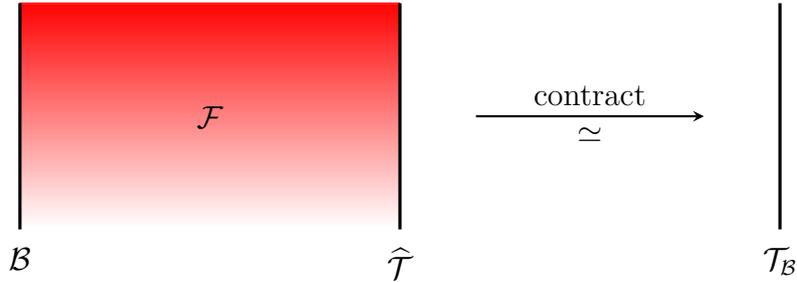

A useful tool to characterize the topological defects of a given $d$-dimensional QFT $\mathcal T$ is to exhibit an 
isomorphism of $\mathcal T$ with a bulk-boundary system in one higher dimension where the bulk is a TQFT --- see Figure \ref{fig:sandwich}. This is the so called \textit{topological symmetry theory} construction. Recall that a $(d+1)$-dimensional (extended) TQFT $\mathcal F$ is a symmetric monoidal functor
\be
\mathcal F : \text{Bord}^S_{d+1} \to \mathscr F
\ee
between the higher category of bordisms $\text{Bord}^S_{d+1}$ (possibly with additional structure $S$) and a suitable target  higher category $\mathscr F$.\footnote{\, Our readers that are not familiar with these concepts can consult e.g.~\cite{Freed} or \cite{Baez:1995xq,Lurie:2009keu} for the original papers.} The key idea to keep in mind is that the higher category of topological defects in an extended TQFT corresponds to the image of the bordism category under the TQFT functor $\mathcal F$. In other words,
\begin{itemize}
\item The target category $\mathscr F$ encodes the higher algebraic data assigned to defects;
\item The TQFT functor $\mathcal F$ maps geometric bordisms (which can be interpreted as networks of defects) into $\mathscr F$, assigning to each piece a corresponding higher algebraic object (like a vector space, category, or higher structure).
\end{itemize}
The topological symmetry theory construction \textit{à la} Freed, Moore and Teleman makes use of the above mathematical framework to characterize the generalized symmetries of a $d$-dimensional QFT $\mathcal T$, in terms of the higher symmetry category of its topological defects. One needs the following ingredients:
\begin{itemize}
\item[\textit{S1.)}] A  $(d+1)$-dimensional (extended) topological theory $\mathcal F$ as above;
\item[\textit{S2.)}]  A topological interface $\mathcal B$ between $\mathcal F$ and the trivial theory $\mathbf{1}_{d+1}$ (i.e.~a gapped topological $d$-dimensional boundary condition);
\item[\textit{S3.)}]  A $d$-dimensional field theory $\widehat{\mathcal T}$  relative to $\mathcal F$;\footnote{\, Our readers can consult, e.g. \cite{Freed:2012bs} to learn about relative field theories.} 
\item[\textit{S4.)}]  An isomorphism $\mathcal T \simeq \mathcal B \otimes_{\mathcal F} \widehat{\mathcal T}$.
\end{itemize}
The \textit{S4.)} isomorphism is realized as follows. Consider placing $\mathcal F$ on a slab $M_{d+1} = [-L,0] \times M_d$, where $M_d$ is a $d$-dimensional spacetime. At $\{-L\} \times M_d$ we insert $\mathcal B$, while on the other side of the slab, at $\{0\} \times M_d$, we insert $\widehat{\mathcal T}$ --- see Figure \ref{fig:sandwich}. Then the isomorphism $\mathcal T \simeq \mathcal B \otimes_{\mathcal F} \widehat{\mathcal T}$ is obtained by sending $L \to 0$: due to the topological nature of $\mathcal B$ and $\mathcal F$ the resulting system is equivalent to a $d$-dimensional field theory.
The TQFT $\mathcal F$ together with the boundary condition $\mathcal B$ forms a \textit{symmetry quiche} $(\mathcal F,\mathcal B)$. To a quiche corresponds the higher category $\text{End}_{\mathcal F}(\mathcal B)$, that encodes the topological defects that can give rise to generalized symmetries of $\mathcal T$ via the isomorphism $\mathcal T \simeq \mathcal B \otimes_{\mathcal F} \widehat{\mathcal T}$ \cite{Freed:2022qnc} (see also \cite{Bhardwaj:2023ayw,Bartsch:2023wvv,Bartsch:2023pzl}). We stress that many inequivalent quiches $(\mathcal F,\mathcal B)$ can give rise to the same $\text{End}_{\mathcal F}(\mathcal B)$ -- we will see examples below.

\medskip

We assume that the theory $\mathcal T$ is indecomposable, meaning that its correlators do not split into disconnected sums. This assumption implies in particular that $\text{End}_{\mathcal F}(\mathcal B)$ does not contain topological points (which would result in a decomposition of $\mathcal T$ into universes).

\medskip

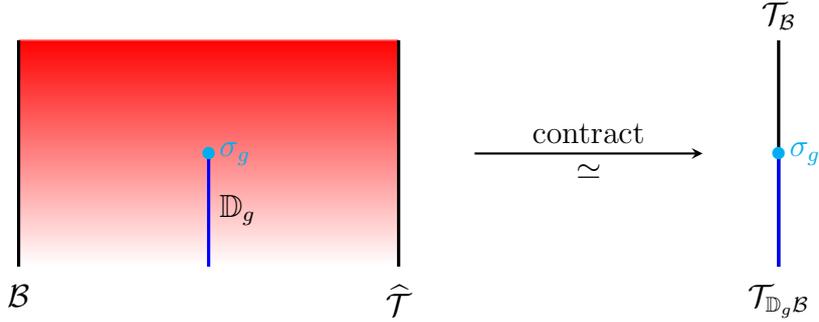
\begin{figure}

\begin{center}
  \begin{tikzpicture}
    \shadedraw [shading=axis,top color =red, bottom color=white, white] (0,0) rectangle (5,3);
    \draw [red] (0,3) -- (5,3);
    \draw [very thick] (0,0) -- (0,3);
    \draw [very thick] (5,0) -- (5,3);
    \draw [very thick,color = blue] (2.5,0) -- (2.5,1.5);
    \node[below] at (0,-0.1){$ \mathcal B$};
    \node[below] at (5,-0.1){$\widehat{\mathcal T}$};
    \node[right] at (2.5,0.75){$\mathbb D_g$};
    \node at (2.5,1.5){\textcolor{cyan}{$\bullet$}};
    \node[right] at (2.5,1.5){\textcolor{cyan}{$\sigma_g$}};
    \draw [very thick] (10,0) -- (10,3);
    \draw [-stealth, thick] (6,1.5) -- (9,1.5);
    \node[above] at (7.5,1.5){contract};
    \node[below] at (7.5,1.5){$\simeq$};
    \node[below] at (10,-0.1){$\mathcal{T}_{\mathbb D_g\mathcal B}$};
    \node[right] at (10,1.5){\textcolor{cyan}{$\sigma_g$}};
    \draw [very thick,color = blue] (10,0) -- (10,1.5);
    \node at (10,1.5){\textcolor{cyan}{$\bullet$}};
    \node[above] at (10,3){$\mathcal{T}_{\mathcal B}$};
  \end{tikzpicture}
\end{center}
  \caption{The boundary changing interfaces. If the theory $\mathcal F$ has a 0-form symmetry group $\mathbb G$, we can represent the latter using topological codimension 1 defects $\mathbb D_g$ for $g\in \mathbb G$. These can transform the boundaries non-trivially and if these can end along a codimension 2 defect $\sigma_g$, this results into a collection of interfaces among the $d$-dimensional QFTs $\mathcal T_{\mathbb D_g\mathcal B}$ for $g\in \mathbb G$.} \label{fig:duality_interfaces}
\end{figure}

\begin{figure}
\begin{center}
  \begin{tikzpicture}
    \shadedraw [shading=axis,top color =red, bottom color=white, white] (0,0) rectangle (5,3);
    \draw [red] (0,3) -- (5,3);
    \draw [very thick] (0,0) -- (0,3);
    \draw [very thick] (5,0) -- (5,3);
    \draw [very thick,color = blue] (2.5,0) -- (2.5,1.5);
    \node[below] at (0,-0.1){$ \mathcal B$};
    \node[below] at (5,-0.1){$\widehat{\mathcal T}$};
    \node[right] at (2.5,0.75){$\mathbb D_g$};
    \node at (2.5,1.5){\textcolor{cyan}{$\bullet$}};
    \node[right] at (2.5,1.5){\textcolor{cyan}{$\sigma_g$}};
    \draw [very thick] (10,0) -- (10,3);
    \draw [-stealth, thick] (6,1.5) -- (9,1.5);
    \node[above] at (7.5,1.5){contract};
    \node[below] at (7.5,1.5){$\simeq$};
    \node[below] at (10,-0.1){$\mathsf{S}_{g^{-1}}: \mathcal{T}_{\mathbb D_g\mathcal B} \simeq \mathcal{T}_{\mathcal B}$};
    \node[right] at (10,1.5){\textcolor{cyan}{$\sigma_g$}};
    \node at (10,1.5){\textcolor{cyan}{$\bullet$}};
    \node[above] at (10,3){$ \mathcal{T}_{\mathcal B}$};
  \end{tikzpicture}
  \end{center}
  \caption{The $\mathbb G$-duality symmetry. When $\mathbb D_g \widehat{\mathcal T} \simeq \widehat{\mathcal T}$, this induces a non-trivial duality isomorphism $\mathsf{S}_{g^{-1}}:  \mathcal{T}_{\mathbb D_g\mathcal B} \simeq \mathcal{T}_{\mathcal B}$. The composition of $\sigma_g$ with $\mathsf{S}_{g^{-1}}$ gives rise to a non-invertible $\mathbb G^{(0)}$-duality symmetry $\mathcal{N}_g = \sigma_g \circ \mathsf{S}_{g^{{-1}}}$ with $g \in \mathbb G^{(0)}$.} \label{fig:duality_symmetries}
\end{figure}
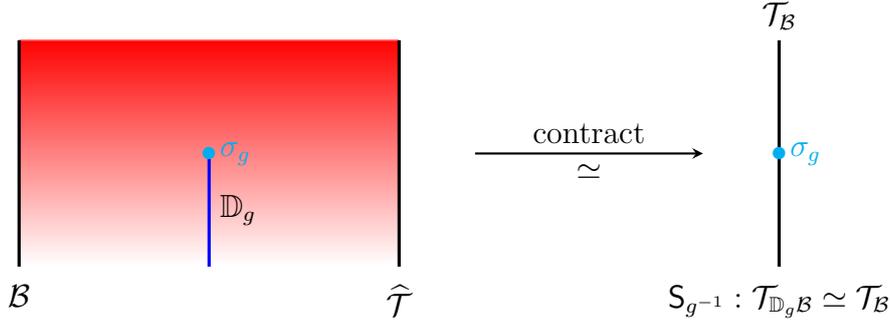

One of the application of the topological symmetry theory is to streamline the construction of non-invertible $\mathbb G$-duality defects, for $\mathbb G$ a finite group, not necessarily abelian. In a general dimension, one can proceed as follows. First of all one should identify the generalized zero-form symmetry of the theory $\mathcal F$, given by its topological defects of codimension one, i.e. $\text{End}(\mathcal F)$, the topological interfaces from $\mathcal F$ to itself. An interesting class of generalized $\mathbb G$-duality defects can be constructed provided we can identify a subcategory of $\text{End}(\mathcal F)$ which is generated by a collection of codimension 1 topological defects $\mathbb D_g(\Sigma)$, labeled by an element $g \in \mathbb G$ and a (possibly stratified) submanifold $\Sigma$ of the $(d+1)$-dimensional bulk, that moreover have the following properties:
  \begin{enumerate}
  \item \textbf{Group grading.}  The fusion of the defects $\mathbb D_g(\Sigma)$ has a selection rule
  \be
  \text{Hom}(\mathbb D_g(\Sigma) \otimes \mathbb D_{h}(\Sigma),\mathbb D_{k}(\Sigma) ) \neq 0 \qquad \Longleftrightarrow \qquad gh = k
  \ee
  \item \textbf{Endability.} The defects all have gapped boundary conditions $\sigma_g$ and $\sigma_g^\vee$ such that
  \be\label{eq:endable}
  \sigma_g(\partial \Sigma) \in \text{Hom}(\mathbb D_g(\Sigma),\mathbf{1}_{\mathcal F}(\Sigma'))  \qquad   \sigma_g^\vee(\partial \Sigma) \in  \text{Hom}(\mathbf{1}_{\mathcal F}(\Sigma'),\mathbb D_g(\Sigma))
  \ee
 where $\Sigma$ and $\Sigma'$ are glued along $\partial \Sigma$ and $\mathbf{1}_{\mathcal F}$ is the transparent interface in $\text{End}(\mathcal F)$. Usually (and often in what follows) the support of the defects is omitted in the notation above, and one would write
\be
\sigma_g \in \text{Hom}(\mathbb D_g,\mathbf{1}_{\mathcal F})  \qquad   \sigma_g^\vee \in  \text{Hom}(\mathbf{1}_{\mathcal F},\mathbb D_g)
\ee 
instead. We will typically adopt this convention in what follows. $\sigma_g$ and $\sigma_g^\vee$ are codimension 2 topological interfaces known as \textit{twist defects} \cite{Barkeshli:2014cna,Teo:2015xla,Kaidi:2022cpf}. We stress here that the fusion of the twist defects is generically non-invertible, meaning that typically 
 \be
 \sigma_g \otimes \sigma_{g}^\vee \simeq \mathcal C_g\,,
 \ee
 where $\mathcal C_g \in \text{End}(\mathbf{1}_{\mathcal F})$ is an element that is not necessarily the transparent interface;
 \item \textbf{Non-trivial boundary action.} The defects $\mathbb D_g$ act \textit{non-trivially} on the topological boundary, via a vertical stacking: 
 \be\mathcal B \neq \mathbb D_g \mathcal B\,.
 \ee
  \end{enumerate}
If $\text{End}(\mathcal F)$ contains such a family, we can exploit the topological symmetry isomorphism to construct \textit{families} of $d$-dimensional field theories  
\be\{\mathcal T_{\mathbb D_g \mathcal B}\}_{g \in \mathbb G},
\ee
as well as topological interfaces interconnecting them (given by the images of $\sigma_g$ along the \textit{S4.)} isomorphism)  -- see Figure \ref{fig:duality_interfaces}. We stress that the latter are not necessarily invertible by construction, and hence do not establish equivalences among these theories. Schematically, we obtain interfaces
\be
\sigma_g^\vee \otimes \sigma_h : \mathcal T_{\mathbb D_h \mathcal B} \to \mathcal T_{\mathbb D_g \mathcal B} \qquad \sigma_h^\vee \otimes \sigma_g : \mathcal T_{\mathbb D_g \mathcal B} \to \mathcal T_{\mathbb D_h \mathcal B}
\ee
Now, \textit{assume} that the relative field theory $\widehat{\mathcal T}$ is left invariant by the action of all $\mathbb D_g$ by vertical stacking
\be
\mathbb D_g \widehat{\mathcal T} \simeq \widehat{\mathcal T}
\ee
for all $g\in \mathbb G$. If this is the case, via the isomorphism \textit{S4.)}, \textit{duality isomorphisms} are induced, parameterized by elements of $\mathbb G$:
\be
\mathsf{S}_g : \mathcal T_{\mathbb D_{h} \mathcal B} \longrightarrow \mathcal T_{\mathbb D_{gh} \mathcal B} \qquad g,h \in \mathbb G\,.
\ee
Exploiting these isomorphisms, a family of non-invertible $\mathbb G$-duality symmetries is obtained by composition
\be
\mathcal N_g :  \mathcal T_{\mathcal B} \xrightarrow{\quad\sigma_g\quad} \mathcal T_{\mathbb D_g \mathcal B}\xrightarrow{\quad\mathsf{S}_{g^{-1}}\quad}   \mathcal T_{\mathcal B} 
\ee
 as described in Figure \ref{fig:duality_symmetries}. The resulting defect is a zero-form symmetry of $\mathcal T_{\mathcal B}$, indeed $\mathcal N_g \in \text{End}(\mathcal T_{\mathcal B})$ by construction. Moreover, we can use the monoidal structure of $\mathcal F$ to evaluate the fusion rules, corresponding to the higher morphisms:
\be
 \begin{aligned}
& \mathcal N_g^{\,\vee} \otimes  \mathcal N_g \simeq \sigma^\vee_g \otimes \sigma_g \in \text{End}(\mathbf{1}_{\mathcal F})\,,\\
&\mathcal N_h^{\,\vee} \otimes  \mathcal N_g \simeq \sigma^\vee_h \otimes \sigma_g \in \text{Hom}(\mathbb D_h,\mathbb D_g)\,,\\
\end{aligned}
\ee
 
 \medskip

Notice that on the LHS the fusion is evaluated using the monoidal structure of topological defects of $\mathcal T_{\mathcal B}$, while on the RHS the fusion is evaluated using the monoidal structure of $\mathcal F$, the two are related by the \textit{S4.)} isomorphism.

\medskip

We stress here that from the perspective of the bulk topological symmetry theory $\mathcal F$, the defects $\mathcal N_g$ do not arise from \textit{genuine} operators. In the terminology introduced by Kapustin and Seiberg in 
\cite{Kapustin:2014gua} non-genuine operators are operators that can only arise along the boundary of other operators, which live in twisted sectors of the theory. Indeed, we see that $\mathcal N_g$ arise at the topological boundary of the defects $\mathbb D_g$. One can then ask whether there is a topological symmetry theory that has these non-invertible defects as genuine topological operators. Kaidi, Ohmori and Zheng answered this question by proposing a construction of a topological symmetry theory where $\mathcal N_g$ arise from genuine topological defects. This is possible whenever there is an orbifold datum corresponding to the family of $\mathbb D_g$ operators  
\cite{Carqueville:2017aoe}, meaning that we can gauge them away. A new bulk TQFT is obtained by orbifolding which we denote $\mathcal F_{\mathbb G}$ in what follows --- see e.g.~the work by Carqueville, Runkel and Schaumann for a more explicit characterization of the orbifold data \cite{Carqueville:2017aoe}.
If the theory $\mathcal F_{\mathbb G}$ is equivalent to a (generalized) Dijkgraaf-Witten theory \cite{Dijkgraaf:1989pz}, the topological defects $\mathcal N_g$ are \textit{grouplike} or not intrinsically non-invertible \cite{Kaidi:2022uux,Kaidi:2022cpf,Sun:2023xxv}.

\medskip

For technical purposes, it is often easier to work with $\mathcal F$ and take the quotient \textit{a posteriori}. However, when discussing the spontaneous symmetry breaking of the non-invertible $\mathbb G$-duality symmetry, using  $\mathcal F_{\mathbb G}$ will be important to analyze the corresponding symmetry preserving RG flows. We will turn to this in Section \ref{sec:RGflows} below.

\section{4+1d topological orders: a lightning review}\label{sec:5dtopordo}

\subsection{Johnson-Freyd--Yu classification}\label{sec:JFY}

The starting point for the analysis we carry out in this work is the classification of topological orders in 4+1 dimensions by Johnson-Freyd and Yu \cite{Johnson-Freyd:2021tbq} (see also \cite{Johnson-Freyd:2020usu}), building on works by Lan, Kong and Wen \cite{Lan:2018vjb,Kong:2020jne}. In this section we present a quick review of these results following closely  \cite{Johnson-Freyd:2021tbq}.

\medskip

In general, a $(d+1)$-dimensional topological order \cite{Lan:2018vjb,Kong:2020jne} is a multifusion $d$-category $\mathcal F$ with trivial center \cite{Johnson-Freyd:2020usu}.\footnote{\, Here we are slightly abusing the notation we introduced in the previous section, labeling with the same letter the TQFT $\mathcal F$ and the higher category of its topological defects of various codimensions (encoded in the image of $\mathcal F$ in $\mathscr F$).} A multifusion $d$-category is the mathematical structure that organizes finite collections of topological defects of codimension $\geq1$ in a field theory in $(d+1)$-dimensions.

This definition needs to be adapted to the possible additional structures within the morphism spaces. The requirement that $\mathcal F$ has a trivial center is an axiomatization of the principle of remote detectability \cite{Lan:2018vjb}, which in a sense can be translated in the fact that the topological order $\mathcal F$ corresponds to the collection of topological defects of a \textit{genuine} $(d+1)$-dimensional topological field theory (as opposed to a \textit{relative} one \cite{Freed:2012bs}) considered up to stacking by invertible phases (retaining the information of the higher structure) \cite{Johnson-Freyd:2020usu}.\footnote{\, More about this in Section \ref{sec:TOornotTO}.} This equivalence is proven in the case of framed TQFTs. From a physics perspective, it is natural to expect that this correspondence can be extended to the case of TQFTs with different structures, where the information about the structure of the TQFT is hidden in the enrichment of the morphism spaces of the multifusion $d$-category --- see \cite{Ferrer:2024vpn} for more details. In the work by Yu and Johnson-Freyd, the topological orders in 4+1 dimensions are classified up to Witt equivalence. Two topological orders $\mathcal F_A$ and $\mathcal F_B$ are Witt equivalent if there is a gapped topological interface $\mathcal I_{AB}$  between them. 
The existence of one such interface is used to establish an equivalence between the topological orders $\mathcal F_A$ and $\mathcal F_B$: starting from $\mathcal F_A$, one can bubble out a sphere of $\mathcal I_{AB}$ with an $\mathcal F_B$ interior and expand it, smoothly transitioning from $\mathcal F_A$ to $\mathcal F_B$. 

\medskip

It is useful to introduce the following four operations on a generic $d$-fusion category $\mathscr A$.
\begin{itemize}
\item The \textbf{Karoubi completion} of $\mathscr A$ is an operation that from $\mathscr A$ forms a new category $\text{Kar}(\mathscr A)$, which is the category that contains all condensates formed via all the allowed higher gaugings/orbifoldings of its topological defects \cite{Gaiotto:2019xmp}. Typically, the collection of topological defects of a given QFT is Karoubi complete. In particular, $\mathcal F$ is Karoubi complete;
\item The \textbf{delooping} of $\mathscr A$, is an operation that extracts from $\mathscr A$ its genuine topological defects of dimension $d-1$, which form a fusion $(d-1)$-category. The latter is obtained as the endomorphisms of the tensor unit of $\mathscr A$ (which is the transparent interface of dimension $d$) and denoted\be \Omega \mathscr A := \text{End}_{\mathscr A}(1_{\mathscr A})\,.\ee This operation can be iterated, and exploited to isolate subcategories of genuine operators of the desired codimension. In general, $\Omega^\ell \mathscr A$ corresponds to the genuine topological defects with support of dimension $d-\ell$. We stress that if $\mathscr A$ is Karoubi complete so are its deloopings.
\item The \textbf{looping} of $\mathscr A$, is an operation that realizes $\mathscr A$ as an $(n+1)$-fusion category $B\mathscr A$, obtained by embedding the same topological defects in one dimension higher. In particular, $\mathscr A \simeq \text{End}_{B\mathscr A}(\mathbf 1_{B\mathscr A})$. Typically, $B\mathscr A$ is not Karoubi complete since additional higher orbifolds are typically possible when increasing the dimensionality of spacetime. For this reason one introduces the following operation.
\item The \textbf{suspension} of $\mathscr A$ is the Karoubi completion of its looping $\Sigma \mathscr A = \text{Kar}(B\mathscr A)$.
\end{itemize}

Let us fix $d=4$ in the rest of this section. Since we are interested in irreducible models, we require that the topological defects of $\mathcal F$ form a \textit{fusion} $4$-category, in particular $\mathcal F$ does not have topological point operators and $\Omega^{4}\mathcal F \simeq \mathbb C$ --- in other words, the transparent interface is simple. Next, we consider the collection of topological lines of $\mathcal F$, which form the fusion 1-category $\Omega^{3}\mathcal F$. The subcategory of $\mathcal F$ generated by the topological lines is $\Sigma^3\Omega^{3}\mathcal F$, which consists of all the topological lines together with their condensates. In the fermionic case, Yu and Johnson-Freyd are able to construct a topological interface $\mathcal I$ which trivializes all of $\Sigma^3\Omega^{3}\mathcal F$, leading to a topological order $\widehat{\mathcal F}$ such that it only has the trivial topological defect line and a transparent fermion, namely it is such that $\Omega^{3}\widehat{\mathcal F }\simeq \Sigma \mathbb C \simeq \text{SVect}_\mathbb{C}$. The construction of the topological interface in question is interesting, albeit somewhat counterintuitive. It is based on a result by Deligne \cite{Deligne} which states that a super fiber functor $f: \Omega^3\mathcal F \to \text{SVect}_{\mathbb C}$ always exists and is unique up to isomorphism. The latter can be suspended to the categories resulting from condensates
\be
\Sigma^3f: \Sigma^3\Omega^3\mathcal F \to \Sigma^3\text{SVect}_{\mathbb C}
\ee
thus establishing a monoidal functor that trivializes the subcategory of $\mathcal F$ generated by its topological lines. The existence of such suspension means that one can always gauge away all the topological defect lines of a 4+1 dimensional (fermionic) topological order.  

The above functor can then be exploited to induce the interface
\be
\mathcal I \simeq \mathcal F \otimes_{\Sigma^3\Omega^3\mathcal F} \Sigma^3\text{SVect}_{\mathbb C}
\ee
which trivializes all the lines of $\mathcal F$. The physical interpretation of $\mathcal I$ is that it performs a half space gauging of the topological defect lines of $\mathcal F$. The instruction of how the latter are trivialized is encoded in $\Sigma^3f$. The definition of the topological order $\widehat{\mathcal F}$ is then
\be
\widehat{\mathcal F} := \text{End}_{\mathcal F}(\mathcal I)\,.
\ee
As a result, $\widehat{\mathcal F}$ and $\mathcal F$ are Witt equivalent. $\widehat{\mathcal F}$ has no topological points and no topological line defects, meaning that $\Omega^4 \widehat{\mathcal F} \simeq \mathbb C$ and $\Omega^3 \widehat{\mathcal F} \simeq \Sigma \mathbb C \simeq \text{SVect}_\mathbb{C}$. The latter consists of a trivial bosonic interface and a transparent fermionic line --- working on spin manifolds the transparent fermionic line trivializes.\footnote{\, In this work we will always assume this is the case.} Consider now the topological surface operators $\Omega^2 \widehat{\mathcal F}$. The final step in the argument of Johnson-Freyd and Yu is that $\widehat{\mathcal F}$ is such that all of its topological membranes of dimension $3$ and $4$ arise from condensations (or higher gaugings) of its topological surface operators \cite{Carqueville:2017aoe,Gaiotto:2019xmp,Roumpedakis:2022aik}, namely that
\be\label{eq:mitico}
\widehat{\mathcal F} \simeq \Sigma^2 \Omega^2 \widehat{\mathcal F}\,.
\ee
If a topological order does not have any non-trivial topological points and line operators, then clearly it has to be determined by its operators of codimension $\geq 2$. However, equation \eqref{eq:mitico} is a much deeper statement, indeed it implies that all higher dimensional topological defects in this case are condensates of the topological defects of codimension 2. This result was conjectured by Lan, Kong and Wen \cite{Lan:2018vjb,Kong:2020jne} and proved by Johnson-Freyd (Theorem 5 of \cite{Johnson-Freyd:2020usu}). This implies that the category of topological surfaces of $\widehat{\mathcal F}$, together with its suspension and condensations, is enough to characterize the topological order up to Witt equivalence.\footnote{\, This is related to the fact that working with topological orders we are considering TQFTs up to stacking with invertible ones.} 

Let us proceed and characterize $\Omega^2   \widehat{\mathcal F}$ and its suspension. Recall that $\Omega^2   \widehat{\mathcal F}$ is (by construction) a fusion 2-category \cite{Douglas:2018qfz}. We can think of its objects as topological surface defects and its morphisms as interfaces, topological lines. An object in a fusion 2-category is indecomposable if it is not a direct sum of other objects. Since $\Omega^3 \widehat{\mathcal F} \simeq \text{SVect}_\mathbb{C}$, the category  $\Omega^2   \widehat{\mathcal F}$ does not contain any topological lines other than the identity and a transparent fermionic line. This simplifies greatly its structure. In particular, \textit{all such fusion 2-categories are grouplike} \cite{Johnson-Freyd:2020ivj} and \textit{sylleptic}, meaning that 
\begin{enumerate}
\item All indecomposables are simple;
\item All indecomposables are invertible;
\item The components of the category (i.e.~the equivalence classes of indecomposables) form an abelian group 
\be
\mathbb A := \pi_0(\Omega^2\widehat{\mathcal F})\,.
\ee
\item The topological surfaces have three transverse directions once embedded in 4+1 dimensions, which gives to these categories a syllepsis, which our readers can think of as an antisymmetric braiding structure (possibly together with a discrete torsion refinement) at a naive level.\footnote{\, A more detailed and explicit description of the structure of surface defects can be found below.}
\end{enumerate}
In general, one expects that a $(d+1)$-dimensional TQFT that has (only) grouplike $p$-dimensional topological defects labeled by elements of a group $\mathbb A$, that moreover have a total of $q$ transverse directions (hence are such that $p + q = d+1$) are classified by $H^{(d+2)}(B^q \mathbb A,U(1))$. In our case, therefore the resulting fusion 2-categories are classified by
\be\label{eq:JFYclass}
H^6(B^3\mathbb A;U(1)) \simeq \text{Hom}(\mathbb Z_2,\mathbb A) \oplus  \text{Hom}(\mathbb A \wedge \mathbb A,U(1)) 
\ee
where the first factor is torsional and trivializes working on Spin manifolds as we do in this work, while the second factor captures the alternating form encoding the braiding of the indecomposable topological surfaces.

\medskip

From a physics perspective, the only theories with these features are the 4+1 dimensional finite gauge theories with gauge group $\mathbb A$, also known as \textit{BF theories} -- see e.g.~\cite{Kapustin:2014gua} for more details about these models. These all admit Dirichlet boundary conditions, which provide a topological interface to the trivial TO (in the fermionic case). Hence, there only one fermionic topological order up to Witt equivalence.

\subsection{A comment on classification of topological symmetry theories}\label{sec:TOornotTO}

To connect with Section \ref{sec:symTFTreview}, the BF theories we have encountered at the end of the previous section are well-known examples of SymTFT for 3+1 dimensional QFTs with finite 1-form symmetry groups --- see e.g.~\cite{Kapustin:2014gua,Gaiotto:2014kfa} for some first examples. More about this connection is reviewed below, as it is important for our study. Thinking about SymTFTs, one can ask whether the classification of topological orders up to Witt equivalence that we have outlined above has a further application in that context. Clearly this is not true: if the Witt equivalences were understood \textit{verbatim} as equivalences of SymTFTs, they would all be trivial, which is clearly false.

\medskip

Of course if $\mathcal F_A$ and $\mathcal F_B$ are TQFTs that can be connected by a topological interface $\mathcal I_{AB}$ we can map a quiche
$(\mathcal F_A,\mathcal B_A)$ to the quiche $(\mathcal F_B,\mathcal B_B)$ where
\be
\mathcal B_B \simeq \mathcal B_A \otimes_{\mathcal F_A} \mathcal I_{AB}\,.
\ee
Considering the orientation reversal, we also obtain an opposite topological interface between $\mathcal F_B$ and $\mathcal F_A$, which we denote $\mathcal I_{BA}$. Without further assumptions,
\be
\mathcal I_{BA} \otimes_{\mathcal F_A} \mathcal I_{AB} \in \text{End}(\mathcal F_{B}) \quad\text{and}\quad  \mathcal I_{AB} \otimes_{\mathcal F_B} \mathcal I_{BA} \in \text{End}(\mathcal F_{A})\,
\ee
are endomorphisms that are \textit{not} proportional to the identity. For example, half space gauging interfaces typically compose to non-trivial condensation defects. If we restrict to \textit{invertible} topological interfaces, we then have the stronger conditions
\be\label{eq:invertible}
\mathcal I_{BA} \otimes_{\mathcal F_A} \mathcal I_{AB} \simeq \textbf{1}_{\mathcal F_B} \qquad \mathcal I_{AB} \otimes_{\mathcal F_B} \mathcal I_{BA} \simeq \textbf{1}_{\mathcal F_A}\,.
\ee
In the invertible case, considering the $\textit{S4.)}$ isomorphisms we have that
\be\label{eq:equivalencesinverto}
\begin{aligned}
\mathcal T_{\mathcal{B}_A} &\simeq \mathcal B_{A} \otimes_{\mathcal F_{A}} \widehat{\mathcal T}\\
&\simeq \mathcal B_{A} \otimes_{\mathcal F_{A}} \mathcal I_{AB} \otimes_{\mathcal F_{B}}  \mathcal I_{BA} \otimes_{\mathcal F_{A}} \widehat{\mathcal T} \\
&\simeq \mathcal B_{B} \otimes_{\mathcal F_{B}} \widehat{\mathcal T}'
\end{aligned}
\ee
where we have defined
\be
\widehat{\mathcal T}' =  \mathcal I_{BA} \otimes_{\mathcal F_A} \widehat{\mathcal T}.
\ee
Therefore, for invertible interfaces, we obtain equivalent bulk-boundary systems giving rise to different equivalent realizations of the \textit{S4.)} isomorphisms.

\medskip

This means that thinking of topological orders (understood in \cite{Johnson-Freyd:2020usu} as $(d+1)$-TQFTs up to stacking of invertible $(d+1)$-TQFTs) only in terms of their Witt equivalence classes is insufficient for the classification of SymTFTs. Witt equivalences are too ``generous'', since non-invertible interfaces are allowed. Our discussion above suggests that \textit{invertible} topological interfaces (dualities) are a better notion of equivalences of topological symmetry theories.

\section{QFTs with 1-form symmetries and their SymTFTs}\label{sec:core}
In this section we derive the structure of SymTFTs and isomorphisms relevant for 3+1 dimensional QFTs with generic anomalous invertible abelian 1-form symmetries, thus extending and systematizing previous results by \cite{Kaidi:2022cpf,Bashmakov:2022uek,Antinucci:2022cdi,Sun:2023xxv,Antinucci:2023ezl,Cordova:2023bja}. We show that indeed the resulting SymTFT is equivalent to a BF gauge theory with finite abelian gauge group $\mathbb A$, that might differ from the 1-form symmetry of the theory depending on the structure of the anomaly. The classification result we have reviewed above says that this is the most general SymTFT that does not have any topological lines. This is the context of our study. In order to exploit this theory to describe non-invertible symmetries, we have to gauge some of its topological interfaces thus producing a SymTFT with lines, which is crucial to characterize phases preserving the non-invertible symmetries.

\subsection{Determining the SymTFT: another angle on 4+1d TO}
\label{sec:determining-symtft}

In this section we give a natural class of SymTFT to characterize the 1-form symmetries of 3+1 dimensional QFTs $\mathcal T$. We will see that the latter are indeed equivalent to the topological orders we considered above, even in presence of anomalies.

\medskip

Suppose the theory $\mathcal T$ has one-form symmetry 
\be
\widehat{\mathbb{A}} ^{(1)} \cong \prod_{i=1}^r \mathbb{Z}_{n_i}^{(1)}\,.
\ee
An $\widehat{\mathbb{A}} ^{(1)}$-background $A$ is a collection of $\mathbb{Z}_{n_i}$ gauge fields, which we represent as flat $\mathrm{U}(1)$ two-form gauge fields $A_i$ (formally e.g.~Hopkins--Singer differential cocycles) such that $n_i A_i$ are pure gauge, or equivalently 
\be
\oint n_i A_i \in \mathbb{Z}\,.
\ee
The pure 't Hooft anomaly for $\widehat{\mathbb{A}} ^{(1)}$ captures the failure of invariance under gauge transformations of $A$ and is given by a five-dimensional topological inflow action $e^{2\pi i \int_{M_5} \alpha(A)}$ where $A$ is extended to $M_5$. The most general such action is of the form
\begin{equation}
  \alpha(A) = \sum_{ij} \alpha_{ij} A_i \dd{A_j} \quad \alpha_{ij} \in \mathbb{Z}_{\gcd(n_i, n_j)}.
\end{equation}
(in particular, there is no cubic anomaly for dimensional reasons).\footnote{\, Here $A_i \dd{A_j}$ is the usual physicist's shorthand for the product in differential cohomology. In particular, although $A_j$ is flat (``$\dd{A_j} = 0$''), $A_i \dd{A_j}$ is a possibly nontrivial flat five-form gauge field with $\oint A_i \dd{A_j} \in \qty(\frac{1}{n_i}\mathbb{Z} \cap \frac{1}{n_j}\mathbb{Z})\Big/\mathbb{Z} = \qty(\frac{1}{\gcd(n_i,n_j)}\mathbb{Z})\Big/\mathbb{Z}$.} As the pairing is antisymmetric (``integration by parts''), it suffices to consider terms with $i \le j$. When $i = j$, antisymmetry means that $\alpha_{ii}$ can be taken to be zero, or $n_i/2$ if $n_i$ is even. Hence we have
\begin{equation}
  \label{eq:inflow-action-alpha-beta}
  \alpha(A) = \sum_{i < j} \alpha_{ij} A_i \dd{A_j}
  + \sum_{\text{$n_i$ even}} \beta_i \frac{n_i}{2} A_i \dd{A_i},
  \quad \alpha_{ij} \in \mathbb{Z}_{\gcd(n_i, n_j)},\ \beta_i \in \mathbb{Z}_2
\end{equation}

While the above argument is somewhat imprecise,\footnote{\ In particular, the identification of $\mathbb{Z}_n$ gauge fields, properly elements of $H^2(M_5; \mathbb{Z}_n)$, with $n$-torsional flat gauge fields valued in $H^2(M_5; \mathbb{R}/\mathbb{Z})$ is not one-to-one in general. However, one can show that it is one-to-one if $H_1(M_5; \mathbb{Z})$ is free of $n$-torsion.} its legitimacy is supported by a rigorous mathematical argument. An inflow action $\alpha$ is a cohomology operation \cite{Kapustin:2014zva}
\be
\alpha\colon H^2(\mbox{--}; \widehat{\mathbb{A}} ^{(1)}) \to H^5(\mbox{--}; \mathbb{R}/\mathbb{Z})
\ee 
or equivalently a \textit{class} 
\be
\alpha \in H^5(B^2 \widehat{\mathbb{A}} ^{(1)}; \mathbb{R}/\mathbb{Z})\,.
\ee
As explained in Appendix \ref{app:H5B2G1}, this group is
\begin{equation}
  \label{eq:H5B2G1}
  H^5\qty\Big(B^2 \prod_i \mathbb{Z}_{n_i}; \mathbb{R}/\mathbb{Z})
  \cong \prod_{i < j} \mathbb{Z}_{\gcd(n_i,n_j)} \times \prod_{\text{$n_i$ even}} \mathbb{Z}_2.
\end{equation}
The coefficients $\alpha_{ij}$ and $\beta_i$ of our inflow action \eqref{eq:inflow-action-alpha-beta} exactly fit this general classification of pure anomalies for $\widehat{\mathbb{A}} ^{(1)}$. Comparing with the classification of topological orders in \eqref{eq:JFYclass}, we recognize that the terms $\alpha_{ij}$ are responsible to the braiding of the topological surfaces in the SymTFT, while the terms $\beta_i$ correspond to the torsional terms that trivialize on Spin manifolds.

\medskip

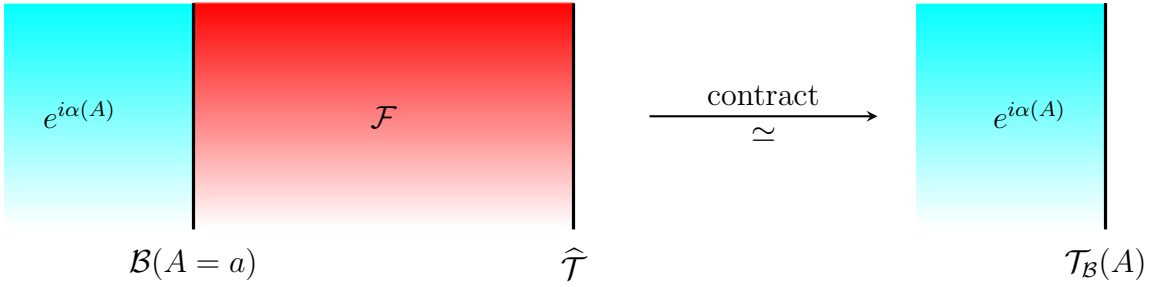
\begin{figure}
  \centering
  \begin{tikzpicture}
    \shadedraw [shading=axis,top color =red, bottom color=white, white] (0,0) rectangle (5,3);
     \shadedraw [shading=axis,top color =cyan, bottom color=white, white] (-2.5,0) rectangle (0,3);
    \draw [red] (0,3) -- (5,3);
    \draw [very thick] (0,0) -- (0,3);
    \draw [very thick] (5,0) -- (5,3);
    \node[below] at (0,-0.1){$ \mathcal B(A=a)$};
    \node[below] at (5,-0.1){$\widehat{\mathcal T}$};
    \node[above] at (2.5,1.2){$\mathcal F$};
     \node[above] at (-1.5,1.2){$e^{i\alpha(A)}$};
    \draw [-stealth, thick] (6,1.5) -- (9,1.5);
    \node[above] at (7.5,1.5){contract};
    \node[below] at (7.5,1.5){$\simeq$};
     \shadedraw [shading=axis,top color =cyan, bottom color=white, white] (9.5,0) rectangle (12,3);
     \node[above] at (11,1.2){$e^{i\alpha(A)}$};
     \draw [very thick] (12,0) -- (12,3);
    \node[below] at (12,-0.1){$\mathcal{T}_{\mathcal B}(A)$};
  \end{tikzpicture}

  \caption{SymTFT for anomalous 1-form symmetries.} \label{fig:sandwichanom}
  \end{figure}

To derive a SymTFT for such a setup, we follow \cite{Kaidi:2022cpf} and introduce dynamical $\mathrm{U}(1)$ gauge fields $a_i, \tilde{a}_i$ in the bulk, Dirichlet boundary conditions fixing $a = A$ on the topological boundary, and a bulk action (see Figure \ref{fig:sandwichanom})
\begin{equation}
  S(a, \tilde{a}) = \int_{M_5}\sum_i n_i a_i \dd{\tilde{a}_i} + \alpha(a)
\end{equation}
Since we are always working on Spin manifolds, all terms $\beta_i$ trivialize in the quantum theory and we can rewrite the SymTFT as
\begin{equation}
  \label{eq:symtft-cTQdc}
  S(a, \tilde{a}) = \int_{M_5}\frac{1}{2} c^T Q \dd{c}
\end{equation}
where $c = (a, \tilde{a})$ and $Q$ is an antisymmetric integer matrix.

\medskip

To deal with BF-type actions of the form \eqref{eq:symtft-cTQdc}, we recall \cite[Theorem~IV.1]{Newman1972}, which states that for any antisymmetric integer matrix $Q$, there is an invertible integer matrix $B$ (which may be found algorithmically) such that $B^T QB$ is block-diagonal with $2\times 2$ blocks $\mqty(0 & N_i \\ -N_i & 0)$, with $N_i$ the invariant factors of $Q$ (in particular $N_i \mid N_{i+1}$). The change of basis $c = Bb$ then expresses $S(a, \tilde{a})$ as a sum of BF terms
\begin{equation}
  S(a, \tilde{a}) = \sum_{i = 1}^r \int_{M_5} N_i b_i \dd{\tilde{b}_i}.
\end{equation}
It will prove useful to further decompose this action as a sum of BF terms where the $N_i$ are prime powers. To this end, we note that if $m$ and $n$ are coprime, a $\mathbb{Z}_{mn}$ BF theory 
\be
\int mn\, b\dd{\tilde{b}}
\ee 
can be expressed as a $\mathbb{Z}_m \times \mathbb{Z}_n$ BF theory by introducing auxiliary fields $e$, $\tilde{e}$ with action 
\be
\int_{M_5} e\dd{\tilde{e}}\,,
\ee 
\textit{i.e.} a trivial $\mathbb{Z}_1$ gauge theory. Since 
\be
\mqty(1 & 0 \\ 0 & mn)
\ee 
is the Smith normal form of 
\be
\mqty(m & 0 \\ 0 & n)
\ee
we can find \textit{unimodular} changes of basis $(e, b) \to (f, g)$ and $(\tilde{e}, \tilde{b}) \to (\tilde{f}, \tilde{g})$ such that
\begin{equation}
  e\dd{\tilde{e}} + mn\, b\dd{\tilde{b}} = m\, f\dd{\tilde{f}} + n\, g\dd{\tilde{g}}.
\end{equation}
Iterating this procedure, thanks to the Chinese Remainder Theorem we can establish the equivalence of this form of the SymTFT to a sum of $\mathbb{Z}_{p^k}$ BF gauge theories. This class of SymTFTs is the starting point for our analysis. 

\medskip

To summarize: \textit{for any given a 3+1 d QFT $\mathcal T$ with 1-form symmetry $\widehat{\mathbb{A}} ^{(1)}$ and 1-form symmetry anomaly $\alpha \in H^5(B^2\widehat{\mathbb{A}}^{(1)},\mathbb R/\mathbb Z)$, there is a corresponding SymTFT $\mathcal F^{\widehat{\mathbb A}^{(1)},\alpha} \simeq \mathcal F^{\mathbb A}$, which is a BF finite gauge theory with gauge group}
\be\label{eq:2}
\mathbb A := \prod_{p \in P} \mathbb A_{p} \qquad \mathbb A_{p} := \prod_{i=1}^\ell \mathbb{Z}_{p^{k_i}}^{r_{i}} ~,
\ee
\textit{with $P$ a subset of primes determined (as sketched above) from the pair $(\widehat{\mathbb{A}} ^{(1)}, \alpha)$.} This indeed corresponds to a topological order, as characterized in Section \ref{sec:JFY} above. 

\medskip

The action of $\mathcal F^{\mathbb A}$ is 
\begin{align}
  \label{eq:3}
  S[M_{5}] = \sum_{p \in P} S_{p}[M_{5}] ~,
\end{align}
where 
\begin{align}
  \label{eq:6}
  S_{p}[M_{5}] = S_{p^{k_{1}},r_{1}}[M_{5}] + S_{p^{k_{2}},r_{2}}[M_{5}] + \dots + S_{p^{k_{\ell}},r_{\ell}}[M_{5}] ~.
\end{align}
and each factor equals
\begin{align}
  \label{eq:4}
  S_{p^{k},r}[M_{5}] = 
   \frac{p^{k}}{2} \int_{M_{5}}\, b^{T} \Omega\, \dd b ~,
\end{align}
with $b = (b_{1} , \dots , b_{2r})$ a collection of suitable $2r$ two-form finite gauge fields and $\Omega$ is a unimodular, integral and antisymmetric bilinear form. Such an antisymmetric bilinear form is unique, i.e.~by a permutation of the basis it can always be brought to the canonical form
\begin{align}
  \label{eq:5}
  \Omega =
  \begin{pmatrix}
    0 & -\mathbbm{1}_{r \times r} \\
    \mathbbm{1}_{r \times r} & 0
  \end{pmatrix} ~.
\end{align}
From now on we assume that $\Omega$ has been brought into such a form. The group 
\be
\mathbb{Z}_{p^{k}}^{2r} \cong \mathbb{Z}_{p^{k}}^{r} \times \left(\mathbb{Z}_{p^{k}}^{r}\right)^\vee
\ee
provides the collection of electromagnetic charges for the topological surface defects in this factor
\be
\Phi(M_2) := \text{exp}\left(2\pi i \oint_{M_2} b \right) \qquad M_{2} \in H_{2}(M_{5} , \mathbb{Z}_{p^k}^{2r})\,.
\ee
The bilinear form $\Omega$ encodes the Dirac pairing among these charges, responsible for the braiding of the corresponding topological surface defects: 
\be
\langle\Phi(M_2) \Phi(M_2')\rangle_{\mathcal F^{\mathbb A}(M_5)} = \text{exp}\left(\frac{2\pi i}{p^k} \ell_{M_5}(M_2,M_2')(b^T \Omega b') \right)  
\ee
In what follows we will denote the above as a braiding
\be
\Phi(M_2) \Phi(M_2') = M_2(M_2') \Phi(M_2')\Phi(M_2) 
\ee
where $M_2 \in H_2(M_5,\mathbb A\times \mathbb A^\vee)$ and the pairing above is our schematic notation for the braiding arising from the various $p^k$ factors in $\mathcal F^{\mathbb A}$. Overall this gives the braiding structure for the topological defects of the whole BF theory $\mathcal F^{\mathbb A}$, that are in turn labeled by $\mathbb A \times \mathbb A^\vee$. Working on simply connected spin manifolds, by the universal coefficient theorem
\be
H_{2}(M_{5} ,\mathbb A \times \mathbb A^\vee) \simeq H_{2}(M_{5} ,\mathbb Z) \otimes (\mathbb A \times \mathbb A^\vee)\,.
\ee
It is well-known that topological boundary conditions are determined by choices of a Lagrangian subgroup $L$ of $\mathbb A \times \mathbb A^\vee$ in this case. By construction all theories $\mathcal B_L \otimes_{\mathcal F^{\mathbb A}} \widehat{\mathcal T}$ have a 1-form symmetry group
\be\label{sec:symtft-one-form}
(\mathbb A \times \mathbb A^{\vee})/L\,.
\ee
Notice that the groups $(\mathbb A \times \mathbb A^{\vee})/L$ and $(\mathbb A \times \mathbb A^{\vee})/L^\prime$ for two lagrangians $L$ and $L^{\prime}$ always have the same number of elements i.e.~$\abs{G}$, however they are not necessarily isomorphic. Lagrangian submanifolds can thus be organized by the orbits of $\mathrm{Sp}(2r,\mathbb{Z})$, and two lagrangians $L$ and $L^{\prime}$ are isomorphic iff they are in the same orbit of $\mathrm{Sp}(2r,\mathbb{Z})$. 

Let us formally obtain these results. It suffices to focus on a factor $\mathbb{Z}_{p^{k}}^{r}$ BF theory. We first observe that any global form $L$ of this theory can be encoded as the mod $p^k$ image of a $2r \times 2r$ integer matrix $\mathcal{M}_{L}$ satisfying,
\begin{align}
  \label{eq:26}
  \mathcal{M}_{L}^{T} \, \Omega \, \mathcal{M}_{L} = p^{k} \, \Omega ~.
\end{align}
To see this, note that since $L$ has at most $2r$ generators, it can in some basis for $\mathbb{Z}_{p^k}^r$ be expressed as $\im(\diag(p^{l_{1}}, \dots, p^{l_{2r}})) \pmod{p^k}$. Since $L$ is isotropic, the basis can be taken to be symplectic, with $p^{l_i} p^{l_{i+r}} = 0 \pmod{p^k}$. For $L$ to be maximal, we must have $l_i + l_{i+r} = k$ (which implies that $L$ has $p^{kr}$ elements). The natural integer lift $\mathcal{M}_D = \diag(p^{l_{1}}, \dots, p^{l_{2r}})$, expressed in an arbitrary symplectic basis, is precisely \eqref{eq:26}: $\mathcal{M}_L = F \mathcal{M}_D B$ with $F, B \in \mathrm{Sp}(2r,\mathbb{Z})$; in other words $\mathcal{M}_D$ is the Smith normal form of $\mathcal{M}_L$ up to permutation of the diagonal elements. Conversely, for any $\mathcal{M}_L$ as in \eqref{eq:26}, the $p^{k}$ on the r.h.s.~ensures that $L = \im\mathcal{M}_L \pmod{p^k}$ is isotropic, and that $\det(\mathcal{M}_{L}) = p^{kr}$ and hence $L$ has $p^{kr}$ elements as required for a maximal isotropic subgroup.

The one-form symmetry corresponding to $L$ can be expressed as
\begin{align}
  \label{eq:27}
  (\mathbb{Z}_{p^k}^{2r} / L )^\vee= \operatorname{coker} \mathcal{M}_L \cong (\mathbb{Z}_{p^{l_{1}}} \times \mathbb{Z}_{p^{k-l_{1}}}) \times \dots \times (\mathbb{Z}_{p^{l_{r}}} \times \mathbb{Z}_{p^{k-l_{r}}}) ~.
\end{align}
The fact that $\mathcal{M}_{L}$ and $\mathcal{M}_{L^{\prime}}$ realize isomorphic one-form symmetries if and only if they are related by the $\mathrm{Sp}(2r,\mathbb{Z})$ action\footnote{\ Implemented on the level of matrices as the two-sided action $\mathcal{M} \mapsto F \mathcal{M} B$, where $B$ is a symplectic change of basis for $L$.}, is simply the fact that every orbit of $\mathrm{Sp}(2r,\mathbb{Z})$ contains a diagonal matrix $\mathcal{M}_{D}$ such that $\mathcal{M}_{D} \pmod{p^{k}}$ is unique up to multiplication by invertible diagonal matrices in $\mathrm{Sp}(2r, \mathbb{Z}_{p^{k}})$. From the form \eqref{eq:27} of the global symmetry it can also be seen that the number of distinct $\mathrm{Sp}(2r,\mathbb{Z})$ orbits is
\begin{align}
  \label{eq:28}
  N_{p,k} = \binom{r + \lfloor \frac{k}{2} \rfloor}{r} ~.
\end{align}
More generally, for a $\mathbb{Z}_{p^{k_{1}}}^{r_{1}} \times \dots \times \mathbb{Z}_{p^{k_{n}}}^{r_{n}}$ BF theory, the number of distinct orbits is,
\begin{align}
  \label{eq:29}
  N_{\{p_{i},k_{i}\}} = \binom{r_{1} + \lfloor \frac{k_{1}}{2} \rfloor}{r_{1}} \dots \binom{r_{n} + \lfloor \frac{k_{n}}{2} \rfloor}{r_{n}} ~.
\end{align}
This generalizes the result of Bergman and Hirano \cite{Bergman:2022otk} for $\mathbb{Z}_{N}$ BF theory.

\subsection{Construction of 3+1d $\mathbb G$-duality defects}
\label{sec:theories-with-mixed}
In this section, we review how duality interfaces acting on the one-form symmetry defects are incorporated in the SymTFT picture, how they can be used to construct non-invertible zero-form symmetries, and under what conditions these can be gauged.

\medskip

We are interested in realizing the scenario we discussed in Section \ref{sec:symTFTreview} starting from a topological symmetry theory $\mathcal F^{\mathbb A}$ with quiche $(\mathcal F^{\mathbb A},\mathcal B_L)$, where $L$ is a Lagrangian subgroup of $\mathbb A \times \mathbb A^\vee$. To this aim we need to identify a collection of codimension 1 topological defects $\mathbb D_g$, graded by $g \in \mathbb G$, that are endable and such that they have a non-trivial action on the boundary conditions $\mathcal B_L$. The action of our BF theory in \eqref{eq:3} has obvious candidates in its $\mathrm{Sp}(\mathbb A \times \mathbb A^\vee)$ symmetry, corresponding to Montonen-Olive electromagnetic rotations of the BF theory 2-form gauge fields (automorphisms of $\mathbb A \times \mathbb A^\vee$ that preserve $\Omega$). These would be interesting, since they have a non-trivial action of the lattice of lines, that typically will not preserve our gapped boundary $\mathcal{B}_L$. But are they endable? According to the Johnson-Freyd--Yu analysis we reviewed in Section \ref{sec:JFY}, we expect that for the 4+1d topological order all higher dimensional defects must be condensations of the surface defects $\Phi(M_2)$. Condensates, being the result of a higher gauging operation, are always endable, since they always admit Dirichlet twist defects on their boundaries \cite{Kaidi:2022cpf}, indeed these are part of the corresponding orbifold datum \cite{Carqueville:2017aoe}.

\medskip

For all finite abelian groups $\mathbb A$, the group $\mathrm{Sp}(\mathbb A \times \mathbb A^\vee)$ can be embedded inside $\mathrm{Sp}(2r,\mathbb{Z})$ for a suitable $r$.\footnote{\ This can be done algorithmically by using the decomposition in \eqref{eq:2}. Although $\mathrm{Sp}(\mathbb A \times \mathbb A)$ is finite for a finite group $\mathbb A$, the group $\mathrm{Sp}(2r,\mathbb{Z})$ is not, in particular $\mathrm{Sp}(\mathbb A \times \mathbb A)$ might not ``lift'' to a finite subgroup of $\mathrm{Sp}(2r,\mathbb{Z})$.} Following \cite{Bashmakov:2022uek}, any topological manipulation can be represented as a $\mathrm{Sp}(2r,\mathbb{Z})$ matrix. For each such a matrix we can find a condensation defect i.e.~for any $g \in \mathrm{Sp}(2r,\mathbb{Z})$ we can construct a codimension one defect $\mathcal{C}_{g}$ such that for all $M_{2} \in H_{2}(M_{5} , \mathbb{Z}_{N}^{2r})$,
\begin{align}
  \label{eq:22}
 \mathcal{C}_{g}(M_{4}) \times \Phi(M_{2}) = \Phi(g^{T}M_{2}) \times \mathcal{C}_g(M_{4}) ~.
\end{align}
Explicitly this defect is given by,
\begin{align}
  \label{eq:23}
  \mathcal{C}_g(M_4) = \sum_{M_{2} \in H_{2}(M_{4} , \mathbb{Z}_{N}^{2r})} \exp(\frac{2\pi i}{2N}\ev{g^{T} M_{2},M_{2}})\Phi((1-g^{T})M_{2}) ~.
\end{align}
This is the unique defect in the BF theory that implements the fusion rules in \eqref{eq:22}. This follows from the fact that the condensation defects supported on $M_{4}$ naturally have the structure of a ring. As a result if there is another defect $\tilde{\mathcal{C}}_{g}$ which has the same fusion rules then for all $M_{2} \in H_{2}(M_{5} , \mathbb{Z}_{N}^{2r})$,
\begin{align}
  \label{eq:24}
  (\tilde{\mathcal{C}}_{g} - \mathcal{C}_{g})\Phi(M_{2}) = 0,
\end{align}
which means that $\tilde{\mathcal{C}}_{g} = \mathcal{C}_{g}$. From \eqref{eq:22} it can also be seen that,
\begin{align}
  \label{eq:25}
  \mathcal{C}_{g} \times \mathcal{C}_{\tilde{g}} = \mathcal{C}_{\tilde{g}g} ~.
\end{align}
Hence the operators $\mathcal{C}_{g}$ give a representation of $\mathrm{Sp}(2r,\mathbb{Z})$ on the operators in the BF theory.

\medskip

As is evident from \eqref{eq:25}, the operators $\mathcal{C}_{g}$ form a fusion ring graded by $\mathrm{Sp}(2r,\mathbb Z)$. Being condensates the corresponding Dirichlet twist defect $c_g$ and $c_g^\vee$ are part of the orbifold datum that defines them.

\medskip

As we have remarked in Section \ref{sec:symTFTreview} since the $\mathcal{C}_{g}$ condensates do not necessarily preserve the lattice $L$ that defines the boundary condition $\mathcal B_{L}$, these do not give rise to symmetries, rather to (non-invertible) topological interfaces. It can happen that a subgroup $\mathbb G$ of $\mathrm{Sp}(\mathbb A \times \mathbb A^\vee)$ for all $g\in\mathbb G$, 
\be\label{eq:actualduality}
\mathcal{C}_g \widehat{\mathcal T} \simeq  \widehat{\mathcal T}
\ee
This subgroup gives rise to the collection of defects we labeled $\mathbb D_g$ in Section \ref{sec:symTFTreview}. In these cases by virtue of Equation \eqref{eq:actualduality}, there are duality isomorphisms that can be composed with the twist defects resulting in (generically non-invertible) $\mathbb G$-duality symmetries. In what follows, we will often abuse notation and denote the subgroup of $\mathrm{Sp}(2r,\mathbb{Z})$ which realizes the action of $\mathbb G$ by the same symbol.

\medskip

The upshot of the discussion above is that $\mathrm{Sp}(2r,\mathbb{Z})$ provides us a universal way of probing the zero from symmetries of the 4+1 dimensional BF theories $\mathcal F^{\mathbb A}$. We stress here, that focusing on subfactors, the action of $\mathrm{Sp}(2r,\mathbb{Z})$ on the $\mathbb{Z}_{p^{k}}^{r}$ BF theories is particularly transparent since mod $p^{k}$ reduction of an $\mathrm{Sp}(2r,\mathbb{Z})$ matrix yields an $\mathrm{Sp}(2r,\mathbb{Z}_{p^{k}})$ matrix.\footnote{\, As described in Appendix \ref{sec:pruf-group-symm}, we can deal with all $k$ in a uniform manner by passing to the limit of $k \to \infty$ and consider an $\mathrm{Sp}(2r,\mathbb{Z})$ matrix as an $\mathrm{Sp}(2r, \hat{\mathbb{Z}}_{p})$ matrix. It is sometimes useful to realize $\mathbb{Z}_{p^{k}}$ symmetry as a subgroup of $\mathbb{Z}(p^{\infty})$ symmetry.}

\subsection{Aspects of the 4+1d TQFT ${\mathcal F}^{\mathbb A}_\mathbb{G}$}\label{sec:hilbert-space-gauged}

For a 3+1 dimensional theory $\mathcal{T}$ with non-invertible $\mathbb G$-duality symmetries described as above, one can consider as SymTFT the orbifold of $\mathcal F^{\mathbb A}$ by the family of $\mathbb D_g$ topological defects for $g \in \mathbb G$. We denote the resulting 4+1d TQFT ${\mathcal F}^{\mathbb A}_\mathbb{G}$. In this Section we discuss some features of this theory. We stress that as pointed out in the references \cite{Cordova:2023bja,Antinucci:2023ezl} it is always possible to stack a $\mathbb G$ 4+1d SPT to $\mathcal F^{\mathbb A}$ before gauging (which is related to the ``second level obstruction'' of \cite{Cordova:2023bja,Antinucci:2023ezl}). Including such an SPT in our analysis is an interesting problem that we leave to a future work. In this work we will restrict ourselves to the theories obtained by gauging without stacking any SPT.

\medskip

Upon gauging $\mathbb G$, the quantum torus algebra of topological surface defects of the theory $\mathcal F^{\mathbb A}$ is indeed extended by further topological defects. We have the twist defects $\sigma_g$ and $\sigma_g^\vee$ that belong to $\text{End}(\mathbf{1}_{\mathcal F^{\mathbb A}})$ and are of codimension two in five dimensions as well as the dual symmetry defects that arise from gauging $\mathbb G$. These are also genuine topological line defects that belong to $\text{End}(\mathbf{1}_{\mathbf{1}_{\mathbf{1}_{\mathcal F^{\mathbb A}}}})$. Since these are topological line defects in a 4+1 dimensional TQFT arising by gauging $\mathbb G$, we expect these operators are organized as the fusion category $\text{Rep}(\mathbb G)$ --- these are Wilson lines of the $\mathbb G$ gauge field.

\medskip

For simplicity, let us discuss the case where $\mathbb G$ is an abelian subgroup of the automorphism of the BF theory (a detailed account for $\mathbb G = \mathbb{Z}_N$ was already given in \cite{Kaidi:2022cpf}). The topological line operators of the theory are the Wilson lines for the one-form $\mathbb G$ gauge fields $\mathcal W(M_1)$, here we view $M_1 \in H_1(M_5; \mathbb G^\vee)$, where $\mathbb G^\vee$ is the Pontryagin dual group. By construction, these have a nontrivial linking pairing with the non-invertible $\mathbb G$ twist defects, which is implemented by the fact that there is a non-trivial braiding
\begin{align}
\label{eq:74}
  \sigma(M_{3}) \mathcal W(M_{1}) = M_{1}(M_{3}) \, \mathcal W(M_{1})\sigma(M_{3})  ~.
\end{align}
where $M_{3} \in H_{3}(M_{5};\mathbb G)$ and $M_{1} \in H_{1}(M_{5};\mathbb G^\vee)$. In addition, there is a crossing action of the $\sigma(M_3)$ defects on the topological defect lines $\Phi(M_2)$. Namely,
\begin{align}
\label{eq:80}
  \sigma(g \otimes M_{3}) \Phi(q \otimes M_{2}) = \Phi((g^{-1}q) \otimes M_{2}) \sigma( g \otimes M_{3}) ~.
\end{align}
where $q \in \mathbb A \times \mathbb A^\vee$ and $g \in \mathbb G$ acts as an $\mathrm{Sp}(\mathbb A \times \mathbb A^\vee)$ automorphism of the lattice. Because of this non-trivial braiding and of the crossing action the structure of the boundary conditions of this theory is enriched. Let us discuss various cases of interest below.

\subsubsection{Topological boundary conditions with two-form symmetry}
\label{sec:topol-bound-with}
Pushing the various braidings above to a codimension 1 boundary, one obtains an enriched quantum torus algebra. In particular, since the quantum torus algebra for topological surface defects is indeed a subalgebra of the algebra of operators described above, we need to choose a lagrangian $L$ to find a topological boundary condition. As before this lagrangian determines the one-form symmetry of the topological boundary condition. However such a choice is not enough to uniquely determine a state in the Hilbert space  $\mathcal{H}_{\mathbb F^\mathbb{A}_\mathbb{G}}(M_4)$. Instead, we must choose boundary conditions for the other topological defects too.

\medskip

Let us first consider the states which are such that $\sigma(M_{3})$ acts trivially on them. Such a state $\ket{L;2}$ satisfies,
\begin{align}
\label{eq:91}
 \sigma(M_{3}) \ket{L ; 2} = \ket{L;2} ~, && \Phi(M_{2})\ket{L;2} = \ket{L;2} ~.
\end{align}
for all $M_{3} \in H_{3}(M_{4},\mathbb{G})$ and $M_{2} \in H_{2}(M_{4},L)$. Considering the action of $\sigma(M_{3}) \Phi(M_{2})$ on $\ket{L;2}$ we see that both \eqref{eq:91} and \eqref{eq:80} can be simultaneously satisfied only if $L$ is invariant under the action of $\mathbb G$.\footnote{In writing the algebra of observables we have implicitly assumed that the $\mathbb G$ symmetry in BF theory is gauged without any SPT. The inclusion of such a 5d SPT will modify the fusion algebra e.g.~there can be an extra phase on right hand side \eqref{eq:80}. In that case in addition to invariant lagrangian we a 4d SPT such that stacking with it absorbs this phase. This is the second level obstruction. We leave its study to the future.}

\medskip

\noindent These states correspond to the gapped boundary conditions for $\mathcal F_\mathbb{G}^{\mathbb A}$ we label $\mathcal B^{(2)}_L$. Theories with a quiche $(\mathcal F_\mathbb{G}^{\mathbb A}, \mathcal B^{(2)}_L)$ are such that the zero form symmetry $\mathbb G$ is gauged, and they have a two-form symmetry $\mathbb G^\vee$, hence their name.

\subsubsection{Topological boundary conditions with zero-form symmetry}
\label{sec:topol-bound-cond-3}
Since the fusion between one-form symmetries and two-form symmetries is trivial, we can always obtain states with Dirichlet boundary conditions for the discrete $\mathbb G$ gauge field $c$. Let us suppose that $\ket{L;0}$ is one such state defined by,
\begin{align}
\label{eq:92}
  \mathcal W(M_{1})\ket{L; 0} = \ket{L; 0}, && \Phi(M_{2}) \ket{L; 0} = \ket{L; 0} ~.
\end{align}
This state also defines a topological boundary condition which has a zero-form symmetry generated by the action of $\sigma(M_{3})$ and one-form symmetry determined by $L$. Because of the Heisenberg (sub-)algebra, the states,
\begin{align}
\label{eq:93}
  \ket{L; g; M_{3}} \equiv \sigma(g \otimes M_{3})\ket{L; 0} ~,
\end{align}
are all linearly independent and carry a one-form symmetry flux $g$ through $M_{3}$. Let us now consider the state obtained by acting by $\Phi((qg^{T}) \otimes M_{2})$ with $M_{2} \subset M_{3}$ and $q \in L$:
\begin{align}
\label{eq:94}
  \Phi((gq) \otimes M_{2})\ket{L; g; M_{3}} &= \Phi((gq) \otimes M_{2}) \sigma(g \otimes M_{3})\ket{L; 0} \nonumber \\
                                         & = \sigma(g \otimes M_{3}) \Phi(q \otimes M_{2}) \ket{L;0} \nonumber \\
                                         &= \sigma(g \otimes M_{3}) \ket{L;0} = \ket{L;g;M_{3}} ~.
\end{align}
Hence, if $q \in gL$, then,
\begin{align}
\label{eq:95}
  \Phi(qM_{2})\ket{L; g; M_{3}} = \ket{L;g ; M_{3}} ~.
\end{align}
Hence, if $\ket{L;0}$ has one-form symmetry corresponding to the lattice $L$ then $\ket{L; g; M_{3}}$ has one-form symmetry corresponding to the lattice $gL$. This in essence just re-derivation of the fact that $\sigma(g \otimes M_{3})$ doesn't correspond to a symmetry on its own but rather must be composed with a topological manipulation to obtain a true (albeit non-invertible) symmetry.

\medskip

These states correspond to the gapped boundary conditions for $\mathcal F_\mathbb{G}^{\mathbb A}$ we label $\mathcal B^{(0)}_L$. Theories with a quiche $(\mathcal F_\mathbb{G}^{\mathbb A}, \mathcal B^{(0)}_L)$ realize a typically non-invertible duality zero-form $\mathbb G$-symmetry.

\medskip

Note that these states with zero-form symmetries always exist. This is to be contrasted with the states with two-form symmetries which need a $\mathbb G$ invariant lagrangian to exist.

\medskip

If the quiche $(\mathbb F^{\mathbb A}_{\mathbb G},\mathbb B^{(0)}_L)$ is formed exploiting a lagrangian $L$ invariant under $\mathbb G$, however, this means that the corresponding theory $\mathcal{T}$ have $\mathbb G$ as a genuine zero-form symmetry, albeit with a mixed anomaly with the one-form symmetry 
\be
(\mathbb A \times \mathbb A^\vee ) / L\,.
\ee 
Following \cite{Kaidi:2021xfk}, non-invertible symmetries can be obtained by gauging such one-form symmetry. Such symmetries were dubbed ``non-intrinsic'' in \cite{Kaidi:2022uux} and ``group-like'' in \cite{Sun:2023xxv}. 

\medskip

The purpose of the current work is to systematically tackle the existence of invariant lagrangians for a generic BF theory building on the methods of \cite{Bashmakov:2022uek} and extending them in various directions. As explained above a generic BF theory can be handled by focusing on $\mathbb{Z}_{p^k}^{r}$ BF theory. For the case of $k = 1$ i.e $\mathbb{Z}_{p}^{r}$ BF theory with $p$ a prime, their existence was studied systematically in \cite{Bashmakov:2022uek}. This case is special since for prime $p$, $\mathbb{Z}_{p}^{2r}$ can be regarded as a vector space over a finite field $\mathbb{F}_{p}$. Hence, the existence or absence of invariant lagrangians for $g \in \mathrm{Sp}(2r,\mathbb{Z}_{p})$ can be readily deduced from the Jordan normal form of $g$. A natural way of dealing with $\mathbb{Z}_{p^k}^{2r}$ theories efficiently is to pass to the limit and consider the Pr\"ufer group symmetry $\mathbb{Z}(p^{\infty})^{2r}$. This is in the spirit of considering the $\mathbb{Q} / \mathbb{Z}$ symmetry \cite{Putrov:2022pua} as ``generating function'' of $\mathbb{Z}_{n}$ symmetries. More concretely this corresponds to considering $\mathbb G$ as a subgroup of $\mathrm{Sp}(2r,\hat{\mathbb{Z}}_{p})$ where $\hat{\mathbb{Z}}_{p}$ is the ring of $p$-adic integers and looking for invariant lagrangians subspaces of $\hat{\mathbb{Z}}^{2r}_{p}$. As we will see, passing to this limit substantially simplifies the analysis of anomalies of non-invertible symmetries in four dimensions by utilizing some foundational tools of $p$-adic analysis and Galois theory. Once $p$-adic invariant lagrangians are known, their knowledge will enable us to recover invariant lagrangians which are invariant mod $p^k$ for all $k$. Details about this derivation are found in Appendix \ref{sec:invar-subsp-polyn}, \ref{sec:invar-lagr-galo}, and \ref{sec:finite-order-elem}, the statement of our main result is in the next Section.

\subsubsection{Topological boundary conditions with mixed zero- and two-form symmetry}
\label{sec:topol-bound-cond-4}
When $\mathbb G$ has non-trivial subgroups, it is possible to have topological boundary conditions for which both zero-form and two form symmetry are non-trivial. If $\mathbb H$ is a proper subgroup of $\mathbb G$, we can define a subgroup $\mathbb{H}^\vee$ of $\mathbb{G}^\vee$ as,
\begin{align}
\label{eq:97}
  h^\vee \in \mathbb{H}^\vee \Leftrightarrow h^\vee(h) = 1 ~ \forall ~ h \in \mathbb H ~.
\end{align}

We can define a topological boundary condition $\ket{L;\mathbb H}$ as,
\begin{align}
  \label{eq:98}
  \sigma(M_{3})\ket{L;\mathbb H} &= \ket{L;\mathbb H} ~, && M_{3} \in H_{3}(M_{4},\mathbb H) ~, \nonumber \\
 \mathcal{W}(M_{1})\ket{L;\mathbb{H}} &= \ket{L;\mathbb H} ~, && M_{1} \in H_{1}(M_{4} , \mathbb{H}^\vee) ~, \nonumber \\
  \Phi(M_{2})\ket{L;\mathbb{H}} &= \ket{L;\mathbb H} ~, && M_{2} \in H_{2}(M_{4}, L) ~.
\end{align}
This boundary condition has zero form symmetry $\mathbb G /\mathbb H$, one-form symmetry $(\mathbb A \times \mathbb A^\vee) / L$ and two form symmetry $\mathbb{G}^\vee / \mathbb{H}^\vee$. 

\medskip

These states correspond to the gapped boundary conditions for $\mathcal F_\mathbb{G}^{\mathbb A}$ we label $\mathcal B^{(\mathbb H)}_L$. As before, due to \eqref{eq:80} this boundary condition can exist only if the lagrangian $L$ is invariant under the action of $\mathbb H$.

\subsection{Main result: coarse classification of $\mathbb{G}$ invariant lagrangians}\label{sec:classificat}

As we have explained above, the existence of group-like duality symmetries is related to the question whether a given non-invertible symmetry is anomalous or not, those are necessary to be able to gauge the symmetry, as well as to understand whether the given symmetry is intrinsically non-invertible or not, see e.g.~\cite{Kaidi:2022uux,Sun:2023xxv,Cordova:2023bja,Antinucci:2023ezl,Antinucci:2024ltv} for further details. The technical core of this study is the following result, establishing a simple criterion for the existence of boundary conditions in $\mathbb{Z}_{N}^{r}$ BF theory which are invariant under the action of a finite cyclic subgroup $\mathbb G \subseteq \mathrm{Sp}(2r,\mathbb{Z})$.

\medskip

\noindent In this special case we obtain the following theorem:
\begin{theorem}
  \label{thm:main}
  Let $N = p_1^{k_1} \dots p_I^{k_I}$ with $p_i$ distinct primes and let $\mathbb G$ be a finite cyclic subgroup of $\mathrm{Sp}(2r,\mathbb{Z})$ generated by $g$. The characteristic polynomial of $g$ can be factored into a product of cyclotomic polynomials $\Phi_{n}(x)$. So let us suppose that
  \begin{align}
    \label{eq:1}
    \det(x - g) = \Phi_{n_{1}}(x)^{m_{1}} \dots \Phi_{n_J}(x)^{m_J} ~.
  \end{align}
  The $\mathbb{Z}_{N}^{r}$ BF theory admits irreducible topological $\mathbb G$-invariant boundary conditions precisely unless one of the following obstructions occurs for some odd power $p_i^{k_i}$ and some odd power $\Phi_{n_j}(x)^{m_j}$:
  \begin{itemize}
    \item $n_j$ is coprime to $p_i$ and $-1 \in \ev{p_i}_{n_j}$,
    \item $p_i = 2$, $n_j = 2 \pmod{4}$ and $-1 \in \ev{2}_{n_j / 2}$.
  \end{itemize}
  Here $\ev{p_i}_{n_j} \subseteq \mathbb{Z}_{n_j}^\times$ denotes the group of powers of $p_i \pmod{n_j}$.
\end{theorem}
The relevant values of $p_i \pmod{n_j}$ for each $n_j \leq 35$ are displayed in \cref{tab:thepowerofvodoo}.  This theorem easily reproduces the results previously available in the literature \cite{Bashmakov:2022uek,Sun:2023xxv,Cordova:2023bja,Antinucci:2023ezl} and provides an easy algorithmic way of determining the existence of invariant topological boundary conditions for any cyclic group.

\medskip

This fact immediately generalizes to any finite abelian subgroup of $\mathrm{Sp}(2r,\mathbb{Z})$ and, as we demonstrate in \cref{sec:non-abel-subgr}, it also streamlines their determination even for the case of non-abelian groups.\footnote{\ The lattice models presented in \cite{Cordova:2023bja} realize examples where $\mathbb G$ is abelian. It would be interesting to generalize further the construction to the case when $\mathbb G$ is non-abelian.} The methods used to obtain it are still more general: they only require control over Jordan canonical form of elements of $\mathbb G$. Deducing Jordan canonical forms is particularly simple for elements of finite order in $\mathrm{Sp}(2r,\mathbb{Z})$. However it possible to determine it even for elements of infinite order of $\mathrm{Sp}(2r,\mathbb Z)$ and apply our methods to classify all possible cyclic finite 0-form symmetries that arise from subgroups of $\mathrm{Sp}(2r,\mathbb Z_N)$ that are not mod $N$ reductions of finite subgroups of $\mathrm{Sp}(2r,\mathbb Z)$. We demonstrate how to proceed in the simple case of $r=1$ in \cref{sec:exampl-subgr-sl2}.

\medskip

An equivalent statement of the condition in \Cref{thm:main} is that for all $p_i$ and $\Phi_{n_j}(x)$ that occur with an odd power, one of the following is true:
\begin{itemize}
  \item $n_j$ is coprime to $p_i$ and $-1 \notin \ev{p_i}_{n_j}$,
  \item $p_i \neq 2$ and $n_j$ is a multiple of $p_i$,
  \item $p_i = 2$ and $n_j$ is a multiple of $4$,
  \item $p_i = 2$, $n_j = 2 \pmod{4}$ and $-1 \notin \ev{2}_{n_j / 2}$.
\end{itemize}
The six bullet points (four above, and the two in the statement of \Cref{thm:main}) correspond to the six types of entries in \Cref{tab:thepowerofvodoo}. Notice that the condition is quite simple for odd $p$, and additional complications arise only for $p = 2$. We remark also that for symplectic matrices, the algebraic multiplicities of eigenvalues $\pm 1$ are always even, that is, $\Phi_1(x) = x - 1$ and $\Phi_2(x) = x + 1$ appear with even powers; therefore \cref{tab:thepowerofvodoo} starts at $n = 3$.

\medskip

\begin{table}
  \centering
  \setlength{\tabcolsep}{1pt}
  \footnotesize
  \begin{tabular}{l|@{\hspace{2pt}}*{34}c}
    3& \cmark & \xmark & \dmark &&&&&&&&&&&&&&&&&&&&&&&&&&&&&&& \\
    4& \cmark & \sqmark & \xmark &&&&&&&&&&&&&&&&&&&&&&&&&&&&&&& \\
    5& \cmark & \xmark & \xmark & \xmark & \dmark &&&&&&&&&&&&&&&&&&&&&&&&&&&&& \\
    6& \cmark & \nmark & \dmark && \xmark &&&&&&&&&&&&&&&&&&&&&&&&&&&&& \\
    7& \cmark & \cmark & \xmark & \cmark & \xmark & \xmark & \dmark &&&&&&&&&&&&&&&&&&&&&&&&&&& \\
    8& \cmark & \sqmark & \cmark && \cmark && \xmark &&&&&&&&&&&&&&&&&&&&&&&&&&& \\
    9& \cmark & \xmark & \dmark & \cmark & \xmark && \cmark & \xmark &&&&&&&&&&&&&&&&&&&&&&&&&& \\
    10& \cmark & \nmark & \xmark && \dmark && \xmark && \xmark &&&&&&&&&&&&&&&&&&&&&&&&& \\
    11& \cmark & \xmark & \cmark & \cmark & \cmark & \xmark & \xmark & \xmark & \cmark & \xmark & \dmark &&&&&&&&&&&&&&&&&&&&&&& \\
    12& \cmark & \sqmark & \dmark && \cmark && \cmark &&&& \xmark &&&&&&&&&&&&&&&&&&&&&&& \\
    13& \cmark & \xmark & \cmark & \xmark & \xmark & \xmark & \xmark & \xmark & \cmark & \xmark & \xmark & \xmark & \dmark &&&&&&&&&&&&&&&&&&&&& \\
    14& \cmark & \ymark & \xmark && \xmark && \dmark && \cmark && \cmark && \xmark &&&&&&&&&&&&&&&&&&&&& \\
    15& \cmark & \cmark & \dmark & \cmark & \dmark && \cmark & \cmark &&& \cmark && \cmark & \xmark &&&&&&&&&&&&&&&&&&&& \\
    16& \cmark & \sqmark & \cmark && \cmark && \cmark && \cmark && \cmark && \cmark && \xmark &&&&&&&&&&&&&&&&&&& \\
    17& \cmark & \xmark & \xmark & \xmark & \xmark & \xmark & \xmark & \xmark & \xmark & \xmark & \xmark & \xmark & \xmark & \xmark & \xmark & \xmark & \dmark &&&&&&&&&&&&&&&&& \\
    18& \cmark & \nmark & \dmark && \xmark && \cmark &&&& \xmark && \cmark &&&& \xmark &&&&&&&&&&&&&&&&& \\
    19& \cmark & \xmark & \xmark & \cmark & \cmark & \cmark & \cmark & \xmark & \cmark & \xmark & \cmark & \xmark & \xmark & \xmark & \xmark & \cmark & \cmark & \xmark & \dmark &&&&&&&&&&&&&&& \\
    20& \cmark & \sqmark & \cmark && \dmark && \cmark && \cmark && \cmark && \cmark &&&& \cmark && \xmark &&&&&&&&&&&&&&& \\
    21& \cmark & \cmark & \dmark & \cmark & \xmark && \dmark & \cmark && \cmark & \cmark && \cmark &&& \cmark & \xmark && \cmark & \xmark &&&&&&&&&&&&&& \\
    22& \cmark & \nmark & \cmark && \cmark && \xmark && \cmark && \dmark && \xmark && \cmark && \xmark && \xmark && \xmark &&&&&&&&&&&&& \\
    23& \cmark & \cmark & \cmark & \cmark & \xmark & \cmark & \xmark & \cmark & \cmark & \xmark & \xmark & \cmark & \cmark & \xmark & \xmark & \cmark & \xmark & \cmark & \xmark & \xmark & \xmark & \xmark & \dmark &&&&&&&&&&& \\
    24& \cmark & \sqmark & \dmark && \cmark && \cmark &&&& \cmark && \cmark &&&& \cmark && \cmark &&&& \xmark &&&&&&&&&&& \\
    25& \cmark & \xmark & \xmark & \xmark & \dmark & \cmark & \xmark & \xmark & \xmark && \cmark & \xmark & \xmark & \xmark && \cmark & \xmark & \xmark & \xmark && \cmark & \xmark & \xmark & \xmark &&&&&&&&&& \\
    26& \cmark & \nmark & \cmark && \xmark && \xmark && \cmark && \xmark && \dmark && \xmark && \xmark && \xmark && \xmark && \xmark && \xmark &&&&&&&&& \\
    27& \cmark & \xmark & \dmark & \cmark & \xmark && \cmark & \xmark && \cmark & \xmark && \cmark & \xmark && \cmark & \xmark && \cmark & \xmark && \cmark & \xmark && \cmark & \xmark &&&&&&&& \\
    28& \cmark & \sqmark & \xmark && \cmark && \dmark && \cmark && \cmark && \cmark && \cmark && \cmark && \xmark &&&& \cmark && \cmark && \xmark &&&&&&& \\
    29& \cmark & \xmark & \xmark & \xmark & \xmark & \xmark & \cmark & \xmark & \xmark & \xmark & \xmark & \xmark & \xmark & \xmark & \xmark & \cmark & \xmark & \xmark & \xmark & \cmark & \xmark & \xmark & \cmark & \cmark & \cmark & \xmark & \xmark & \xmark & \dmark &&&&& \\
    30& \cmark & \ymark & \dmark && \dmark && \cmark &&&& \cmark && \cmark &&&& \cmark && \cmark &&&& \cmark &&&&&& \xmark &&&&& \\
    31& \cmark & \cmark & \xmark & \cmark & \cmark & \xmark & \cmark & \cmark & \cmark & \cmark & \xmark & \xmark & \xmark & \cmark & \xmark & \cmark & \xmark & \cmark & \cmark & \cmark & \xmark & \xmark & \xmark & \xmark & \cmark & \xmark & \xmark & \cmark & \xmark & \xmark & \dmark &&& \\
    32& \cmark & \sqmark & \cmark && \cmark && \cmark && \cmark && \cmark && \cmark && \cmark && \cmark && \cmark && \cmark && \cmark && \cmark && \cmark && \cmark && \xmark &&& \\
    33& \cmark & \xmark & \dmark & \cmark & \cmark && \cmark & \xmark && \cmark & \dmark && \cmark & \cmark && \cmark & \xmark && \cmark & \cmark &&& \cmark && \cmark & \cmark && \cmark & \xmark && \cmark & \xmark && \\
    34& \cmark & \nmark & \xmark && \xmark && \xmark && \xmark && \xmark && \xmark && \xmark && \dmark && \xmark && \xmark && \xmark && \xmark && \xmark && \xmark && \xmark && \xmark & \\
    35& \cmark & \cmark & \cmark & \cmark & \dmark & \cmark & \dmark & \cmark & \cmark && \cmark & \cmark & \cmark &&& \cmark & \cmark & \cmark & \xmark &&& \cmark & \cmark & \xmark && \cmark & \cmark && \cmark && \cmark & \cmark & \cmark & \xmark  \\
    \hline
    \diagbox[dir=NE]{$n$}{$p$} & 1 & 2 & 3 & 4 & 5 & 6 & 7 & 8 & 9 & 10 & 11 & 12 & 13 & 14 & 15 & 16 & 17 & 18 & 19 & 20 & 21 & 22 & 23 & 24 & 25 & 26 & 27 & 28 & 29 & 30 & 31 & 32 & 33 & 34
  \end{tabular}
  \caption{Existence (blue) and non-existence (red) of $g$-invariant lagrangian subgroups of $\mathbb{Z}_{p^k}^{2r}$ when $k$ is odd and the characteristic polynomial of $g$ is $\Phi_n(x)^m$ with $m$ odd. Values of $p$ are mod $n$. Check marks and crosses show the condition $-1 \notin \ev{p}_n$ when $n$ and $p$ are coprime. Dots mark odd primes dividing $n$. For $p = 2$, double dots indicate $n$ divisible by $4$, while circled dots and crosses show the condition $-1 \notin \ev{p}_{n/2}$ for $n = 2 \pmod{4}$. Spots where $p$ cannot be prime are left empty.}
  \label{tab:thepowerofvodoo}
\end{table}

To apply \Cref{thm:main} directly, knowledge of the isomorphism class of $\mathbb G$ is not sufficient; one needs to know its representation in $\mathrm{Sp}(2r, \mathbb{Z})$ (in fact, only its conjugacy class in $\mathrm{GL}(2r, \mathbb{C})$). One can, however, draw some conclusions with only minimal information about the representation. Notice that the numbers appearing in \eqref{eq:1} must satisfy
\begin{equation}
  \label{eq:mj-constraint}
  \sum_{j = 1}^J m_j \varphi(n_j) = 2r,
\end{equation}
where $\varphi(n_j) = \operatorname{deg}\Phi_{n_j}$ is the Euler totient function, and furthermore that each $n_j$ must divide the order of $g$. For simplicity, suppose $g^q = 1$ where $q$ is a prime. Then $\det(x - g)$ is of the form $\Phi_1(x)^{m_1} \Phi_q(x)^{m_q}$ where $\Phi_1(x) = x - 1$ and $\Phi_q(x) = \frac{x^q - 1}{x - 1}$, and \eqref{eq:mj-constraint} becomes $m_1 + m_q(q - 1) = 2r$. The obstructions in \cref{tab:thepowerofvodoo} apply when $m_q = \frac{2r - m_1}{q - 1}$ is odd.\footnote{\ Since $g$ is a symplectic matrix, $m_1$ is always even.} For example, this is true if $g$ acts freely ($m_1 = 0$) and $\frac{2r}{q - 1}$ is odd, or if $g \neq 1$ ($m_1 \neq 2r$) and $q - 1 > r$, which forces $m_q = 1$. Such conditions, although expressed in terms of the numbers $r$ and $q$ as far as possible, must always include some assumption on the matrix form of $g$.

\medskip

The derivation of this result is extremely technical and summarized in Appendix 
\ref{sec:invar-subsp-polyn}, \ref{sec:invar-lagr-galo}, and \ref{sec:finite-order-elem}, where we also present several detailed examples as consistency checks for our methods, in particular recovering and/or extending several results in the literature (in particular \cite{Sun:2023xxv,Cordova:2023bja,Antinucci:2023ezl}). 

\medskip

In Appendix \ref{sec:non-abel-subgr} we generalize our techniques to exhibit invariant lagrangians also in the cases $\mathbb G$ is a non-abelian, explicitly realized by several class $\mathcal S$ theories \cite{Bashmakov:2022uek,Antinucci:2022cdi}.

\section{Applications: symmetry preserving RG flows}\label{sec:RGflows}

\subsection{Some general remarks}

Let us consider the general setup in Section \ref{sec:symTFTreview}, where we are interested in a $d$-dimensional field theory that has an isomorphism \textit{S4.)}
\be\label{eq:quicazz}
\mathcal T \simeq \mathcal B \otimes_\mathcal F \otimes \widehat{\mathcal T}
\ee
that realizes a non-invertible symmetry $\mathcal N_g$ as we outlined there.

\medskip

It is an interesting question to understand what happens if $\mathcal T$ has RG-flows that are symmetry preserving. An example to consider which will be relevant later is the case
\be
\mathcal T + \lambda \int_d \mathcal O
\ee
where $\mathcal T$ is a CFT and $\mathcal O$ is a relevant deformation that is such that $[\mathcal N_g, \mathcal O] = 0$.

\medskip

For the IR theory one has the following options: $\mathcal T_{IR}$ can be gapless or gapped. In the latter case the symmetry can be preserved or undergo spontaneous symmetry breaking, depending on the dynamics of the theory in question. Further, if the symmetry is preserved it is interesting to understand whether the resulting IR theory is trivially gapped, meaning that it is an SPT, or non-trivially gapped, meaning that it supports a non-trivial TO, a topological order.

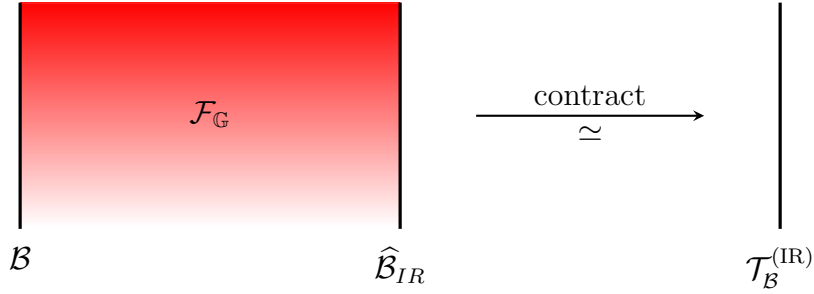
\begin{figure}
\centering
$$
  \begin{gathered}
    \begin{tikzpicture}
    \shadedraw [shading=axis,top color =red, bottom color=white, white] (0,0) rectangle (5,3);
    \draw [red] (0,3) -- (5,3);
    \draw [very thick] (0,0) -- (0,3);
    \draw [very thick] (5,0) -- (5,3);
    \node[below] at (0,-0.1){$ \mathcal B$};
    \node[below] at (5,-0.1){$\widehat{\mathcal B}_{IR}$};
    \node[above] at (2.5,1.2){$\mathcal F_\mathbb G$};
    \draw [very thick] (10,0) -- (10,3);
    \draw [-stealth, thick] (6,1.5) -- (9,1.5);
    \node[above] at (7.5,1.5){contract};
    \node[below] at (7.5,1.5){$\simeq$};
    \node[below] at (10,-0.1){$\mathcal{T}_{\mathcal B}^{(\textrm{IR})}$};
  \end{tikzpicture}
  \end{gathered}
$$
\caption{SymTFT and RG flow: the case $\mathcal T_{IR}$ is gapped.}\label{fig:RGsymTFT}
\end{figure}

\medskip

Along the flow, since the symmetry is unbroken, the topological symmetry theory construction is preserved \cite{Gaiotto:2019xmp,DelZotto:2022ras}, and the quiche $(\mathcal F,\mathcal B)$ in \eqref{eq:quicazz} is left unchanged. If we consider an RG flow that preserves the symmetry leading to a \textit{gapped} phase in the IR, it is tautological that we have an isomorphism
\be
\mathcal T_{IR} \simeq \mathcal B \otimes_{\mathcal F} \mathcal B_{IR} 
\ee
where $\mathcal B_{IR}$ is a gapped boundary for $\mathcal F$. The fact that the non-invertible symmetry is preserved along the RG flow implies that
\be
\mathbb D_g \mathcal B_{IR} \simeq \mathcal B_{IR}\qquad \forall \, g \in \mathbb G.
\ee
meaning that we are after invariant boundaries $\mathcal B_{IR}$. Clearly if one can prove that
\begin{enumerate}
\item The theory $\mathcal F$ does not admit irreducible symmetry preserving topological boundaries, and
\item The non-invertible symmetry is not spontaneously broken,
\end{enumerate}
then the IR endpoint of the corresponding symmetry preserving RG flow cannot be gapped and it must be gapless -- an instance of \textit{symmetry protected gaplessness} as discussed by \cite{Choi:2021kmx,Choi:2022zal,Cordova:2023bja}.

\medskip

In the above discussion we have considered the original symmetry theory $\mathcal F$, however, we should also keep into account that there is a \textit{different} SymTFT where the non-invertible symmetry defects arise as genuine operators, the orbifolded SymTFT $\mathcal F_{\mathbb G}$. Clearly, the irreducible boundary conditions of $\mathcal F$ that are symmetry preserving uplift to irreducible boundary conditions of $\mathcal F_{\mathbb G}$. However, if a collection of boundary conditions in $\mathcal F$ is closed with respect to the action of $\mathbb D_g$, meaning that
\begin{equation}
    \mathbb D_g (\mathcal B_1 \oplus \cdots \oplus \mathcal B_\ell) \simeq \mathcal B_1 \oplus \cdots \oplus \mathcal B_\ell 
\end{equation}
holds, the latter uplifts to a symmetry preserving boundary condition of $\mathcal F_\mathbb{G}$ corresponding to a gapped phase where the non-invertible symmetry undergoes \textit{spontaneous symmetry breaking}.\footnote{\, We thank Kantaro Ohmori for discussions about this.} In this latter case, the structure of the various vacua interconnected by the action of the non-invertible domain walls is captured by $\mathcal B \otimes_\mathcal F \mathcal B_K$, with $K=1,...,\ell$.

\medskip

In the remaining part of this section, we use the tools developed above to study these questions in the context of 3+1 dimensional QFTs with non-invertible symmetries.

\subsection{Non-invertible symmetry preserving RG flows in 3+1 dimensions}
\label{sec:lagr-subl-anom}

In the $d=3+1$ case we have a bulk SymTFT that is typically a topological order $\mathcal F^{\mathbb A}$. In the previous section we have outlined a classification for a given subgroup $\mathbb G$ of $\mathrm{Sp}(\mathbb A \times \mathbb A^\vee)$ of the corresponding invariant Lagrangian subgroups $L$ , leading to symmetry preserving boundary conditions $\mathcal B_L$. This classification has an immediate application here, after \cite{Cordova:2023bja,Antinucci:2023ezl}.

\medskip

Consider a 3+1 dimensional theory $\mathcal T$ such that it has an \textit{S4.)} isomorphism
\be
\mathcal T \simeq \mathcal B_{L_e} \otimes_{\mathcal F^\mathbb{A}} \widehat{\mathcal T}\,.
\ee
and a non-invertible $\mathbb G$-duality symmetry $\mathcal N_g$.  The possible symmetry preserving gapped phases for this theory are obtained by computing
\be
\mathcal T_{IR} \simeq \mathcal B_{L_e} \otimes_{\mathcal F^\mathbb{A}} \mathcal B_{L}
\ee
where $L$ is a $\mathbb G$-invariant Lagrangian sublattice out of our classification.

\medskip

The theory is trivially gapped provided $\mathcal B_{L_e} \otimes_{\mathcal F^\mathbb{A}} \mathcal B_{IR} \simeq \Sigma^4 \mathbb C$ as topological orders, meaning that the IR is an invertible TQFT. If this is not the case, the theory captures a non-trivial 3+1 dimensional topological order whose structure is determined by $\mathcal B_{L_e} \otimes_{\mathcal F^\mathbb{A}} \mathcal B_{IR}$.

\subsubsection{Classification of trivially gapped $\mathbb G$-symmetric phases}\label{sec:anom-1-form}

Due to the correspondence between the lagrangian subgroups $L \subset \mathbb{Z}_{p^k}^{2r}$ and topological boundary conditions of $\mathbb{Z}_{p^k}^{r}$ theory, a $\mathbb G$-invariant boundary condition, or in other words a gapped phase in which $\mathbb G$ acts trivially, requires a $\mathbb G$-invariant lagrangian sublattice.\footnote{\ It will be interesting to study the case of non-abelian groups systematically and understand the consequences of spontaneous symmetry breaking in that case. Gauging a non-normal subgroup of $\mathbb G$ results in a (higher version) of $\mbox{Rep}(\mathbb G)$ symmetry and generalizes the construction of \cite{Bhardwaj:2017xup}. We leave the study of this case to the future.} Taking this one step further as in \cite{Cordova:2023bja}, we may search for obstructions to trivially gapped realizations of a full symmetry structure comprising 1-form symmetry as well as non-invertible 0-form symmetry.

\medskip

\begin{figure}
  \centering
  \begin{equation*}
    \newcommand{\rx}{5}
    \newcommand{\gx}{4}
    \begin{tikzpicture}[baseline=0,scale=0.8]
      \draw (0,-2,-2) -- (0,-2,2) -- (0,2,2) -- (0,2,-2) -- cycle;
      \draw[blue] (\rx,-1,-2) -- (\rx,-1,2);
      \draw (\rx,-2,-2) -- (\rx,-2,2) -- (\rx,2,2) -- (\rx,2,-2) -- cycle;
      \foreach \p in {(0,-2,-2), (0,-2,2), (0,2,2), (0,2,-2)}
      \draw[dotted] \p -- +(\rx,0,0);
      \node[below] at (0,-2,2) {$\mathcal B_{L_\mathrm{e}}$};
      \node[blue,right] at (\rx,-1,-1) {$\in \mathbb G$};
      \node[below] at (\rx,-2,2) {$\mathcal B_L$};

      \begin{scope}[red,dotted]
        \draw[thick] (0,0,0) -- (\rx,0,0);
        \node[above] at (.45*\rx,0,0) {\footnotesize $\in L_\mathrm{e} \cap L$};
        \node[below] at (.45*\rx,0,0) {\scriptsize disallowed};
      \end{scope}
      \begin{scope}[blue]
        \draw[canvas is yz plane at x=.15*\rx, thick] (0,0) circle (0.8);
        \node at (.35*\rx,-1,0)
        {\footnotesize $\in \mathbb{Z}_{p^k}^{2r}/L_\mathrm{e}$};
      \end{scope}
    \end{tikzpicture}
    \quad\cong
    \begin{tikzpicture}[baseline=0,scale=0.8]
      \draw (0,-2,-2) -- (0,-2,2) -- (0,2,2) -- (0,2,-2) -- cycle;
      \node[below] at (0,-2,2) {trivial};

      \begin{scope}[blue]
        \draw[canvas is yz plane at x=0,thick,dotted] (0,0) circle (0.8);
        \node[below right, outer sep=5pt] at (0,-0.4,0) {$1$};
        \draw[thick,dotted] (0,-1,-2) -- (0,-1,2);
      \end{scope}
    \end{tikzpicture}
  \end{equation*}
  \caption{Topological symmetry theory construction of a trivially gapped phase 
  \cite{Cordova:2023bja}.
  }
  \label{fig:Le}
\end{figure}
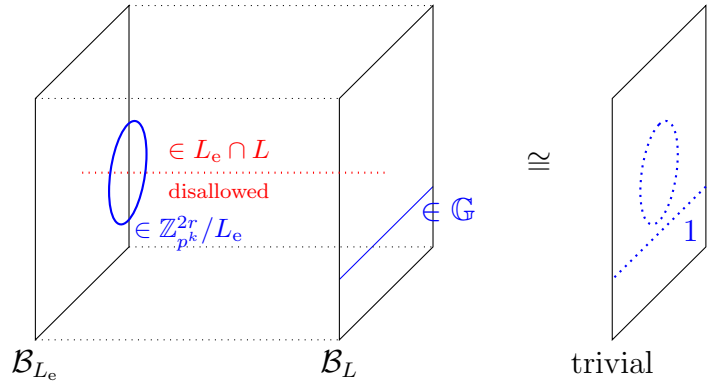

Let us focus on the case $\mathbb A$ has a single factor $\mathbb A_p \simeq \mathbb{Z}_{p^k}^r)$. Giving a reference Lagrangian $L_e$ fixes the one-form symmetry arising from the quiche $(\mathcal F^{\mathbb{A}},\mathcal{B}_{L_e})$ to be
\be
F^{(1)} = \mathbb{Z}_{p^k}^{2r} / L_\mathrm{e}\,.
\ee 
If we have a theory with such a symmetry structure together with a non-invertible $\mathbb G$-duality symmetry, it can flow to a symmetry-preserving trivially gapped phase only if there is a $\mathbb{G}$-invariant lagrangian $L$ that further satisfies 
\be
L_\mathrm{e} \cap L = \{0\},
\ee
so that the absolute IR theory does not contain any nontrivial line operators with $F^{(1)}$ charge; see \cref{fig:Le}. Following \cite{Cordova:2023bja}, we dub one such lagrangian `magnetic'.

\medskip

In practice, one can decide existence of such a lagrangian by constructing the (typically few) $\mathbb{G}$-invariant lagrangians by our method, and checking their intersection with $L_\mathrm{e}$ one by one. Let us proceed with some more systematic observations below.

\paragraph{The $r=1$ case.} Let us begin with the case of $r = 1$ (see also Appendix \ref{sec:exampl-subgr-sl2} for a review), as exemplified by the \be
\qty(\mathrm{SU}(N)/\mathbb{Z}_K)_n\ee
gauge 
theories, as it is simple enough that it can be easily described in general. 

\medskip

Consider a theory with one-form symmetry given by $L_\mathrm{e} \subset \mathbb{Z}_{p^k}^2$ (of the form \eqref{eq:between-the-lines-lagr} with $N = p^k$). If $L_\mathrm{e}$ is torsional, so that its Smith form is $\diag(p^\ell, p^{k-\ell})$ with $1 \leq \ell \leq k/2$, then $p^{k-\ell} \mathbb{Z}_{p^k}^2 \subset L_\mathrm{e}$, so that $L_\mathrm{e}$ nontrivially intersects every nontrivial subgroup.
\textit{Therefore only theories with free $L_\mathrm{e}$ can possibly be trivially gapped.}

\medskip

Among $\mathfrak{su}(N)$ gauge theories, the only global forms are 
\be
\mathrm{SU}(p^k) 
\ee
\be
\mathrm{PSU}(p^k)_n = \qty(\mathrm{SU}(p^k)/\mathbb{Z}_{p^k})_n 
\ee
and
\be 
\qty(\mathrm{SU}(p^k)/\mathbb{Z}_{p^l})_n\,,
\ee
where $\gcd(n,p) = 1$.

\medskip

For $g \neq \pm 1 \in \mathrm{SL}(2,\mathbb{Z})$, the number of $g$-invariant free (also cyclic) lagrangians $L$, that is, independent eigenvectors over $\mathbb{Z}_{p^k}$, is either 0 (when there is symmetry-enforced gaplessness), 1 (when $g \sim \scriptsize\mqty(\pm 1 & 1 \\ 0 & \pm 1)$ is a Jordan block) or 2 (when $g$ diagonalizes). In the first case, there is nothing to discuss. In the last case, the two lagrangians trivially intersect, so at least one of them trivially intersects $L_\mathrm{e}$ and we have a trivially gapped realization. In the middle case, we find an obstruction to a trivially gapped phase precisely when $L_\mathrm{e}$ is the single invariant lagrangian $L$. Any other free $L_\mathrm{e}$ trivially intersects $L$. For example, consider the $\mathcal{N} = 4$ SYM gauge theory with group
\be
\qty(\mathrm{SU}(2)/\mathbb{Z}_2)_1 = \mathrm{SO}(3)_-\,.
\ee 
In this case, 
\be
L_\mathrm{e} = \Span_{\mathbb{Z}_2}(1, 1)
\ee
is the only $\mathsf{S}$-invariant lagrangian, which guarantees that this theory cannot be trivially gapped (of course, this is not surprising as all $\mathcal{N} = 4$ SYM theories are conformal, but it illustrates the principle).

\medskip

To take another example, \cite[sec.~2.2.2]{Cordova:2023bja} treats the case $g = \mathsf{ST}^n$ with the particular choice of $L_\mathrm{e} = \Span_{\mathbb{Z}_N}(1, 0)$ for the electric lattice. As $(1, 0)$ is never an eigenvector of $\mathsf{ST}^n$, $L_\mathrm{e}$ is not invariant, so by the above reasoning the symmetry admits a trivially gapped realization whenever $\mathsf{ST}^n$ has any eigenvector in $\mathbb{Z}_N^2$ (even if $\mathsf{ST}^n$ does not diagonalize). An eigenvector exists if the characteristic polynomial $x^2 - nx + 1$ has a root in $\mathbb{Z}_N$, and one may check that this indeed reproduces Equations (2.26) and (2.27) of \cite{Cordova:2023bja}.

\paragraph{The $r>1$ case.} Let us now move on to general $r$. Remarkably, it is still true that any one-form symmetry that is not $\mathbb{Z}_{p^k}^r$ is anomalous, independently of the zero-form symmetry $\mathbb G$. In other words (see \eqref{eq:27}), if $L_\mathrm{e}$ is torsional, then one always finds non-trivial phases in the IR, because a torsional lagrangian intersects every other lagrangian nontrivially. To show this, write any two lagrangians $L_\mathrm{e}, L \subset \mathbb{Z}_{p^k}^{2r}$ as
\begin{equation}
  L_\mathrm{e} = \Span_{\mathbb{Z}_{p^k}}(p^{\alpha_1} x_1, \dots, p^{\alpha_R} x_R),
  \qquad
  L = \Span_{\mathbb{Z}_{p^k}}(p^{\beta_1} y_1, \dots, p^{\beta_{R'}} y_{R'})
\end{equation}
where $(x_1, \dots, x_R)$ and $(y_1, \dots, y_{R'})$ are two sets of linearly independent elements; this is the Smith normal form discussed in \cref{sec:determining-symtft}. The number of generators of a lagrangian is at least $r$, so $R, R' \geq r$. Equality holds precisely for free lagrangians. Clearly we have
\begin{equation}
  L_\mathrm{e} \cap L \supset
  p^{k-1} \Span_{\mathbb{Z}_{p^k}}(x_1, \dots, x_R) \cap
  p^{k-1} \Span_{\mathbb{Z}_{p^k}}(y_1, \dots, y_{R'}).
\end{equation}
Using the isomorphism $p^{k-1} \mathbb{Z}_{p^k}^{2r} \cong \mathbb{Z}_p^{2r}$, the right hand side corresponds to the intersection of vector spaces $\Span_{\mathbb{Z}_p}(x_1, \dots, x_R) \cap \Span_{\mathbb{Z}_p}(y_1, \dots, y_{R'})$.\footnote{\ These spaces still have dimension $R$ and $R'$ respectively. A nontrivial linear relation $f(x_1, \dots, x_R) = 0$ mod $p$ would imply the relation $p^{k-1} f(x_1, \dots, x_R) = 0$ mod $p^k$, but $(x_1, \dots, x_R)$ are by definition linearly independent.} The codimension of this intersection is at most $(2r - R) + (2r - R')$, which is $2r$ only when $R = R' = r$, that is, when $L_\mathrm{e}$ and $L$ are both free. In short, \textit{a torsional lagrangian must always intersect all other lagrangians for dimensional reasons.}

\medskip

Consider then a free lagrangian $L_\mathrm{e} \subset \mathbb{Z}_{p^k}^{2r}$; we search for free $\mathbb{G}$-invariant lagrangians $L$ with $L_\mathrm{e} \cap L = \{0\}$. If $k > 1$, they can be lifted to invariant vector subspaces of $\mathbb{Q}_p^{2r}$, while for $k = 1$, $\mathbb{Z}_p^{2r} = \mathbb{F}_p^{2r}$ is already a vector space --- see Appendix \ref{sec:invar-subsp-polyn}. Hence we work over the field $F = \mathbb{F}_p$ or $\mathbb{Q}_p$ as appropriate.

\medskip

We will not decide existence of a magnetic Lagrangian $L$ in full generality, but restrict ourselves to the particularly simple case where $g \in \mathbb{G}$ is diagonalizable over $F$. If this is the case, we shall show that a magnetic lagrangian always exists. 

\medskip

There is a basis of $F^{2r}$ of eigenvectors that come in pairs $v_{i,0},\, v_{i,1}$ such that their eigenvalues are inverses and $\ev{v_{i,0}, v_{j,1}} = \delta_{ij}$. For a binary string $\alpha$ of length $s \leq r$, define a $g$-invariant isotropic subspace $V_{\alpha_1\dots\alpha_s} = \Span(v_{1,\alpha_1}, \dots, v_{s,\alpha_s})$. Those with $s = r$ are $g$-invariant lagrangians. We will show that not all of these $2^r$ lagrangians can intersect $L_\mathrm{e}$. Assume that $L_\mathrm{e}$ intersects all $V_{\alpha_1\dots\alpha_{s+1}}$ for fixed $s$. Then it contains two nonzero vectors $u_0 + a v_{s+1,0} \in V_{\alpha_1\dots\alpha_s 0}$ and $u_1 + b v_{s+1,1} \in V_{\alpha_1\dots\alpha_s 1}$, with $u_0, u_1 \in V_{\alpha_1\dots\alpha_s}$. As $L_\mathrm{e}$ is isotropic, $0 = \ev{u_0 + a v_{s+1,0}, u_1 + b v_{s+1,1}} = ab$, so that at least one of the two vectors is in fact in $V_{\alpha_1\dots\alpha_s}$. This shows that $L_\mathrm{e}$ (nontrivially) intersects every $V_{\alpha_1\dots\alpha_s}$. By downward induction on $s$, this shows that if $L_\mathrm{e}$ intersects every invariant lagrangian, it must also nontrivially intersect $V_\emptyset = \{0\}$, which is contradictory. In conclusion, $L_\mathrm{e}$ must trivially intersect some $L_m = V_{\alpha_1\dots\alpha_r}$, hence a magnetic Lagrangian always exists.

\subsubsection{A simple remark about gapless SPTs}
While the anomalies discussed above are obstructions to SPTs for the symmetry in question, we may also consider so-called gapless SPTs (gSPTs) \cite{Scaffidi:2017ppg,Bhardwaj:2024qrf,Antinucci:2024ltv}, which are associated to isotropic subgroups $S \subset \mathbb{Z}_{p^k}^{2r}$, not necessarily maximal (c.f.~the condensable algebras of \cite{Bhardwaj:2024qrf}), that intersect $L_\mathrm{e}$ trivially. Our viewpoint allows us to make the following remark: If $L_\mathrm{e}$ consists of only torsional elements, that is $p\mathbb{Z}_{p^k}^{2r} \subset L_\mathrm{e}$ (which is equivalent to $L_\mathrm{e}$ being torsional for $r=1$, but stronger for $r>1$), then we have $p^{k-1} \mathbb{Z}_{p^k}^{2r} \subset L_\mathrm{e}$ so that $L_\mathrm{e}$ intersects every nontrivial subgroup of $\mathbb{Z}_{p^k}^{2r}$ nontrivially. The only admissible $S$ is therefore the trivial group. This is associated to the so-called \emph{canonical} gSPT \cite{Bhardwaj:2024qrf}, characterized by $F^{(1)}$ acting faithfully on a set of non-topological lines. 

\medskip

Phrased differently, in a theory with the $\mathbb{Z}_{p^k}^{2r}$ BF SymTFT, in a global form where the 1-form symmetry contains no factor $\mathbb{Z}_{p^k}$, any IR phase contains either a charged topological line (so the 1-form symmetry is spontaneously broken), or else non-topological lines of every charge. 

\medskip

The straightforward interpretation of this is that the BF term encodes a mixed anomaly for each pair $\mathbb{Z}_{p^{l_i}} \times \mathbb{Z}_{p^{k - l_i}}$ in \eqref{eq:27}, which must be matched by the appropriate charged lines. We will explore further consequences of these ideas in future work.

\subsection{Application to deformations of class $\mathcal{S}$ theories}
\label{sec:appl-gapl-deform}

In this section we use the technology built up in the last sections to find non-invertible symmetry preserving deformations of selected class $\mathcal{S}$ theories to models with less supersymmetries.\footnote{\, We refer our readers that are not familiar with class $\mathcal S$ theories to \cite{Tachikawa:2013kta,Tachikawa:2015bga} for very nice thorough reviews.} We will focus on class $\mathcal{S}$ theories of type $\mathfrak{a}_{N-1}$, associated to a closed Riemann surface of genus $g$. As an application of our classification of symmetry preserving Lagrangians, it is easy to identify IR phases that must exhibit either spontaneous breaking or be gapless in the infrared. In this section we present some first simple examples, and we reserve to return to a systematic analysis in the future.\footnote{\, In particular, we limit ourselves to a coarse analysis, and we expect a more interesting and refined phase structure to arise, refining our classification -- see footnote \ref{foot:coarse}.}

\medskip

In short, we use class $\mathcal{S}$ theories with a non-invertible symmetry (see e.g. \cite{Bashmakov:2022jtl,Bashmakov:2022uek,Antinucci:2022cdi}) as the theoretical laboratory of our introduction. By this we mean that we look for a symmetry preserving relevant deformation of the fixed point which triggers a flow to an IR phase. In particular, this IR phase can break partially or in its entirety the supersymmetries of the class $\mathcal S$ model. Using our result it is easy to identify examples that cannot admit symmetry preserving trivially gapped IR phases. Then the IR is either gapless or the symmetry is spontaneously broken. Of course, using the topological symmetry theory, we can only study these possibilities abstractly and cannot decide the dynamical question of what phases are realized in a given QFT. For example, one should exhibit specific order parameters and demonstrate that these pick VEVs corresponding to a SSB in the IR.

\medskip

In this way, we can use the existence of highly supersymmetric conformal theories to constrain the structure of the vacua of theories with less and even no supersymmetry.

\subsubsection{Vacua of $\mathcal{N}=1^{*}$ theories and spontaneous symmetry breaking}
\label{sec:vacua-n=1-theories}
As a first example we recast the results of \cite{Damia:2023ses} about $\mathcal{N}=1^{*}$ theories in the language of SymTFT.

\medskip

$\mathcal{N}=1^{*}$ theories are deformations of $\mathcal{N}=4$ SYM obtained by giving masses to the three $\mathcal{N}=1$ chiral multiplets in $\mathcal{N}=4$ vector multiplet. We are interested in $\mathcal{N} = 4$ theory at $\tau = i$ which has a duality symmetry corresponding to the $\mathsf{S}$ transformation and $\tau=e^{\frac{2\pi i}{3}}$ which has a triality symmetry. These models have been analyzed in detail in \cite{Damia:2023ses}, hence here we are very brief.

\medskip

The mass terms for the hypermultiplets are not invariant under the duality and triality symmetry. However, these non-invertible symmetries can be composed with suitable $R$-symmetries to realize non-invertible symmetries which are preserved by the mass terms. These non-invertible symmetries are described by the same SymTFT as the original duality and triality symmetries.

\paragraph{The case of non-invertible duality symmetries.} The duality symmetry for $\mathcal{N}=1^{*}$ theory at $\tau = i$, with $\mathfrak{su}(N)$ gauge algebra is described starting from a $\mathbb{Z}_{N}$ BF theory and then orbifolding it with respect to $\mathbb G = \mathbb{Z}_{4}$, generated by the $\mathsf{S}$ transformation. 

\medskip The values of $N$ for which there are trivially gapped phases are described in Appendix \ref{sec:exist-invar-lagr-1} and the first few values of $N$ for which these exist are tabulated in the first row of Table \ref{tab:SL2Z-N}. Let us discuss different cases:
\paragraph{$N=2$:}
Here, the $\mathsf{S}$-symmetry reduces to a $\mathbb{Z}_2$ symmetry rather than $\mathbb{Z}_4$, since $\mathsf{S}^2 = -1 = 1 \bmod{2}$. It exchanges the labels $A$ and $B$ of the $\mathbb{Z}_2^A \times \mathbb{Z}_2^B$ group of simple surface defects of the $\mathbb{Z}_2$ BF theory. Before gauging $\mathsf{S}$, the BF theory has three simple surface defects $\Phi_A$, $\Phi_B$ and $\Phi_{A+B}$, and three simple boundary conditions: $\mathcal B_A^{(0)}$, $\mathcal B_B^{(0)}$ and $\mathcal B_{A+B}^{(0)}$. The latter is $\mathsf{S}$-invariant while the other two are exchanged by $\mathsf{S}$.

There are four topological boundary conditions, namely:\footnote{\, Notice that compared to \cite{Damia:2023ses} there is an extra topological boundary condition. For us this is there in principle as one of the allowed possibilities, however not all of them must be realized. In this case such boundary condition corresponds to gauging of $\mathsf{S}$-duality, and is not realized. Except for this subtlety our analysis agrees with the results in \cite{Damia:2023ses}.}
\begin{align}
\label{eq:96}
  \mathcal B_A^{(0)} ~, && \mathcal B_B^{(0)} ~, &&  \mathcal B_{A+B}^{(0)} ~, &&  \mathcal B_{A+B}^{(2)} ~,
\end{align}
corresponding to states $\ket{A;0}$, $\ket{B;0}$, $\ket{A+B;0}$ and $\ket{A+B;2}$.
In a theory with noninvertible $\mathsf{S}$-symmetry, the physical boundary must flow to a $\mathsf{S}$-invariant boundary condition. The irreducible such boundary conditions are
\begin{align}
  \mathcal B_A^{(0)} + \mathcal B_B^{(0)} ~, && \mathcal B_{A+B}^{(0)} ~, && \mathcal B_{A+B}^{(2)} ~,
\end{align}
corresponding to states $\ket{A;0} + \ket{B;0}$, $\ket{A+B;0}$ and $\ket{A+B;2}$.

\medskip

For this example, let us consider different possibilities for three global forms with $0$-form symmetry in detail:
\begin{itemize}
\item If the global form for the theory corresponds to the state $\mathcal B_A^{(0)}$ in the Hilbert space of the gauged SymTFT, and the dynamical boundary flows to either of $\mathcal B_A^{(0)} + \mathcal B_B^{(0)}$ or $\mathcal B_{A+B}^{(0)}$, the operators $\mathcal W(M_{1})$ can end on both boundaries, giving rise to a topological local operator, which we also call $\mathcal{W}$, in the four-dimensional theory. Furthermore, the theory has a $\mathbb{Z}_2$ zero-form symmetry generated by $\sigma(M_{3})$. Since $\mathcal W(M_{1})$ links with $\sigma(M_{3})$ in the bulk, according to \eqref{eq:74}, $\mathcal{W}$ is charged: $\sigma \mathcal{W} = -\mathcal{W} \sigma$. Therefore $\mathcal{W}$ acts as the order parameter parameterizing the spontaneous breaking of $\mathsf{S}$.

  If the dynamical boundary is $\mathcal B_{A+B}^{(0)}$, the one-form symmetry, with $\Phi_A$ as order parameter, is trivial in both vacua. However, if the dynamical boundary is $\mathcal B_A^{(0)} + \mathcal B_B^{(0)}$, the one-form symmetry is spontaneously broken in the $\mathcal B_A^{(0)}$ vacuum (as the topological surface operator $\Phi(A \otimes M_{2})$ with $M_{2} \in H_{2}(M_{4},\mathbb{Z})$ can end on both boundaries, and is charged under $\Phi_B$), but trivial in the $\mathcal B_B^{(0)}$ vacuum ($\Phi_A$ does not take a VEV, and $\Phi_B$ acts trivially). As \eqref{eq:80} shows, the broken non-invertible symmetry exchanges these physically distinct vacua: $\sigma$ is a domain wall between $\mathcal B_A^{(0)}$ and $\mathcal B_B^{(0)}$.

  On the other hand, if the dynamical boundary flows to $\mathcal B_{A+B}^{(2)}$ we obtain a trivially gapped state.
  \item If the global form for the theory is $\mathcal B_B^{(0)}$ the analysis is exactly the same except for interchanging $A$ and $B$.
  \item If the global form for the theory is $\mathcal B_{A+B}^{(0)}$ and the dynamical boundary flows to $\mathcal B_A^{(0)} + \mathcal B_B^{(0)}$, the $0$-form symmetry is again spontaneously broken, exchanging the physically equivalent vacua $\mathcal B_A^{(0)}$ and $\mathcal B_B^{(0)}$ (the 1-form symmetry, with $\Phi_{A+B}$ as order parameter, being trivial in both). If the dynamical boundary flows to $\mathcal B_{A+B}^{(0)}$ both zero- and one-form symmetry are spontaneously broken. There is no interplay between the two symmetries, as the line order parameter $\Phi_{A+B}$ is $\mathsf{S}$-invariant; $\sigma$ is a domain wall between physically equivalent vacua.
    Lastly, if the dynamical boundary flows to $\mathcal B_{A+B}^{(2)}$, the zero-form symmetry is trivial but the one-form symmetry is spontaneously broken, as $\Phi_{A+B}$ takes a VEV.
\end{itemize}
This analysis refines the results in \cite{Damia:2023ses} by taking into account the possibility of gauging the $\mathsf{S}$ symmetry. Let us briefly note how to translate our notation to the one used there. The map between the global forms is the obvious one: $\mathcal B_A^{(0)}$ corresponds to $\mathrm{SU}(2)$, $\mathcal B_B^{(0)}$ to $\mathrm{PSU}(2)_{0}$ and $\mathcal B_{A+B}^{(0)}$ to $\mathrm{PSU}(2)_{1}$. To this list of global forms we add $\mathcal B_{A+B}^{(2)}$ which is obtained by gauging the invertible $\mathsf{S}$ symmetry in $\mathcal B_{A+B}^{(0)}$ and hence this global form possesses a dual $2$-form symmetry.

\medskip

It is slightly less obvious to identify the three vacua they consider with these topological boundary conditions. Nevertheless the correspondence is: $\mathcal B_A^{(0)} \to H, \mathcal B_B^{(0)} \to C^{(0)}$ and $\mathcal B_{A+B}^{(2)} \to C^{(1)}$. Using this correspondence it is straightforward to reproduce the tables in Equation (3.16) of \cite{Damia:2023ses}.

\medskip

For example, for the $SU(2)$ global form we find that in the $H$ vacuum the $\mathbb Z_2^{(1)}$ symmetry is spontaneously broken resulting in a topological order with $\mathbb Z_2^{(1)}$ form symmetry. In the $C^{(0)}$ vacuum there is no overlap and one obtains a trivially gapped theory, while in the  $C^{(1)}$ vacuum there is overlap and one again obtains a topological order with $\mathbb Z_2^{(1)}$ form symmetry. The $H$ vacuum and the $C^{(0)}$ vacuum are swapped by the action of the non-invertible symmetry defects, as indeed $\sigma_\textsf{S}\mathcal B_A^{(0)} = \mathcal B_B^{(0)}$ in this case. The fact that the two vacua have different features follows by the non-invertible nature of $\sigma_\textsf{S}$. Because of this action, these two vacua correspond to the SSB phase, while the $C^{(1)}$ is in a trivially gapped symmetry preserving phase.

\medskip

This leaves the question of $\mathcal B_{A+B}^{(2)}$. From the analysis of \cite{Damia:2023ses} we conclude that this vacuum is not realized by $\mathcal{N} = 1^{*}$ theories. It would be an interesting endeavor to construct such a theory. Heuristically we can expect that a theory with such a vacuum is not a gauge theory in the normal sense since the running of the gauge coupling breaks the $\mathsf{S}$ symmetry and hence must lift this vacuum. A theory that flows to this vacuum is as a result stuck at the self dual point. This makes its construction an interesting avenue for further study.

\paragraph{$N$ a power of 2:}
If $N=2^{2k}$ then there is precisely one $\mathsf{S}$-invariant lagrangian i.e. $L_{0} = 2^{k}\mathbb{Z}_{2^{2k}}^{2}$. Similarly for $N=2^{2k+1}$, there is once again a single invariant lagrangian $L_{0}$ which is torsional with global symmetry $\mathbb{Z}_{2^{k}} \times \mathbb{Z}_{2^{k+1}}$. Hence in both cases there is one TBC $\ket{L_{0}; 0}$ which realizes $\mathsf{S}$ as an invertible symmetry and one TBC $\ket{L_{0}; 2}$ in which it is gauged. However since the lagrangian $L_{0}$ intersects non-trivially with any other lagrangian, if the dynamical theory flows to $\ket{L_{0}; 2}$, the one-form symmetry must be spontaneously broken. Hence there are no trivially gapped phases.

\paragraph{$N = p^{k}$ with $p = 1 \mod 4$ a prime:}
If $N$ is prime of a prime $p$ congruent to $1 \mod 4$, the $\mathsf{S}$ transformation is diagonalizable and there are two $\mathsf{S}$-invariant lagrangians corresponding the span of two eigenvectors $u_{\lambda}$ and $u_{\lambda^{-1}}$ of $\mathsf{S}$ where $\lambda^{2} = -1 \mod N$. Hence there are two topological boundary conditions which realize $\mathsf{S}$ as (possibly with mixed anomalies with one-form symmetry) invertible symmetry which we label $\ket{\lambda; 0}$ and $\ket{\lambda; 0}$ and two boundary conditions in which $\mathsf{S}$ is gauged and which we label as $\ket{\lambda; 2}$ and $\ket{\lambda^{-1}; 2}$. All other lagrangians form doublets under $\mathsf{S}$-transformation and these lagrangians realize non-invertible symmetry. If the global form of the theory has non-invertible symmetry and the dynamical boundary flows to $\ket{\lambda; 2}$ we obtain a trivially gapped state.

If $N = p^{2k}$ there is an additional invariant lagrangian $L_{0} = p^{k}\mathbb{Z}_{p^{2k}}^{2}$. However since it is torsional, it intersects non-trivially with any other lagrangian and hence if the theory flows to $\ket{L_{0}; 2}$ in IR the one form symmetry must be broken.

\paragraph{$N=p^{2k+1}$ with $p = 3 \mod 4$ a prime:}
If $N$ is an odd power of a prime congruent to $3 \mod 4$, there are no $\mathsf{S}$-invariant lagrangians and all lagrangians form doublets under $\mathsf{S}$-duality. So either $\mathsf{S}$-duality is spontaneously broken by the RG flow or the theory flows to a gapless phase.

\paragraph{$N=p^{2k}$ with $p = 3 \mod 4$ a prime:}
If $N$ is even power of a prime congruent to $3 \mod 4$ there is a torsional $\mathsf{S}$-invariant lagrangian $L_{0} = p^{k}\mathbb{Z}_{p^{2k}}^{2}$. However a theory flowing to $\ket{L_{0};2}$ must spontaneously break $1$-form symmetry.

\paragraph{General Case:}
Generally a theory with non-invertible symmetry can flow to a trivially gapped phase if there is a free $\mathsf{S}$-invariant lagrangian in $\mathbb{Z}_{N}^{2}$. If $N = \prod_{i}p_{i}^{k_{i}}$, such a lagrangian exists iff it exist for each of $\mathbb{Z}_{p_{i}^{k_{i}}}$. In all other cases either one-form or zero-form symmetry or both must be spontaneously broken.

\paragraph{The case of triality.} At $\tau = e^{\frac{2\pi i}{3}}$, the triality symmetry of $\mathcal{N}=1^{*}$ theory is described by $\mathbb{Z}_{N}$ BF theory with $\mathsf{ST}$ gauged. We will be more brief in describing the structure of the vacua for this theory since the case is quite similar to the case of duality.

For $N=3$, there are five topological boundary conditions. Three of these $\ket{A; 0}$, $\ket{B; 0}$ and $\ket{A-B; 0}$ form a triplet under $\mathsf{ST}$-duality and possess non-invertible symmetry. $\ket{A+B; 0}$ realizes $\mathsf{ST}$ as an invertible symmetry while in $\ket{A+B; 2}$ this symmetry is gauged so it possesses a two form symmetry. If the global form corresponds to one of theories in the triplet we obtain a gapped phase if the dynamical boundary flows to $\ket{A+B; 2}$. In all other cases either zero form or one-form symmetry is spontaneously broken.

If $N=p^{2k+1}$ with $p = 1 \mod 6$ a prime, then $\mathsf{ST}$ is diagonalizable and there two TBCs which realizes $\mathsf{ST}$ as an invertible symmetry which we denote by $\ket{u_{\lambda}; 0}$ and $\ket{u_{\lambda^{-1}}; 0}$ with $\lambda$ an eigenvalue satisfying $\lambda^{2}-\lambda+1 = 0$. Dual to these there are two TBCs with $2$-form symmetry $\ket{u_{\lambda}; 2}$ and $\ket{u_{\lambda^{-1}}; 2}$. The rest of the TBCs form triplets under $\mathsf{ST}$. A theory with such a global form and $\mathbb{Z}_{N}$ one-form symmetry realizes a trivially gapped phase if the dynamical boundary flows to $\ket{\lambda; 2}$ or $\ket{\lambda^{-1}; 2}$.

Generally trivially gapped phases exist if $N = 3^{s}N^{\prime}$ where $s \in \{0,1\}$ and every prime factor of $N^{\prime}$ is congruent to $1 \mod 6$. For all other cases either there is spontaneous symmetry breaking (of zero form or one-form symmetry or both) in the IR or else the theory flows to a gapless phase.

\subsubsection{Deformations of class $\mathcal{S}$ theories}
\label{sec:more-deform-class}
In this section we construct some more deformations of class $\mathcal{S}$ theories which break supersymmetry but preserve non-invertible defects of higher orders. Recall that class $\mathcal{S}$ theory obtained by compactifying $6d, (2,0)$ theory has non-invertible duality defects corresponding to elements of the mapping class group stabilizing $\Sigma_g$ \cite{Gukov:2020btk,Bashmakov:2022jtl,Bashmakov:2022uek,Antinucci:2022cdi}.

\medskip

This stabilizer plays the role of $\mathbb G$ for this class of examples. To find a mass deformation that commutes with the action of the corresponding non-invertible $\mathbb G$--duality defect is a straightforward exercise in geometry. We use the fact that whenever the Riemann surface $\Sigma_{g}$ has a small non-contractible closed curve, there is a duality frame such that we can use associate to it a weakly coupled $\mathcal{N} = 2$ vector multiplet of type $\mathfrak{a}_{N-1}$.\footnote{\ This can be seen cutting open the Riemann surface $\Sigma_{g}$ along this curve to obtain a surface $\Sigma_{g-1,2}$ with two boundaries and hence $\mathfrak{a}_{N-1} \times \mathfrak{a}_{N-1}$ flavor symmetry. In the standard class $\mathcal{S}$ construction the theory on $\Sigma_{g}$ is obtained by gauging the diagonal of $\mathfrak{a}_{N-1} \times \mathfrak{a}_{N-1}$ symmetry with gauge coupling given by the length of the boundary \cite{Gaiotto:2009we,Gaiotto:2009hg}.} If $M$ is an element of mapping class group that stabilizes $\Sigma_{g}$ and moreover it is such that there is a collection of such small closed curves which are permuted among another by $M$, such permutation encodes the action of the $\mathbb G$--duality defect on the corresponding components of the associated $\mathcal N=2$ vector multiplets. It is easy to write down a mass term for the scalars as well as the fermions in the corresponding $\mathcal N=2$ vector multiplets corresponding to this collection of long tubes, which is left invariant by the action of the corresponding $\mathbb G$--duality defect.

\medskip

In the remaining part of this section we give some examples. The resulting theories can be $\mathcal N=1$ or $\mathcal N=0$. In the $\mathcal N=1$ cases we believe these correspond to some of the $\mathcal N=1$ theories of class $\mathcal S$ \cite{Bah:2012dg} (which then have non-invertible symmetries).

\medskip

Let us consider the genus $g$ Riemann surface obtained by identifying the diagonally opposite edges of a $4g+2$ side polygon in hyperbolic space. The polygon is centered at the origin of the Poincaré disk and we restrict ourselves to polygons such that all of their vertices lie at the same distance $r$ from the origin. Such a polygon is determined by the angles $\alpha_{i}$ subtended by (Euclidean) lines from consecutive vertices to the center. Since we need to identify the opposite edges, $\alpha_{i+2g+1} = \alpha_{i}$. The independent angles have one relation between them, i.e.~they sum to $\pi$. After the identification, two distinct vertices remain. \footnote{\ Both vertices must be surrounded by a total angle $2\pi$. Alternatively, the angle deficit $(18 - 2)\pi - \sum_i \theta_i$ is the hyperbolic area, which by the Gauss--Bonnet theorem is $-2\pi\chi = 12\pi$, so $\sum_i \theta_i = 4\pi$. If one relaxes this condition, one obtains surfaces with conical defects. For certain values of the deficit angles, such defects are believed to correspond to regular punctures in the class $\mathcal{S}$ construction \cite{Bobev:2019ore}. Extending our analysis for models in this class is an interesting exercise we leave for future work.}

A Riemann surface constructed in this way has a symmetry whenever there is some periodicity in $\alpha_{i}$. E.g., the periodicity  $\alpha_{i+2g+1} = \alpha_{i}$ means that there is always a $\mathbb{Z}_{2}$ symmetry corresponding to charge conjugation. Smaller periodicity results in bigger symmetries. The case of smallest periodicity, i.e.~when all the angles are equal, results in $\mathbb{Z}_{4g+2}$ symmetry described in \cite{Bashmakov:2022jtl}. For a larger periodicity only a subgroup of this symmetry survives.

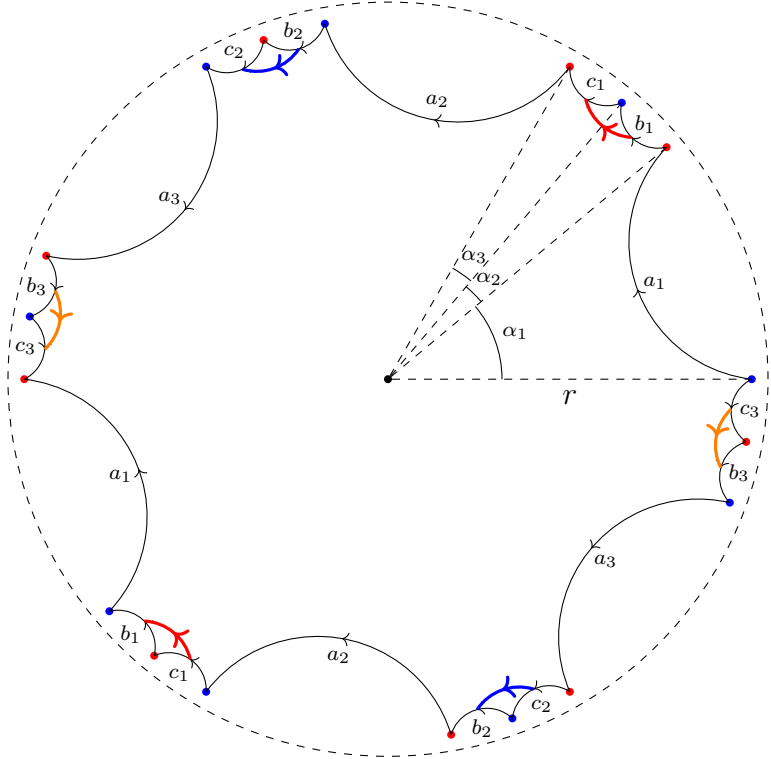
\begin{figure}
  \centering
  \begin{tikzpicture}[scale=5]
    \draw[dashed] (0,0) circle[radius=1];
    \node[smalldot] (O) at (0,0) {};
    \newcommand{\euclRad}{0.957158638496845}
    \node[smalldot,blue] (p0) at (0:\euclRad) {};
    \node[smalldot,red] (p1) at (40:\euclRad) {};
    \node[smalldot,blue] (p2) at (50:\euclRad) {};
    \node[smalldot,red] (p3) at (60:\euclRad) {};
    \node[smalldot,blue] (p4) at (100:\euclRad) {};
    \node[smalldot,red] (p5) at (110:\euclRad) {};
    \node[smalldot,blue] (p6) at (120:\euclRad) {};
    \node[smalldot,red] (p7) at (160:\euclRad) {};
    \node[smalldot,blue] (p8) at (170:\euclRad) {};
    \node[smalldot,red] (p9) at (180:\euclRad) {};
    \node[smalldot,blue] (p10) at (220:\euclRad) {};
    \node[smalldot,red] (p11) at (230:\euclRad) {};
    \node[smalldot,blue] (p12) at (240:\euclRad) {};
    \node[smalldot,red] (p13) at (280:\euclRad) {};
    \node[smalldot,blue] (p14) at (290:\euclRad) {};
    \node[smalldot,red] (p15) at (300:\euclRad) {};
    \node[smalldot,blue] (p16) at (340:\euclRad) {};
    \node[smalldot,red] (p17) at (350:\euclRad) {};
    \draw[->-] (p0) arc[start angle=263.1445,end angle=136.8555,radius=0.3669];
    \node at (0.7031,0.2559) {\scriptsize$a_1$};
    \draw[->-] (p1) arc[start angle=283.4277,end angle=166.5723,radius=0.0979];
    \node at (0.6766,0.6766) {\scriptsize$b_1$};
    \draw[->-] (p2) arc[start angle=293.4277,end angle=176.5723,radius=0.0979];
    \node at (0.5488,0.7838) {\scriptsize$c_1$};
    \draw[->-] (p3) arc[start angle=-36.8555,end angle=-163.1445,radius=0.3669];
    \node at (0.1299,0.7369) {\scriptsize$a_2$};
    \draw[->-] (p4) arc[start angle=-16.5723,end angle=-133.4277,radius=0.0979];
    \node at (-0.2477,0.9243) {\scriptsize$b_2$};
    \draw[->-] (p5) arc[start angle=-6.5723,end angle=-123.4277,radius=0.0979];
    \node at (-0.4044,0.8672) {\scriptsize$c_2$};
    \draw[->-] (p6) arc[start angle=23.1445,end angle=-103.1445,radius=0.3669];
    \node at (-0.5732,0.4810) {\scriptsize$a_3$};
    \draw[->-] (p7) arc[start angle=43.4277,end angle=-73.4277,radius=0.0979];
    \node at (-0.9243,0.2477) {\scriptsize$b_3$};
    \draw[->-] (p8) arc[start angle=53.4277,end angle=-63.4277,radius=0.0979];
    \node at (-0.9532,0.0834) {\scriptsize$c_3$};
    \draw[-<-] (p9) arc[start angle=83.1445,end angle=-43.1445,radius=0.3669];
    \node at (-0.7031,-0.2559) {\scriptsize$a_1$};
    \draw[-<-] (p10) arc[start angle=103.4277,end angle=-13.4277,radius=0.0979];
    \node at (-0.6766,-0.6766) {\scriptsize$b_1$};
    \draw[-<-] (p11) arc[start angle=113.4277,end angle=-3.4277,radius=0.0979];
    \node at (-0.5488,-0.7838) {\scriptsize$c_1$};
    \draw[-<-] (p12) arc[start angle=143.1445,end angle=16.8555,radius=0.3669];
    \node at (-0.1299,-0.7369) {\scriptsize$a_2$};
    \draw[-<-] (p13) arc[start angle=163.4277,end angle=46.5723,radius=0.0979];
    \node at (0.2477,-0.9243) {\scriptsize$b_2$};
    \draw[-<-] (p14) arc[start angle=173.4277,end angle=56.5723,radius=0.0979];
    \node at (0.4044,-0.8672) {\scriptsize$c_2$};
    \draw[-<-] (p15) arc[start angle=203.1445,end angle=76.8555,radius=0.3669];
    \node at (0.5732,-0.4810) {\scriptsize$a_3$};
    \draw[-<-] (p16) arc[start angle=223.4277,end angle=106.5723,radius=0.0979];
    \node at (0.9243,-0.2477) {\scriptsize$b_3$};
    \draw[-<-] (p17) arc[start angle=233.4277,end angle=116.5723,radius=0.0979];
    \node at (0.9532,-0.0834) {\scriptsize$c_3$};
    \coordinate (x1-0) at (0.6413,0.6413);
    \coordinate (x2-0) at (-0.6413,-0.6413);
    \coordinate (y1-0) at (0.5202,0.7429);
    \coordinate (y2-0) at (-0.5202,-0.7429);
    \draw[->-,very thick,red] (x1-0) arc[start angle=-93.0829,end angle=-166.9171,radius=0.1316];
    \draw[-<-,very thick,red] (x2-0) arc[start angle=86.9171,end angle=13.0829,radius=0.1316];
    \coordinate (x1-1) at (-0.2347,0.8760);
    \coordinate (x2-1) at (0.2347,-0.8760);
    \coordinate (y1-1) at (-0.3833,0.8219);
    \coordinate (y2-1) at (0.3833,-0.8219);
    \draw[->-,very thick,blue] (x1-1) arc[start angle=-33.0829,end angle=-106.9171,radius=0.1316];
    \draw[-<-,very thick,blue] (x2-1) arc[start angle=146.9171,end angle=73.0829,radius=0.1316];
    \coordinate (x1-2) at (-0.8760,0.2347);
    \coordinate (x2-2) at (0.8760,-0.2347);
    \coordinate (y1-2) at (-0.9034,0.0790);
    \coordinate (y2-2) at (0.9034,-0.0790);
    \draw[->-,very thick,orange] (x1-2) arc[start angle=26.9171,end angle=-46.9171,radius=0.1316];
    \draw[-<-,very thick,orange] (x2-2) arc[start angle=206.9171,end angle=133.0829,radius=0.1316];
    \draw[dashed] (O) -- (p0) node[midway,below] {$r$};
    \draw[dashed] (O) -- (p1);
    \draw[dashed] (O) -- (p2);
    \draw[dashed] (O) -- (p3);
    \tkzMarkAngle[size=.3](p0,O,p1)
    \tkzLabelAngle[pos=.35](p0,O,p1){\scriptsize$\alpha_1$}
    \tkzMarkAngle[size=.32](p1,O,p2)
    \tkzLabelAngle[pos=.37](p1,O,p2){\scriptsize$\alpha_2$}
    \tkzMarkAngle[size=.34](p2,O,p3)
    \tkzLabelAngle[pos=.39](p2,O,p3){\scriptsize$\alpha_3$}
  \end{tikzpicture}
  \caption{A genus 4 surface constructed by identifying opposite sides of an 18-gon. In the limit $\alpha_2 = \alpha_3 \ll \frac{2\pi}{6}$, the three non-contractible cycles shown are small, giving rise to weakly coupled $\mathcal{N} = 2$ vector multiplets.}
  \label{fig:18-gon}
\end{figure}

Let us now specialize to $g=4$ in which case the polygon has 18 sides. If we consider the subclass of polygons such that the angles are $3$-periodic we obtain a two-parameter family of Riemann surfaces. The free parameters are three angles $\alpha_{1},\alpha_{2},\alpha_{3}$ with the relation
\begin{align}
  \label{eq:71}
  \alpha_{1} + \alpha_{2} + \alpha_{3} = \frac{2\pi}{6} ~.
\end{align}
This whole family has a $\mathbb{Z}_{6}$ isometry giving rise to a $\mathbb{Z}_{6}$ non-invertible symmetry in the corresponding class $\mathcal{S}$ theory. We work in the limit where $\alpha_{0} \equiv \alpha_{2} = \alpha_{3} \ll \frac{2\pi}{6}$. In this limit we can find a trio of closed curves as shown in \cref{fig:18-gon}. Cutting open the Riemann surface along this curve we obtain genus 1 surfaces with six boundaries. Hence gluing these back together we obtain three weakly coupled $\mathcal{N} = 2$ vector multiplets each at the same gauge coupling $\tau$. By tuning $\alpha_{0}$ we can make this coupling arbitrarily small.

In the $\mathcal{N}=1$ language, each of these $\mathcal{N}=2$ adjoint valued vector multiplets contains a vector multiplet $V_{s}$ and a chiral multiplet $\varphi_{s}$, with $s \in \{1,2,3\}$. The chiral multiplet $\varphi_{s}$ contains a scalar $\phi_{s}$ and a left hand hand fermion $\psi_{s}$. While the vector multiplet $V_{s}$ splits into gauge potential $A_{s}$ and a right handed fermion $\overline{\chi}_{s}$. Under the $\mathbb{Z}_{6}$ symmetry, it is easy to see that\footnote{\ Since the cube of the symmetry acts as $-1$ on the homology of Riemann surface, it corresponds to charge conjugation.}
\begin{gather}
  \label{eq:72}
  V_{1} \to V_{2} \to V_{3} \to V_{1} ~, \nonumber \\
  \varphi_{1} \to \varphi_{2} \to \varphi_{3} \to \varphi^{\dagger}_{1} ~.
\end{gather}
In view of this action it is straightforward to write down $\mathbb{Z}_6$-invariant mass term that break supersymmetry partially to $\mathcal{N} = 1$. It is given by the superpotential,
\begin{align}
  \label{eq:73}
  \sum_{s=1}^{3}\tr(\varphi_{s}^{2}).
\end{align}
There are also mass terms that break all supersymmetry. These correspond to mass terms involving just bosons or just fermion in the potential obtained from the superpotential in \eqref{eq:76}. These are,
\begin{align}
  \sum_{s = 1}^{3}(\psi_{s}\psi_{s} + \overline{\psi}_{s}\overline{\psi}_{s})~, && \sum_{s=1}^{3} \phi_{s}^{2} ~.
\end{align}
Adding such mass terms preserves the $\mathbb{Z}_{6}$ non-invertible duality symmetry. The resulting theory has trivially gapped boundary conditions only when there are $\mathbb{Z}_{6}$-invariant lagrangians. When this happens it can be easily determined by using our techniques. The generator of $\mathbb{Z}_{6}$ is the cube of the $\mathbb{Z}_{18}$ symmetry given in \cite{Bashmakov:2022jtl} and its characteristic polynomial is
\begin{align}
  \label{eq:75}
  P(x) = (x+1)^{2}(x^{2} - x + 1)^{3} = \Phi_{2}^{2}(x)\Phi_{6}^{3}(x) ~,
\end{align}
a product of the square of the second cyclotomic polynomial and the cube of the sixth cyclotomic polynomial. Hence using our analysis, the deformed $\mathfrak{a}_{N-1}$ theory can flow to a non-invertible $\mathbb{Z}_{6}$ symmetry preserving gapped phase only if the prime decomposition of $N$:
\begin{align}
  \label{eq:76}
  N = \prod_i p_i^{k_i},
\end{align}
is such that for all $i$ one of the following is true,
\begin{itemize}
  \item $p_i = 3$,
  \item $p_i = 1 \mod 3$,
  \item $k_i = 0 \mod 2$.
\end{itemize}
The first few $N$ for which this holds are\footnote{\ We analyze $\Phi_6(x) = x^2 - x + 1$ as the characteristic polynomial of $\mathsf{ST}$ in \cref{sec:example-ST,sec:exist-invar-lagr-1}, and \eqref{eq:18-gon-obstructed-N} is the complement of the second row of \cref{tab:SL2Z-N}, namely the numbers that cannot be written as $x^2 + xy + y^2$, or equivalently as $k^2 N'$ where $-3$ is a quadratic residue mod $4N'$.}
\begin{equation}
  \label{eq:18-gon-obstructed-N}
  2, 5, 6, 8, 10, 11, 14, 15, 17, 18, 20, 22, 23, 24, 26, 29, 30, 32, 33, 34, 35, 38, 40, \dots.
\end{equation}
In all these cases the main question is whether the symmetry is spontaneously broken or gapless. The $N$ for which there are trivially gapped vacua which preserve both $0$-form and $1$-form symmetry are of the form $N=3^{s}N^{\prime}$ where $s \in \{0,1\}$ and every prime factor of $N^{\prime}$ is congruent to $1 \mod 3$. For all other value the RG flow must spontaneously break some symmetry or else lead to a gapless phase. Notice that the $N$ for which trivially gapped vacua exist are precisely those for which such vacua exist with triality symmetry described in Section \ref{sec:vacua-n=1-theories}. This is because in both cases the structure of symmetry breaking is determined by factorization of $\Phi_{3}(x)$.

\medskip

This setup can easily be generalized. E.g.~when $4g+2$ is a multiple of $3$, we can consider the polygons such that the angles are three-periodic. This always gives us a two-parameter family of Riemann surfaces with a $\mathbb{Z}_{2h}$ symmetry with $h = \frac{4g+2}{6}$ and our construction carries over straightforwardly.

\medskip

Expanding this setup further, there is nothing special about periodicity $3$. Indeed whenever $d$ is a divisor of $2g+1$,\footnote{\ $d$ is necessarily odd since $2g+1$ is odd.} we get a $(d-1)$-parameter family of Riemann surfaces corresponding to polygons in which angles have periodicity $d$ and which have $\mathbb{Z}_{2h}$ symmetry with $h = \frac{2g+1}{d}$. In this case it is again possible to find a collection of $h$ curves which give rise to $h$ vector multiplets. In these multiplets we can find $h$ $\mathcal{N}=1$ chiral multiplets $\varphi_{i}$ such that the effect of the $\mathbb{Z}_{2h}$ symmetry is,
\begin{align}
  \label{eq:77}
  \varphi_{1} \to \varphi_{2} \to \cdots \to \varphi_{h} \to \varphi_{1}^{\dagger} ~.
\end{align}
Hence, we can write down an $\mathcal{N}=1$ mass term
\begin{align}
  \label{eq:78}
  \sum_{s = 1}^{h} \tr(\varphi_{s}^{2}) ~,
\end{align}
and non-supersymmetric mass terms
\begin{align}
  \label{eq:79}
  \sum_{s=1}^{h} \tr(\psi_{s} \psi_s + \overline{\psi}_{s}\overline{\psi}_{s}) \qq{or} \sum_{s=1}^{h} \tr(\phi_{s}^{2}).
\end{align}

\begin{table}[t]
  \centering
  \small
  \renewcommand{\arraystretch}{1.1}
  \begin{tabular}{r|r|lllll}
    $g$ & $2g + 1$ & $\det(x - f_{4g + 2}^d)$ \\
    \hline
    4 & 9 & $f_{18}^{3}$: $\Phi_{2}^{2}\Phi_{6}^{3}$ \\
    7 & 15 & $f_{30}^{3}$: $\Phi_{2}^{2}\Phi_{10}^{3}$ & $f_{30}^{5}$: $\Phi_{2}^{4}\Phi_{6}^{5}$ \\
    10 & 21 & $f_{42}^{3}$: $\Phi_{2}^{2}\Phi_{14}^{3}$ & $f_{42}^{7}$: $\Phi_{2}^{6}\Phi_{6}^{7}$ \\
    12 & 25 & $f_{50}^{5}$: $\Phi_{2}^{4}\Phi_{10}^{5}$ \\
    13 & 27 & $f_{54}^{3}$: $\Phi_{2}^{2}\Phi_{6}^{3}\Phi_{18}^{3}$ & $f_{54}^{9}$: $\Phi_{2}^{8}\Phi_{6}^{9}$ \\
    16 & 33 & $f_{66}^{3}$: $\Phi_{2}^{2}\Phi_{22}^{3}$ & $f_{66}^{11}$: $\Phi_{2}^{10}\Phi_{6}^{11}$ \\
    17 & 35 & $f_{70}^{5}$: $\Phi_{2}^{4}\Phi_{14}^{5}$ & $f_{70}^{7}$: $\Phi_{2}^{6}\Phi_{10}^{7}$ \\
    19 & 39 & $f_{78}^{3}$: $\Phi_{2}^{2}\Phi_{26}^{3}$ & $f_{78}^{13}$: $\Phi_{2}^{12}\Phi_{6}^{13}$ \\
    22 & 45 & $f_{90}^{3}$: $\Phi_{2}^{2}\Phi_{6}^{3}\Phi_{10}^{3}\Phi_{30}^{3}$ & $f_{90}^{5}$: $\Phi_{2}^{4}\Phi_{6}^{5}\Phi_{18}^{5}$ & $f_{90}^{9}$: $\Phi_{2}^{8}\Phi_{10}^{9}$ & $f_{90}^{15}$: $\Phi_{2}^{14}\Phi_{6}^{15}$ \\
    24 & 49 & $f_{98}^{7}$: $\Phi_{2}^{6}\Phi_{14}^{7}$ \\
    25 & 51 & $f_{102}^{3}$: $\Phi_{2}^{2}\Phi_{34}^{3}$ & $f_{102}^{17}$: $\Phi_{2}^{16}\Phi_{6}^{17}$
  \end{tabular}
  \caption{Characteristic polynomial of the $2g \times 2g$ matrix $f_{4g+2}^d$, where $d$ is a nontrivial divisor of $2g + 1$.}
  \label{tab:4g+2}
\end{table}

Since these mass terms are invariant under $\mathbb{Z}_{2h}$ symmetry, the RG flow triggered by adding one of them can flow to a trivially gapped theory only if there is a $\mathbb{Z}_{2h}$ invariant boundary condition in the SymTFT. Whether such a boundary condition exists or not can be straightforwardly determined from the characteristic polynomial of the generator of $\mathbb{Z}_{2h}$ symmetry. This generator is $f_{2h} = f_{4g+2}^d$ where $f_{4g+2}$ is the generator of $\mathbb{Z}_{4g+2}$ symmetry given in \cite{Bashmakov:2022jtl}. Explicitly, the first homology of a surface as in \cref{fig:18-gon} is generated by cycles $C_1, \dots, C_{2g+1}$ appearing as straight lines through the origin, subject to the relation $C_1 - C_2 + C_3 - \cdots + C_{2g+1} = 0$, and $f_{4g+2}$ acts as
\begin{equation}
  C_1 \mapsto C_2 \mapsto \cdots \mapsto C_{2g+1} \mapsto -C_1.
\end{equation}
From this description, one can deduce the characteristic polynomial of $f_{4g + 2}^d$: With the convention $C_{j + 2g + 1} = -C_j$, $f_{4g + 2}$ has eigenvectors $v = \sum_{j = 1}^{4g+2} \lambda^j C_j$ where $\lambda^{2g+1} = -1$, and furthermore $\lambda \neq -1$ since the corresponding eigenvector is zero by the relation mentioned above. These odd $(4g + 2)$-th roots of unity are the $2g$ eigenvalues of $f_{4g + 2}$, all nondegenerate. The eigenvalues of $f_{4g + 2}^d$ are then $\lambda^d$, counted with multiplicity. If $d$ is a divisor of $2g + 1$, they all have multiplicity $d$, apart from $-1$ which now has multiplicity $d - 1$. Partitioning them into primitive roots of unity, we find
\begin{equation}
  \det(x - f_{4g+2}^d) = \frac{1}{\Phi_2(x)}
  \qty\Big(\prod_{u \mid \frac{2g+1}{d}} \Phi_{2u}(x))^d.
\end{equation}
\Cref{tab:4g+2} shows a sample of these characteristic polynomials. In particular, the cyclotomic polynomial for every even divisor of $\frac{4g + 2}{d}$, except for $2$, appears with an odd power. \Cref{thm:main} or \cref{tab:thepowerofvodoo} then gives the conditions for which $N$ the deformed $\mathfrak{a}_{N - 1}$ theory with $f_{4g + 2}^d$ $\mathbb{Z}_{\frac{4g + 2}{d}}$ noninvertible symmetry must either spontaneously break symmetry or be gapless.

Apart from the genus 4 case we have already described, the next example comes at genus 7, where we get two families of theories by deforming $\mathfrak{a}_{N-1}$ class $\mathcal{S}$ theories with $\mathbb{Z}_{10}$ and $\mathbb{Z}_6$ noninvertible symmetries, respectively. Consider for instance the first of these. As $\Phi_{10}$ occurs with an odd power, consulting \cref{tab:thepowerofvodoo} we find that the theory cannot be trivially gapped if there is an odd-power prime factor of $N$ which is not $1 \pmod{10}$ or $5$. The first few $N$ for which the theory must spontaneously break non-invertible symmetry or are gapless are
\begin{equation}
  2, 3, 6, 7, 8, 10, 12, 13, 14, 15, 17, 18, 19, 21, 22, 23, 24, 26, 27, 28, 29, 30, 32, 33, 34, \dots
\end{equation}

\section*{Acknowledgments}
We thank Mahbub Alam, Clay Córdova, Justin Kaidi, Shani Nadir Meynet, and especially Kantaro Ohmori, and Theo Johnson-Freyd for discussions. ERG thanks especially Matteo Dell'Acqua for discussing the final proof in \cref{sec:anom-1-form} and Shani Nadir Meynet for discussions on \cref{sec:determining-symtft} and appendix \ref{app:H5B2G1}. MDZ and AH thank the organizers of the Solvay Workshop on Symmetries, Anomalies, and Dynamics of Quantum Field Theories (April 2024, Brussels) and ERG thanks the organizers of the Young Researchers School on Topological aspects of low-dimensional quantum physics (April 2024, Maynooth) for hospitality while this work was under completion.

\section*{Declarations}

\paragraph{Funding.} The work of AH, ERG, and MDZ has received funding from the European Research Council (ERC) under the European Union’s Horizon 2020 research and innovation program (grant agreement No.~851931). MDZ also acknowledges support from the Simons Foundation (grant No.~888984, Simons Collaboration on Global Categorical Symmetries), and from the Swedish Research Council (grants No.~2023-05590 and No.~2022-06593).

\paragraph{Conflict of interest.} The authors have no competing interests to declare that are relevant to the content of this
article. The authors have no financial or proprietary interests in any material discussed in this article.

\paragraph{Data availability statement.} This manuscript has no associated data.

\appendix

\section{Classification of anomalies via $H^5(B^2 \widehat{\mathbb{A}}^{(1)}; \mathbb{R}/\mathbb{Z})$}
\label{app:H5B2G1}
In this appendix, we give a detailed derivation of the expression \eqref{eq:H5B2G1} for the classification of the anomalies of finite group-like 1-form symmetries in four dimensions.

In this appendix, we use $\oplus$ in place of $\times$ to denote the product (direct sum) of abelian groups, and reserve $\times$ for the product of topological spaces.

First, we set the stage for the non-expert reader. Recall from \cref{sec:determining-symtft} that a pure 't Hooft anomaly of an $\widehat{\mathbb{A}}^{(1)}$ 1-form symmetry in four dimensions is described by an inflow action $e^{2\pi i \int_{M_5} \alpha(A)}$ in terms of a background gauge field $A$ for a non-anomalous $\mathbb{Z}_n$ symmetry on $M_5$. A non-anomalous background is an element $A \in H^2(M_5; \widehat{\mathbb{A}}^{(1)})$. The inflow action is a function (not necessarily a group homomorphism) $\alpha\colon H^2(M_5; \widehat{\mathbb{A}}^{(1)}) \to H^5(M_5; \mathbb{R}/\mathbb{Z})$. For $\alpha$ to be defined coherently on all $M_5$ it should be natural in $M_5$. Such a function is known as a cohomology operation \cite{mosher2008cohomology}. For an abelian group $G$, there is a one-to-one correspondence $H^p(X; G) \cong [X, B^p G]$, where $[\mbox{--},\mbox{--}]$ denotes homotopy classes of maps and $B^p G = K(G, p)$ is an Eilenberg--MacLane space: a space determined up to homotopy equivalence by having only one nontrivial homotopy group $\pi_p(B^p G) \cong G$. Our cohomology operation $\alpha$ is therefore a natural transformation $[\mbox{--}, B^2 \widehat{\mathbb{A}}^{(1)}] \to [\mbox{--}, B^5 (\mathbb{R}/\mathbb{Z})]$, or by the Yoneda lemma a class in $[B^2 \widehat{\mathbb{A}}^{(1)}, B^5 (\mathbb{R}/\mathbb{Z})] \cong H^5(B^2 \widehat{\mathbb{A}}^{(1)}; \mathbb{R}/\mathbb{Z})$.\footnote{\ Here we understand $\mathbb{R}/\mathbb{Z}$ as a discrete group, not a Lie group.} It is this last cohomology group we wish to determine.

To this end, we first show that
\begin{equation}
  \label{eq:BpG-cohomology-to-homology}
  H^i(B^p G; \mathbb{R}/\mathbb{Z}) \cong H_i(B^p G).
\end{equation}
By the universal coefficient theorem \cite[Theorem~3.2]{hatcher2002algebraic}, there is a (split) short exact sequence
\begin{equation}
  \label{eq:UCT}
  0 \to \Ext(H_{i-1}(B^p G), \mathbb{R}/\mathbb{Z}) \to H^i(B^p G; \mathbb{R}/\mathbb{Z})
  \to \Hom(H_i(B^p G), \mathbb{R}/\mathbb{Z}) \to 0.
\end{equation}
When $G$ is a finite abelian group, so is $H_i(B^p G)$ for $p,i > 0$ \cite[Lemma~5.10]{hatcherSpectral}. Using the properties $\Ext(\bigoplus_i H_i, K) \cong \bigoplus_i \Ext(H_i, K)$ and $\Ext(\mathbb{Z}_h, K) \cong K/hK$, we see that $\Ext(H_{i-1}(B^p G), \mathbb{R}/\mathbb{Z}) = 0$ for $i > 1$ since it reduces to a direct sum of terms $\Ext(\mathbb{Z}_h, \mathbb{R}/\mathbb{Z}) \cong (\mathbb{R}/\mathbb{Z})/h(\mathbb{R}/\mathbb{Z}) = 0$. For $i = 1$, $\Ext(H_0(B^p G), \mathbb{R}/\mathbb{Z})$ is also $0$ since $H_0(B^p G) \cong \mathbb{Z}$ is free. Therefore \eqref{eq:UCT} gives an isomorphism $H^i(B^p G; \mathbb{R}/\mathbb{Z}) \cong \Hom(H_i(B^p G), \mathbb{R}/\mathbb{Z})$. The latter group is the Pontryagin dual $H_i(B^p G)^\vee$, which is isomorphic to $H_i(B^p G)$ itself since it is a finite group; this shows \eqref{eq:BpG-cohomology-to-homology}.

By \eqref{eq:BpG-cohomology-to-homology}, the goal is now to compute $H_5(B^2 \widehat{\mathbb{A}}^{(1)})$ where $\widehat{\mathbb{A}}^{(1)}$ is finite abelian. The core calculation was done by Eilenberg and MacLane \cite[sections~18,~21,~22]{EilenbergMacLaneII}, who determined the homology groups of $B^2 \mathbb{Z}_n$ in low degree as follows:
\begin{equation}
  \label{eq:eilenberg-maclane-homology}
  H_0 \cong \mathbb{Z}, \quad
  H_1 = 0, \quad
  H_2 \cong \mathbb{Z}_n, \quad
  H_3 = 0, \quad
  H_4 \cong \mathbb{Z}_{\gcd(2,n)n}, \quad
  H_5 \cong \mathbb{Z}_{\gcd(2,n)}.
\end{equation}
Decomposing $\widehat{\mathbb{A}}^{(1)} = \bigoplus_i \mathbb{Z}_{n_i}$, we have $B^2 \widehat{\mathbb{A}}^{(1)} \simeq \bigtimes_i B^2 \mathbb{Z}_{n_i}$ since $\pi_i(X \times Y) \cong \pi_i(X) \times \pi_i(Y)$ \cite[Proposition~4.2]{hatcher2002algebraic}. We can therefore compute the homology groups of $B^2 \widehat{\mathbb{A}}^{(1)}$ using the Künneth formula \cite[Theorem~3B.6]{hatcher2002algebraic}
\begin{equation}
  H_i(X \times Y) \cong \bigoplus_j \qty\Big[(H_j(X) \otimes H_{i-j}(Y))
  \oplus \Tor(H_j(X), H_{i-j-1}(Y))]
\end{equation}
together with the properties $\Tor(A,B) \cong \Tor(B,A)$, $\Tor(\bigoplus_i A, B) \cong \bigoplus_i \Tor(A,B)$, $\Tor(\mathbb{Z},A) = 0$ and $\Tor(\mathbb{Z}_m, \mathbb{Z}_n) \cong \mathbb{Z}_{\gcd(m,n)}$ \cite[Proposition~3A.5]{hatcher2002algebraic} as well as $\mathbb{Z}_m \otimes \mathbb{Z}_n \cong \mathbb{Z}_{\gcd(m,n)}$ and $\mathbb{Z} \otimes A \cong A$. It is a straightforward exercise to apply the formula inductively to show that the homology groups of $B^2 \widehat{\mathbb{A}}^{(1)}$ are given by
\begin{align}
  H_0 &\cong \mathbb{Z}, \quad
  H_1 = 0, \quad
  H_2 \cong \bigoplus_i \mathbb{Z}_{n_i} \cong \widehat{\mathbb{A}}^{(1)}, \quad
  H_3 = 0, \\
  H_4 &\cong \bigoplus_{i<j} \mathbb{Z}_{\gcd(n_i,n_j)} \oplus \bigoplus_i \mathbb{Z}_{\gcd(2,n_i)n_i}, \\
  H_5 &\cong \bigoplus_{i<j} \mathbb{Z}_{\gcd(n_i,n_j)} \oplus \bigoplus_i \mathbb{Z}_{\gcd(2,n_i)}.
\end{align}
The last of these equations is \eqref{eq:H5B2G1}.

\section{Invariant subspaces and polynomial factorization}
\label{sec:invar-subsp-polyn}

\subsection{Pr\"{u}fer group symmetry}
\label{sec:pruf-group-symm}
Let us now consider the case of $\mathbb A_{p}$ starting with the simple case of $\mathbb A_{p} = \mathbb{Z}_{p^{k}}^{r_{1}} \times \mathbb{Z}_{p^{k+2l}}^{r_{2}}$ and the corresponding BF theory,
\begin{align}
  \label{eq:18}
  S = 2\pi i \qty(\frac{p^{k}}{2} \int_{M_{5}} b_{1}^{T}\,\Omega\,\dd b_{1} + \frac{p^{k+2l}}{2} \int_{M_{5}} b_{2}^{T} \,\Omega\,\dd b_{2}) ~.
\end{align}
Let's focus on the operator
\begin{align}
  \label{eq:19}
  \Phi_{1}(M_{2}) = \exp(2\pi i \int_{M_{2}} b_{1}) ~, &&  M_{2} \in H_{2}(M_{4} = \partial M_{5} , \mathbb{Z}_{p^{k}}^{2r_{1}}) ~.
\end{align}
For a simply connected $M_{4}$ we can choose an integer lift $\tilde{M}_{2}$ of $M_{2}$ and consider the operator $\tilde{\Phi}_{1}(p^{l} \tilde{M}_{2})$ in $\mathbb{Z}_{p^{k+2l}}^{r_1}$ BF theory. Given two such operators $\tilde{\Phi}_{1}(p^{l} \tilde{M}_{2})$ and $\tilde{\Phi}_{1}(p^{l} \tilde{M}_{2}^{\prime})$ we obtain,
\begin{align}
  \label{eq:20}
  \tilde{\Phi}_{1}(p^{l} \tilde{M}_{2})\tilde{\Phi}_{1}(p^{l} \tilde{M}_{2}^{\prime}) = \exp(\frac{2\pi i}{p^{k}}\ev{\tilde{M}_{2} , \tilde{M}_{2}^{\prime}}) \tilde{\Phi}_{1}(p^{l} \tilde{M}_{2}^{\prime})\tilde{\Phi}_{1}(p^{l} \tilde{M}_{2})
\end{align}
where the phase $\exp(\frac{2\pi i}{p^{k}}\ev{\tilde{M}_{2} , \tilde{M}_{2}^{\prime}})$ depends only on the mod $p^{k}$ reduction of $\tilde{M}_{2}$ (i.e. $M_{2}$) and $\tilde{M}_{2}^{\prime}$ (i.e.~$M_{2}^{\prime}$) and it is independent of their integral lifts $\tilde{M}_{2}$ and $\tilde{M}_{2}^{\prime}$. Hence the map $M_{2} \mapsto p^{l} \tilde{M}_{2}$ provides an embedding of the operators of the $\mathbb{Z}_{p^{k}}^{r_{1}}$ theory into the ones of the $\mathbb{Z}_{p^{k+2l}}^{r_{1}}$ theory. Using this idea we can embed the BF theory with action \eqref{eq:18} into a $\mathbb{Z}_{p^{k+2l}}^{r_{1}+r_{2}}$ theory. In this way we can always lift a lagrangian subgroup of $\mathbb{Z}_{p^{k}}^{2r_{1}} \times \mathbb{Z}_{p^{k+2l}}^{2r_{2}}$ to a subgroup of $\mathbb{Z}^{2(r_{1}+r_{2})}$,\footnote{\ Such a lift is not unique, only its mod $p^{k}$ reduction is. But we are only interested in this reduction. Note also that he integral lift itself cannot be chosen to be lagrangian: a lagrangian $\mathbb{Z}_{p^{k}}$ submodule can be lifted to an integral lagrangian iff it is free i.e.~it admits a basis.} which we can reduce mod $p^{k+2l}$ to obtain a (non-unique) lagrangian subgroup of $\mathbb{Z}_{p^{k+2l}}^{2(r_{1}+r_{2})}$.
More to the point, we can uniquely go the other way: Given a lagrangian subgroup $L$ of $\mathbb{Z}_{p^{k+2l}}^{2r_1}$, on the subgroup $L_l$ of multiples of $p^l$, the intersection pairing is $\mathbb{Z}_{p^k}$-valued as in \eqref{eq:20}. Reducing $L_l$ mod $p^{k + l}$, we obtain a lagrangian subgroup of $p^l \mathbb{Z}_{p^{k + l}}^{2r_1} \cong \mathbb{Z}_{p^k}^{2r_1}$. Hence a lagrangian subgroup of
$\mathbb{Z}_{p^{k+2l}}^{2(r_{1}+r_{2})}$ gives rise to one of $\mathbb{Z}_{p^{k}}^{2r_{1}} \times \mathbb{Z}_{p^{k+2l}}^{2r_{2}}$.

This discussion can be generalized to any number of factors in $\mathbb A_{p}$ and the upshot is that for any $\mathbb A_{p}$ it is enough to consider $\mathbb{Z}_{p^{k}}^{2r}$ theory with large enough $k$ and $r$. As a result of this structure it is tempting to consider the inverse limit,
\begin{align}
  \label{eq:21}
  \mathbb{Z}_{p} &\leftarrow \mathbb{Z}_{p^{3}} \leftarrow \mathbb{Z}_{p^{5}} \leftarrow \dots \leftarrow \mathbb{Z}_{p^{2k+1}} \leftarrow \dots ~, \nonumber \\
  1 &\leftarrow \mathbb{Z}_{p^{2}} \leftarrow \mathbb{Z}_{p^{4}} \leftarrow \dots \leftarrow \mathbb{Z}_{p^{2k}} \leftarrow \dots ~.
\end{align}
In both cases the inverse limit is the group of $p$-adic integers $\hat{\mathbb{Z}}_{p}$. In this limit we consider $\Omega$ as a symplectic form over $\hat{\mathbb{Z}}_{p}$. This symplectic form encodes a symmetry $\mathbb{Z}(p^{\infty})^{r}$ BF theory where the Pr\"ufer group $\mathbb{Z}(p^{\infty})$ is the Pontryagin dual of $\hat{\mathbb{Z}}_{p}$.\footnote{\ This BF theory is not a topological field theory in the strict sense since the Hilbert spaces assigned to four-manifolds are not finite dimensional} It is a subgroup of $\mathbb{Q} / \mathbb{Z}$ and consists of all elements of $\mathbb{Q} / \mathbb{Z}$ with order a power of $p$. As we will see this connection to $p$-adic numbers provides a key ingredient for a systematic analysis of anomalies of non-invertible symmetries.

\subsection{Lagrangians over the $p$-adics}

In this section, we introduce some foundational facts about lagrangians over the $p$-adic numbers and see that they are sufficient to classify the invariant lagrangians in the case $r = 1$.

Let us discuss some general properties of an invariant lagrangian for an element $g \in \mathrm{Sp}(2r,\mathbb{Z})$. For any ring $R$ the we can regard $g$ as an element of $\mathrm{Sp}(2r,R)$ i.e. $g$ acts on the free module $R^{2r}$ with the invariant form $\Omega$ regarded as an $R$-bilinear form. For us the ring $R$ will $\mathbb{Z}_{p^k}$. Let $L$ be an invariant lagrangian of $g$ over $\mathbb{Z}_{p^k}$. As explained around Equation \eqref{sec:symtft-one-form} the image of $L$ under the map $x \to p^l x$ gives us a lagrangian over $\mathbb{Z}_{p^{k-2l}}$ which we call $L^{\prime}$.\footnote{\ On the level of matrices, $\mathcal{M}_{L'}$ is obtained as follows: In the diagonal form $\mathcal{M}_D$ of $\mathcal{M}_L$, replace every pair $(p^{l_i}, p^{l_{i+r}})$ such that $l_i < l$ with $(p^l, p^{k+l})$ (this corresponds to restricting to multiples of $p^l$ and reducing mod $p^{k+l}$). The resulting matrix $\widetilde{M}_L$ is divisible by $p^l$; set $\mathcal{M}_{L'} = \widetilde{\mathcal{M}}_L / p^l$.} Furthermore if $L$ is invariant, so is $L^{\prime}$. There are two important cases to consider,
\begin{itemize}
  \item If $k = 2\ell+1$ is odd, we can choose $l = \ell$. In this case $L^{\prime}$ is an invariant lagrangian over $\mathbb{Z}_{p}$. Hence an invariant lagrangian over $\mathbb{Z}_{p^k}$ can exist only if it exists over $\mathbb{Z}_{p}$. This is a simple but powerful constraint on the existence of invariant lagrangian since it tells that their non-existence over $\mathbb{Z}_{p}$ is enough to confirm their non-existence over $\mathbb{Z}_{p^{2\ell+1}}$ for all $\ell \in \mathbb{Z}$. Following \cite{Bashmakov:2022uek}, an invariant lagrangian over $\mathbb{Z}_{p}$ exists iff we can find degree $r$ polynomials $P(x),Q(x) \in \mathbb{Z}_{p}[x]$ such that,
    \begin{align}
      \label{eq:30}
      \det(x - g) = P(x)Q(x) ~.
    \end{align}
  \item If $k = 2\ell+2$ is even, we can again choose $k = \ell$. Hence, an invariant lagrangian $L$ over $\mathbb{Z}_{p^k}$ can exist only if it exists over $\mathbb{Z}_{p^{2}}$. Over $\mathbb{Z}_{p^{2}}$ one invariant lagrangian always exists given simply by $p \mathbb{Z}_{p^{2}}^{2r}$. Notice that if $L$ has an element of order greater than $p^{\ell+1}$, using this map we obtain a lagrangian $\tilde{L}$ with an element of order $p^{2}$. Hence the only lagrangian that maps to $p \mathbb{Z}_{p^{2}}^{2r}$ is $L_{p,\ell} = p^{\ell+1} \mathbb{Z}_{p^{2\ell+1}}^{2r}$.

    Any other invariant lagrangian can be decomposed as $L^{(0)} \oplus p L^{(1)}$ where $L^{(0)}$ and $L^{(1)}$ are both free submodules satisfying $L^{(0)} \cap L^{(1)} = \{0\}$ and $L^{(0)}$ is nonempty. For any element $v$ in $L$ we can write,
    \begin{align}
      \label{eq:31}
      g v = \tilde{v} + p \tilde{w}, && \tilde{v} \in L^{(0)}, \tilde{w} \in L^{(1)} ~.
    \end{align}

    Let $\tilde{L}^{(0)}$ be any integral lift of $L^{(0)}$. Then using \eqref{eq:31} it can be observed that $\tilde{L}^{(0)} \mod p$ is an invariant subspace over $\mathbb{Z}_{p}$. Moreover it depends on $L^{(0)}$ and is independent of the integral lift $\tilde{L}^{(0)}$. For such an invariant subspace to exist, the characteristic polynomial of $g$ must factorize modulo $p$.
\end{itemize}

Hence under the map described above, any invariant lagrangian $L$ of $g$ is either $L_{p,\ell}$ or requires a factorization of characteristic polynomial $\det(x - g)$ into monic polynomials over $\mathbb{Z}_{p}$.

\subsection{$p$-adic invariant subspaces}
\label{sec:p-adic-lagrangians}
More generally whenever $\det(x-g) = P(x)Q(x)$ with $P(x)$ and $Q(x)$ monic polynomials over $\mathbb{Z}_{p^k}$, the subspaces $\ker(P(x))$ and $\ker(Q(x))$ are $g$-invariant as a consequence of Cayley-Hamilton theorem.\footnote{\ They are mere invariant subspaces, not lagrangians. We will need to work a bit more to construct invariant lagrangians from knowledge of the factorization of characteristic polynomial.} Modulo $p^{k^{\prime}}$ reduction of $P(x)$ and $Q(x)$ gives us factorization of the characteristic polynomial over $\mathbb{Z}_{p^{k^{\prime}}}$. This once again leads us to consider the limit of $k \to \infty$ or more precisely the structure of $g$ as an element of $\mathrm{Sp}(2r, \hat{\mathbb{Z}}_{p})$ i.e.~a $p$-adic symplectic transformation.

\subsubsection{$p$-adic numbers}
\label{sec:p-adic-numbers}
A $p$-adic integer $n_{p}$ can be represented as a formal $p$ series i.e.
\begin{align}
  \label{eq:32}
  n_{p} = c_{0} + c_{1}p + c_{2} p^{2} + \dots
\end{align}
where $0 \leq c_{i} < p$. If the series above has only finitely many non-zero entries, we obtain an integer expressed in base $p$ and as a result $p$-adic integers contain integers as a subset.

Alternatively $n_{p}$ can also be expressed as a sequence $n_{p,1} , n_{p,2}  , \dots $ such that for $i < j$, $n_{p,j} = n_{p,i} \mod p^{i}$. In this representation integers correspond to sequence that stabilize i.e. there is $i_{\max}$ such that for $i > i_{\max}, n_{p,i} = n_{p,i_{\max}}$.

More important for us will be the algebraic structure of $\hat{\mathbb{Z}}_{p}$. The $p$-adic numbers form a principal integral domain. This means in particular that up to multiplication by units, every polynomial in $\hat{\mathbb{Z}}_{p}[x]$ has a unique factorization. Another consequence is that we can construct their field of fractions which is conventionally labeled as $\mathbb{Q}_{p}$. An element of $\mathbb{Q}_{p}$ can also be represented as a $p$-series with some negative powers of $p$ allowed i.e. $q_{p} \in \mathbb{Q}_{p}$ can be expressed as,
\begin{align}
  \label{eq:33}
  q_{p} = c_{k}p^{k} + c_{k+1}p^{k+1} + c_{k+2}p^{k+2} + \dots
\end{align}
with $0 \leq c_{i} < p$ and $c_{k} \neq 0$. The number $p^{-k}$ is called the $p$-adic absolute value $\abs{q_{p}}_p$ of $q_{p}$ and $k$ is referred to as the valuation of $q_{p}$. Hence $\hat{\mathbb{Z}}_{p}$ is the subring of $\mathbb{Q}_{p}$ consisting of elements with non-negative valuation.

Given a factorization of $\det(x - g) \in \hat{\mathbb{Z}}_{p}[x]$ i.e $\det(x-g) = \mathcal{P}(x)\mathcal{Q}(x)$ with $\mathcal{P}(x),\mathcal{Q}(x)$ monic $p$-adic polynomials we obtain a factorization of $\det(x-g)$ modulo $p^k$ for all $k$, i.e.
\begin{align}
  \label{eq:34}
  &\det(x - g)  = P_{p^k}(x) Q_{p^k}(x)  \mod p^k ~, \nonumber \\
  \mbox{with } &P_{p^k}(x) = \mathcal{P}(x) \mod p^k, ~ ~  Q_{p^k}(x) = \mathcal{Q}_{p^k}(x) \mod p^k ~.
\end{align}
Similarly $\ker(P_{p^k}(g)) = \ker(\mathcal{P}(g)) \mod p^k$ is a $g$-invariant free module over $\mathbb{Z}_{p^k}$ and the same holds for $\ker(Q_{p^k}(g))$.

\subsubsection{Hensel's lemma}
\label{sec:hensels-lemma}
As a result of the discussion above to construct invariant subspaces of $g$ we need $p$-adic factorization of $\det(x-g)$. The elementary result in this direction which will be immensely useful to us is called Hensel's lemma. It tells us that for a polynomial which is irreducible over integers, $p$-adic factorization is mostly equivalent to factorization over $\mathbb{F}_{p}$. Factorization of a polynomial $P(x)$ over $\mathbb{Z}_{p^k}$ implies a factorization of $P(x) \mod p \in \mathbb{F}_{p}[x]$. Hensel's lemma states that the converse also mostly holds. It can be stated in very general terms (see e.g.~\cite{MR1697859}) but we will restrict ourselves to a version which is directly useful to us.

Let $P(x) \in \mathbb{Z}[x]$ be an irreducible monic polynomial with integer coefficients which factors mod $p$ as,
\begin{align}
  \label{eq:35}
  P(x) = Q_{1}(x)Q_{2}(x) \mod p ~,
\end{align}
which $Q_{1}(x),Q_{2}(x) \in \mathbb{F}_{p}[x]$ monic polynomials which are relatively prime. Then there exist polynomials $\mathcal{Q}_{1}(x),\mathcal{Q}_{2}(x) \in \hat{\mathbb{Z}}_{p}[x]$ such that as a $p$-adic polynomial,
\begin{align}
  \label{eq:36}
  P(x) = \mathcal{Q}_{1}(x) \mathcal{Q}_{2}(x) ~.
\end{align}
Let us unpack this statement a little bit and explain how $\mathcal{Q}_{1},\mathcal{Q}_{2}$ are constructed: given a $Q_{1,p^k}$ and $Q_{2,p^k}$ such that,
\begin{align}
  \label{eq:37}
  P(x) = Q_{1,p^k}(x) Q_{2,p^k}(x) \mod p^k
\end{align}
it is always possible to find $Q_{1,p^{k+1}}$ and $Q_{2,p^{k+1}}$ such that,
\begin{align}
  \label{eq:38}
  P(x) = Q_{1,p^{k+1}}(x) Q_{2,p^{k+1}}(x) \mod p^{k+1} && Q_{i,p^{k+1}} = Q_{i,p^k} \mod p^k
\end{align}
As a result starting with the factorization over $\mathbb{F}_{p}$ we can iterate this process to construct better and better approximation to $\mathcal{Q}_{1}$ and $\mathcal{Q}_{2}$. Notice that although there is no unique factorization of polynomials over $\mathbb{Z}_{p^k}$, the unique factorization over $\hat{\mathbb{Z}}_{p}[x]$ ensures that $Q_{i,p^{k+1}}$ are all uniquely determined by $Q_{1}$ and $Q_{2}$.

\subsection{Example: subgroups of $\mathrm{SL}(2,\mathbb{Z})$ and $\mathfrak{su}(N)$ gauge theories}
\label{sec:exampl-subgr-sl2}
Let us make the reasoning above as concrete as possible by working out in detail the simplest case of $r = 1$. One setting where this is realized physically is in $\mathfrak{su}(N)$ gauge theories, where a global form is specified by a lagrangian subgroup of $\mathbb{Z}_N^2$. Following \cite{Aharony:2013hda}, such a subgroup can be expressed as
\begin{equation}
  \label{eq:between-the-lines-lagr}
  L = \Span_{\mathbb{Z}_N}\!\qty\big((K,0), (n, N/K))
\end{equation}
where $K$ divides $N$; the absolute theory is denoted $\qty\big(\mathrm{SU}(N)/\mathbb{Z}_K)_n$ with $n \in \mathbb{Z}_K$. Its one-form symmetry is seen via the Smith normal form to be $\mathbb{Z}_G \times \mathbb{Z}_{N/G}$ where $G = \gcd(K, N/K, n)$. In $\mathcal{N} = 4$ supersymmetric Yang--Mills theories, the whole of $\mathrm{SL}(2, \mathbb{Z})$ is realized as dualities, an element $g = {\scriptsize\mqty(a & b \\ c & d)} \in \mathrm{SL}(2,\mathbb{Z})$ mapping the lagrangian and complexified gauge coupling $\tau$ as $(L, \tau) \mapsto (gL,g\cdot\tau)$ where $g\cdot\tau = \frac{a\tau + b}{c\tau + d}$, and also transforming the local degrees of freedom of the theory. At self-dual couplings $g\cdot\tau = \tau$, this is a duality between different global forms of the same theory (it is a self-duality of the relative theory), and so has a chance of being a symmetry: if $gL = L$, the transformation is an invertible symmetry by itself; otherwise, one can construct a noninvertible symmetry by composing it with a topological manipulation, which acts as $g^{-1}$ on $L$ but does not act on $\tau$ or the local degrees of freedom; see e.g.~\cite{Kaidi:2022uux}. Below, we will use these $\mathcal{N} = 4$ SYM theories as examples, but we highlight that the methods work for any theories with the requisite 1-form symmetries and dualities.

As explained earlier, it is enough to consider $N = p^k$ where $p$ is a prime. Proceeding in general, we pick an element $g \in G^{(0)} \subset \mathrm{Sp}(2, \mathbb{Z}) = \mathrm{SL}(2, \mathbb{Z})$ and search for boundary conditions invariant under the cyclic subgroup generated by $g$. Coupled to the SymTFT $\frac{p^k}{2} \int_{M_5} b^T \Omega \dd{b}$, such a $g$-invariant boundary condition is specified by a $g$-invariant lagrangian subgroup $L \subset \mathbb{Z}_{p^k}^2$. As explained in \ref{sec:determining-symtft}, we can represent $L$ as $\im \mathcal{M}_L \pmod{p^k}$ where $\mathcal{M}_L$ is a $2 \times 2$ integer matrix such that $\mathcal{M}_L^T \Omega \mathcal{M}_L = p^k \Omega$.\footnote{\ For $r = 1$, this is equivalent to $\det(\mathcal{M}_L) = p^k$.} In other words, we specify two generating vectors $u, v \in \mathbb{Z}^2$ with $\ev{v,u} = p^k$. The Smith normal form of $\mathcal{M}_L$ is $\diag(p^\ell, p^{k-\ell})$ for some $\ell \leq \frac{k}{2}$, meaning that $u$ and $v$ can be chosen such that $u$ is a multiple of $p^\ell$ and $v$ is a multiple of $p^{k-\ell}$ ($u$ has order $p^{k-\ell}$ and $v$ has order $p^\ell$ over $\mathbb{Z}_{p^k}$). This makes it obvious that $L$ has $p^k$ elements. The $\mathrm{Sp}(2,\mathbb{Z})$ orbit of $\mathcal{M}_L$ is determined by the integer $\ell$, and the 1-form symmetry is $\mathbb{Z}_{p^\ell} \times \mathbb{Z}_{p^{k-\ell}}$. The requirement that $L$ be invariant under $g$, $gL = L$, explicitly means that $g\mathcal{M}_L = \mathcal{M}_L B$ for some invertible $B$ encoding a change of basis for $L$.

If $\ell = 0$, $L$ is freely generated by a single $u$ of order $p^k$, and the 1-form symmetry is $\mathbb{Z}_{p^k}$; this is realized in the $\qty\big(\mathrm{SU}(p^k)/\mathbb{Z}_{p^l})_n$ theories where $l = 0$, $l = k$ or $n$ is coprime to $p$. We begin by searching for these free lagrangian subgroups. The condition of $g$-invariance simplifies to $u$ being an eigenvector of $g$. Thus, there are free $g$-invariant lagrangians if and only if the characteristic polynomial of $g$ factorizes into linear factors
\begin{equation}
  \label{eq:2x2-char-poly}
  \det(x - g) = x^2 - (\tr g)\,x + 1 = (x - \lambda) (x - \lambda^{-1})
\end{equation}
over $\mathbb{Z}_{p^k}$.

A root of the characteristic polynomial over $\mathbb{Z}_{p^k}$ reduces to a root over $\mathbb{Z}_p$, and these roots are easy to characterize. If $p = 2$, the characteristic polynomial is irreducible if $\tr g = 1$ and factors as $(x - 1)^2$ if $\tr g = 0$; then $1$ is a repeated root. If $p \neq 2$, the eigenvalues are given by the quadratic formula
\begin{equation}
  \lambda^{\pm 1} = \frac{\tr g \pm \sqrt{d}}{2}
\end{equation}
where $d = (\tr g)^2 - 4$. They exist if $d$ is a quadratic residue modulo $p$, that is, if the Legendre symbol $\qty(\frac{d}{p}) = d^{\frac{p - 1}{2}} \pmod{p}$ is not $-1$.

Conversely, in most cases we can uniquely lift a root in $\mathbb{Z}_p$ to $\mathbb{Z}_{p^k}$ using Hensel's lemma, which is conveniently viewed as a $p$-adic version of Newton's method. Here, given a root $\lambda$ of the characteristic polynomial \eqref{eq:2x2-char-poly} modulo $p^k$, it provides in the simplest case an explicit lift
\begin{equation}
  \lambda' = (\lambda^2 - 1) (2\lambda - \tr g)^{-1}
\end{equation}
(where the inverse is taken mod $p^{2k}$), which is the unique root modulo $p^{2k}$ such that $\lambda' = \lambda \pmod{p^k}$; iterating this procedure one obtains a $p$-adic root $\hat{\lambda}$. This simple version of Hensel lifting works unless $2\lambda - \tr g = 0 \pmod{p}$, which occurs when $\lambda = \pm 1$ is a repeated eigenvalue over $\mathbb{Z}_p$. We analyze the roots in the latter case in \cref{sec:rep-roots}. In the examples below, it will suffice to rule out roots by noting that no roots exist modulo some small power of $p$.

Now we come to the torsional invariant lagrangians $L$ in $\mathbb{Z}_{p^k}^2$, that is, those with $\ell \geq 1$. We begin with some general observations. We can choose generators $u = p^\ell u'$ and $v = p^{k-\ell}v'$ with $\ev{v',u'} = 1$ and $\ell \leq k - \ell$. Then note that $p^\ell v = 0$ and $p^\ell L = \Span(p^\ell u)$ is $g$-invariant, hence $p^\ell u = p^{2\ell} u'$ is either zero or an eigenvector of $g$. Thus we have two cases: If $2\ell = k$, $L$ consists only of multiples of $p^\ell$ and therefore $L = p^\ell \mathbb{Z}_{p^{2\ell}}^2$ (conversely, this is automatically a $g$-invariant lagrangian for any $g$), realized in the $\qty\big(\mathrm{SU}(p^{2\ell})/\mathbb{Z}_{p^\ell})_{p^\ell}$ theory. If $2\ell < k$, $p^\ell L$ is the embedding into $\mathbb{Z}_{p^k}^2$ of a free invariant lagrangian in $\mathbb{Z}_{p^{k-2\ell}}^2$, as explained in \cref{sec:pruf-group-symm}.

Taking this further, in fact knowledge of $p^\ell u$ is enough to reconstruct $L$: Suppose $p^\ell u = p^{2\ell} a w_1$, with $\{w_1, w_2\}$ a basis of $\mathbb{Z}_{p^k}^2$ and $a$ coprime to $p$. Then we may expand $u = p^\ell a w_1 + p^{k-\ell} b w_2$ and $v = p^{k-\ell} (c w_1 + d w_2)$, with $d$ coprime to $p$. By virtue of this, we can perform a series of changes of basis for $L$ (column operations)
\begin{equation}
  \label{eq:SL2Z-explicit-basis-change}
  \mathcal{M}_L
  = \mqty(p^\ell a & p^{k-\ell} c \\ p^{k-\ell} b & p^{k-\ell} d)
  \to \mqty(p^\ell a & 0 \\ p^{k-\ell} b & p^{k-\ell} d')
  \to \mqty(p^\ell a & 0 \\ 0 & p^{k-\ell} d')
  \to \mqty(p^\ell & 0 \\ 0 & p^{k-\ell})
\end{equation}
realizing explicitly the Smith normal form of $\mathcal{M}_L$. This shows that
\begin{equation}
  \label{eq:plu-reconstruct}
  L = \Span_{\mathbb{Z}_{p^k}}\!\!\qty(p^\ell w_1, p^{k-\ell} w_2).
\end{equation}
In particular, if $\{w_1, w_2\}$ is a basis of eigenvectors of $g$, the $g$-invariant lagrangians are precisely those of the form
\begin{equation}
  \label{eq:SL2Z-standard-torsional}
  L = \Span_{\mathbb{Z}_{p^k}}\!\!\qty(p^l w_1, p^{k-l} w_2)
\end{equation}
(where we no longer restrict $2l \leq k$ to allow for $p^\ell u \propto w_2$ as well as $w_1$).

\subsubsection{Example: $\mathsf{S}$}

Let us exemplify this reasoning for $g = \mathsf{S} = \mqty(0 & -1 \\ 1 & 0)$, which satisfies $\mathsf{S}^2 = -1$. For example, in $\mathfrak{su}(N)$ $\mathcal{N} = 4$ SYM, $\mathsf{S}$-duality maps $\tau \mapsto -1/\tau$, so at $\tau = \pm i$, this is a (possibly noninvertible) symmetry.

First, we determine when the characteristic polynomial $\det(x - \mathsf{S}) = x^2 + 1$ factors, implying the existence of free lagrangians. Examining the Lagrange symbol shows that roots exist in $\mathbb{Z}_p$ iff $p = 1 \pmod{4}$ or $p = 2$. For $p \neq 2$, a root $\lambda$ in $\mathbb{Z}_{p^k}$ can be lifted to $\lambda' = (\lambda^2 - 1) / 2\lambda$ in $\mathbb{Z}_{p^{2k}}$, so the characteristic polynomial factors over the $p$-adics.

For $p = 3 \pmod{4}$, although there are no free $\mathsf{S}$-invariant lagrangians, we must still consider torsional ones. A lagrangian over $\mathbb{Z}_{p^k}$ with $k$ odd would have to reduce to a lagrangian over $\mathbb{Z}_p$, and so cannot exist. For $k = 2\ell$ even, it must instead reduce to $p \mathbb{Z}_{p^2}^2$; as we have seen in the general discussion, this means that there is a single possibility $p^\ell \mathbb{Z}_{p^{2\ell}}^2$, corresponding to the $\qty\big(\mathrm{SU}(p^{2\ell})/\mathbb{Z}_{p^\ell})_{p^\ell}$ theory.

For $p = 1 \pmod{4}$, there are two distinct free $\mathsf{S}$-invariant lagrangians, spanned by $(\lambda, 1)$ and $(\lambda, -1)$. As a numerical example, starting from $3^2 = -1 \pmod{5}$, Hensel's lemma provides a $5$-adic square root of $-1$: $\lambda = \dots 12013233_5$, so $(\lambda, \pm 1)$ span two $\mathsf{S}$-invariant lagrangians over $\hat{\mathbb{Z}}_5$. The most general $\mathsf{S}$-invariant lagrangian over $\mathbb{Z}_{p^k}$ is $\Span_{\mathbb{Z}_{p^k}}\!\!\qty(p^l (\lambda, 1), p^{k-l} (\lambda, -1))$. Thus for each one-form symmetry $\mathbb{Z}_{p^\ell} \times \mathbb{Z}_{p^{k-\ell}}$, there are two invariant lagrangians, except for the case $2\ell = k$ where there is only one. The corresponding gauge theories are $\qty\big(\mathrm{SU}(p^k)/\mathbb{Z}_{p^{k-\ell}})_{\pm p^\ell\lambda}$ with $2\ell \leq k$.

For $p = 2$, there is only one free invariant lagrangian over $\mathbb{Z}_2$, spanned by $(1, 1)$, which corresponds to the $\qty(\mathrm{SU}(2)/\mathbb{Z}_2)_1$ or $\mathrm{SO}(3)_-$ theory of \cite{Aharony:2013hda}. Since there is no square root of $-1$ modulo $4$, and hence not modulo any higher power of $2$, there are no free invariant lagrangians over $\mathbb{Z}_{2^k}$ with $k > 1$. For $k = 2\ell$ even, there is a single torsional lagrangian $2^\ell \mathbb{Z}_{2^{2\ell}}^2$; the $\qty(\mathrm{SU}(4^\ell)/\mathbb{Z}_{2^\ell})_{2^\ell}$ theory. For $k$ odd, a torsional lagrangian $L$ with 1-form symmetry $\mathbb{Z}_{2^\ell} \times \mathbb{Z}_{2^{k-\ell}}$ must reduce to a free lagrangian over $\mathbb{Z}_{2^{k - 2\ell}}$; hence $k - 2\ell = 1$ is the only possibility. This in fact uniquely determines $L$: With $u$, $v$ as above, $2^{\ell} u$ must be an eigenvector of order $2$; this fixes $2^{\ell} u = 2^{k-1} (1, 1)$. Hence in the basis $\{(1, 1), (1, 0)\}$ for $\mathbb{Z}_{2^{2\ell + 1}}^2$, the argument for \eqref{eq:plu-reconstruct} applies, and we see that $L = \Span_{\mathbb{Z}_{2^{2\ell + 1}}}\qty\big(2^\ell (1, 1), 2^{\ell + 1} (1, 0))$; this is the unique lagrangian over $\mathbb{Z}_{2^{2\ell + 1}}$, realized in the $\qty(\mathrm{SU}(2^{2\ell + 1})/\mathbb{Z}_{2^{\ell + 1}})_{2^\ell}$ theory, and has $\mathbb{Z}_{2^\ell} \times \mathbb{Z}_{2^{\ell + 1}}$ one-form symmetry.

\subsubsection{Example: $\mathsf{T}$}
Next, consider the case of $g = \mathsf{T} = \mqty(1 & 1 \\ 0 & 1)$, with characteristic polynomial $(x - 1)^2$. Here there is a repeated eigenvalue $\lambda = 1$ over any $\mathbb{Z}_{p^k}$ and there is one free $\mathsf{T}$-invariant lagrangian spanned by $(1, 0)$. In the standard basis, the same argument as above (beginning by observing that $p^\ell u$ must be a multiple of $(1,0)$) shows that any tentative $\mathsf{T}$-invariant lagrangian must be of the form
\begin{equation}
  \mathcal{M}_L = \mqty(p^\ell & 0 \\ 0 & p^{k-\ell}).
\end{equation}
It is easily checked that all of these are indeed invariant as long as $2\ell \leq k$; thus there is a single $\mathsf{T}$-invariant lagrangian in each orbit with 1-form symmetry $\mathbb{Z}_{p^\ell} \times \mathbb{Z}_{p^{k-\ell}}$. The corresponding gauge theory is $\qty(\mathrm{SU}(p^k)/\mathbb{Z}_{p^\ell})_0$. In this case, however, there is no $\mathcal{N} = 4$ SYM theory that realizes $\mathsf{T}$ as a symmetry, as the action on the coupling $\tau \mapsto \tau + 1$ has no fixed point.

\subsubsection{Example: $\mathsf{ST}$}
\label{sec:example-ST}
Next, we analyze $g = \mathsf{ST} = \mqty(0 & -1 \\ 1 & 1)$, satisfying $(\mathsf{ST})^3 = -1$. In $\mathcal{N} = 4$ SYM, it maps the coupling $\tau \mapsto -1/(\tau + 1)$, so it is a (possibly noninvertible) symmetry at $\tau = e^{\pm 2\pi i/3}$.

The characteristic polynomial is $\det(x - \mathsf{ST}) = x^2 - x + 1$. It is irreducible over $\mathbb{Z}_2$, while for $p \neq 2$, roots are given by
\begin{equation}
  \lambda^{\pm 1} = \frac{1 \pm \sqrt{-3}}{2}
\end{equation}
and exist when $-3$ is a quadratic residue. For $p = 3$, $\lambda = -1$ is a double root, corresponding to the single eigenvector $(1, 1)$ (the $\qty(\mathrm{SU}(3)/\mathbb{Z}_3)_1$ theory, cf.~\cite[fig.~4]{Aharony:2013hda}), while one can check that no root exists mod $9$. For $p > 3$, properties of the Legendre symbol including quadratic reciprocity show that $-3$ is a quadratic residue mod $p$ iff $p = 1 \pmod{3}$. In this case, there are two distinct roots $\lambda^\pm$, for $\lambda \in \mathbb{Z}_{p^k}$ we have the Hensel lift $\lambda' = (\lambda^2 - 1)/(2\lambda - 1) \in \mathbb{Z}_{p^{2k}}$ providing a $p$-adic factorization. We conclude that free $\mathsf{ST}$-invariant lagrangians are of the form $\Span_{\mathbb{Z}_3} (1, 1)$ or $\Span_{\mathbb{Z}_{p^{3j+1}}}(1, -\lambda^{\pm 1})$.

Turning to torsional lagrangians, the situation is quite analogous to the case of $\mathsf{S}$: For $p = -1 \pmod{3}$ there are only the cases $p^{k/2} \mathbb{Z}_{p^k}^2$ where $k$ is even, while for $p = 1 \pmod{3}$ a general lagrangian is of the form $\Span_{\mathbb{Z}_{p^k}}(p^l (1, -\lambda), p^{k-l} (1, -\lambda^{-1}))$. It can be checked that this corresponds to the gauge theory $\qty(\mathrm{SU}(p^k)/\mathbb{Z}_{p^{k-\ell}})_{-p^\ell \lambda^\mp}$ where $2\ell \leq k$. For $p = 3$, when $k$ is even $p^{k/2} \mathbb{Z}_{p^k}^2$ is the only possibility, while for $k = 2\ell + 1$ odd, the same argument as for $\mathsf{S}$ shows that the unique possibility is $L = \Span_{\mathbb{Z}_{3^{2\ell + 1}}}(3^\ell (1, 1), 3^{\ell + 1}(1, 0))$ with $\mathbb{Z}_{3^\ell} \times \mathbb{Z}_{3^{\ell + 1}}$ 1-form symmetry, realized in $\qty(\mathrm{SU}(3^{2\ell + 1})/\mathbb{Z}_{3^{\ell + 1}})_{3^\ell}$ gauge theory.

\subsubsection{Existence of invariant lagrangians over $\mathbb{Z}_N$}
\label{sec:exist-invar-lagr-1}
We now take a step back from finding explicit lagrangians to address the coarser question of existence: For which $N$ do there exist $g$-invariant lagrangians over $\mathbb{Z}_{N}$? In fact, the arguments of this section allow us to answer this question completely given only the integer $\tr g$. First, let $N = p_1^{k_1} \cdots p_I^{k_I}$ be the prime factorization of $N$. Since an invariant lagrangian over $\mathbb{Z}_N$ reduces to one over each $\mathbb{Z}_{p_i^{k_i}}$, and a collection of lagrangians for each $i$ can be lifted to one over $\mathbb{Z}_N$, it suffices to consider prime powers. Furthermore, it is generally true (not only for $r = 1$ as in this section) that invariant lagrangians over $\mathbb{Z}_{p^k}$ exist if and only if $k$ is even or invariant lagrangians exist over $\mathbb{Z}_p$: For $k$ even there is $p^{k/2} \mathbb{Z}_{p^k}^{2r}$, while for $k = 2\ell + 1$ odd and $\tilde{L}$ a lift of an invariant lagrangian over $\mathbb{Z}_p$, there is $p^\ell \tilde{L} \oplus p^{\ell+1} \mathbb{Z}_{p^k}^{2r}$ (see also \cref{eq:68}).

\begin{table}[h]
  \centering
  \begin{tabular}{c|p{.8\textwidth}}
    $\abs{\tr g}$ & $N$ with $g$-invariant lagrangians \\
    \hline
    0 & {\footnotesize 2, 4, 5, 8, 9, 10, 13, 16, 17, 18, 20, 25, 26, 29, 32, 34, 36, 37, 40, 41, 45, 49, 50, 52, 53, 58, 61, 64, 65, 68, 72, 73, 74, 80, 81, 82, 85, 89, 90, 97, 98, 100, 101, 104, 106, 109, 113, 116, 117, 121, 122, 125, 128, 130, 136, 137, 144, 145, 146, 148, 149, \dots} \\
    \hline
    1 & {\footnotesize 3, 4, 7, 9, 12, 13, 16, 19, 21, 25, 27, 28, 31, 36, 37, 39, 43, 48, 49, 52, 57, 61, 63, 64, 67, 73, 75, 76, 79, 81, 84, 91, 93, 97, 100, 103, 108, 109, 111, 112, 117, 121, 124, 127, 129, 133, 139, 144, 147, 148, \dots} \\
    \hline
    2 & {\footnotesize every $N$} \\
    \hline
    3 & {\footnotesize 4, 5, 9, 11, 16, 19, 20, 25, 29, 31, 36, 41, 44, 45, 49, 55, 59, 61, 64, 71, 76, 79, 80, 81, 89, 95, 99, 100, 101, 109, 116, 121, 124, 125, 131, 139, 144, 145, 149, \dots} \\
    \hline
    4 & {\footnotesize 2, 3, 4, 6, 8, 9, 11, 12, 13, 16, 18, 22, 23, 24, 25, 26, 27, 32, 33, 36, 37, 39, 44, 46, 47, 48, 49, 50, 52, 54, 59, 61, 64, 66, 69, 71, 72, 73, 74, 75, 78, 81, 83, 88, 92, 94, 96, 97, 98, 99, 100, 104, 107, 108, 109, 111, 117, 118, 121, 122, 128, 131, 132, 138, 141, 142, 143, 144, 146, 147, 148, 150, \dots} \\
    \hline
    5 & {\footnotesize 3, 4, 5, 7, 9, 12, 15, 16, 17, 20, 21, 25, 27, 28, 35, 36, 37, 41, 43, 45, 47, 48, 49, 51, 59, 60, 63, 64, 67, 68, 75, 79, 80, 81, 83, 84, 85, 89, 100, 101, 105, 108, 109, 111, 112, 119, 121, 123, 125, 127, 129, 131, 135, 140, 141, 144, 147, 148, \dots}
  \end{tabular}
  \caption{Existence of invariant lagrangian subgroups of $\mathbb{Z}_N^2$ for $g \in \mathrm{SL}(2, \mathbb{Z})$.}
  \label{tab:SL2Z-N}
\end{table}

For $r = 1$, we have seen that $g$-invariant lagrangians over $\mathbb{Z}_p$ exist precisely when $p \neq 2$ and $d = (\tr g)^2 - 4$ is a quadratic residue in $\mathbb{Z}_p$, or when $p = 2$ and $\tr g$ is even. Hence, the $N$ for which $g \in \mathrm{SL}(2,\mathbb{Z})$ has invariant lagrangians over $\mathbb{Z}_N$ are those that can be written as products of the following: $2$ if $\tr g$ is even, odd primes $p$ with $\qty(\frac{d}{p}) \neq -1$, and $p^2$ for any prime $p$. Equivalently, they are the numbers of the form $N = k^2 N'$ such that $d$ is a quadratic residue mod $4N'$. \Cref{tab:SL2Z-N} gives the admissible $N \leq 150$ for $\abs{\tr g} \leq 5$.

This condition has an interpretation in the theory of integral binary quadratic forms: For $d \neq 0$, it is equivalent to $N$ being represented by some quadratic form $ax^2 + bxy + cy^2$ with discriminant $b^2 - 4ac = d$ \cite[Prop.~4.1]{Buell1989}.\footnote{\ The proposition refers to primitive representations ($\gcd(x, y) = 1$). A generic represented number is of the form $k^2 N'$ where $N'$ is primitively represented; this reproduces our condition.} For $\abs{\tr g} = 0, 1, 3$, all such forms are equivalent\footnote{\ Up to a possible sign change between positive and negative definite forms.}, i.e.~related by $\mathrm{SL}(2, \mathbb{Z})$ transformations of $(x, y)$, hence the corresponding entries of \cref{tab:SL2Z-N} are exactly the $N$ that can be expressed as $x^2 + (\tr g)\, xy + y^2$ for some $x, y \in \mathbb{Z}$. For all other values of $\abs{\tr g}$ (at least those $\leq 1000$) there are multiple classes,\footnote{\ The equivalence classes can be listed in Sage \cite{sagemath} with \texttt{BinaryQF\char`_reduced\char`_representatives(d)}.} so it is not a priori enough to consider a single quadratic form.

Our classification generalizes results obtained for specific $g$ in the
literature. Lagrangians invariant under $g = \mathsf{S}$ were studied in
\cite{Sun:2023xxv}; the row $\tr g = 0$ (listing the sums of two squares)
reproduces table 1 of that paper. For $g = \mathsf{ST}^n$, which has trace $n$,
the existence of invariant lagrangians $L$ was addressed in
\cite{Cordova:2023bja} by solving the condition $g \mathcal{M}_L = \mathcal{M}_L
g$ \cite[eq.~2.10]{Cordova:2023bja}, which reduces to solving $x^2 + nxy + y^2 =
N$ for $x, y \in \mathbb{Z}$. This is a sufficient condition for the existence
of an invariant $L$ ($g \mathcal{M}_L = \mathcal{M}_L B$), but as we have seen
above, only guaranteed to be necessary for $n = 0, 1, 3$. Indeed, the
corresponding rows of \cref{tab:SL2Z-N} agree with equations 2.20, 2.21 and 2.23 of that paper.\footnote{\ Apart from some minor omissions for $n = 0, 1$, beginning with $N = 81$.} For $n = 2$, \cite[eq.~2.22]{Cordova:2023bja} lists only the perfect squares, but we have seen that invariant lagrangians exist for every $N$ since the characteristic polynomial is $(x - 1)^2$ over the integers; indeed, for $g = \mathsf{ST}^2$, $\Span_{\mathbb{Z}_N} (1, -1)$ is always an invariant lagrangian. Also for higher $n$, we expect our results to diverge. For e.g.~$n = 4$, we have already found invariant lagrangians over $\mathbb{Z}_2$ and $\mathbb{Z}_3$, where $\mathsf{ST}^4$ reduces to $\mathsf{S}$ and $\mathsf{ST}$ respectively, but neither $2$ nor $3$ can be expressed as $x^2 + 4xy + y^2$.\footnote{\ This can be seen by reducing mod $3$ and mod $4$, respectively.}

\section{Invariant lagrangians and Galois theory}
\label{sec:invar-lagr-galo}
In the last section we have seen that $p$-adic numbers provide us a powerful handle on the existence of $g$-invariant subspaces over base ring $\mathbb{Z}_{p^k}$ for all $k$. In this section we will further develop the methods to find invariant lagrangians. These techniques are powerful enough for us to completely analyze the case of finite subgroups of $\mathrm{Sp}(2r,\mathbb{Z})$. This is an important class of automorphisms of BF theory e.g.~these are the only subgroups that can be symmetries of a theory of class $\mathcal{S}$. However we stress that our methods are not limited the this class, they are generally applicable. The class of finite subgroups of $\mathrm{Sp}(2r,\mathbb{Z})$ merely provides a particularly tractable yet very rich example of their applicability.

\subsection{Structure of symplectic matrices over a field}
\label{sec:struct-sympl-matr}
Here we restate the results of \cite{Bashmakov:2022uek} in a manner that is applicable to any field $F$. For us the fields of interest will primarily be $\mathbb{F}_{p}$, $\mathbb{Q}_{p}$ and $\mathbb{Q}$. Let $g \in \mathrm{Sp}(2r,F)$ and $P(x) = \det(x-g)$ be its characteristic polynomial. There is always an extension $E$ of $F$ such that over $E$, $P(x)$ splits into linear factors i.e.
\begin{align}
  \label{eq:39}
  P(x) = \prod_{\lambda}(x - \lambda)^{m_{\lambda}}
\end{align}
where $\lambda$ are the distinct eigenvalues of $g$ and $m_{\lambda}$ the multiplicity of $\lambda$. The subspaces $\ker(g - \lambda)^{m}$ with $m \leq m_{\lambda}$ are invariant under the action of $g$. Furthermore $\ker(g - \lambda)^{m}$ and $\ker(g - \lambda^{\prime})^{m^{\prime}}$ have no common vectors except $0$ if $\lambda \neq \lambda^{\prime}$. Also for $m^{\prime} < m$, $\ker(g - \lambda)^{m^{\prime}} \subseteq \ker(g - \lambda)^{m}$. This is true for for all matrices. Since $g$ is symplectic we are also interested in how vectors in these subspaces intersect with each other. If $u_{\lambda} \in \ker(g - \lambda)^{m}$ and $u_{\lambda^{\prime}} \in \ker(g - \lambda^{\prime})^{m^{\prime}}$, it can be seen that,
\begin{align}
  \label{eq:40}
  \ev{u_{\lambda} , u_{\lambda^{\prime}}} = 0 ~, && \mbox{unless } \lambda\lambda^{\prime} = 1.
\end{align}
i.e.~if two vectors intersect each other they must belong to invariant subspaces labeled by eigenvalues which are inverses of each other. Since the symplectic form $\Omega$ is non-degenerate, a vector must intersect some other vector. Hence if $\lambda$ is an eigenvalue of $g$, $\lambda^{-1}$ must also be eigenvalue. As a result $P(x)$ is always a palindromic polynomial i.e. the coefficient of $x^{i}$ is the same as the coefficient of $x^{2r-i}$.

This also means that $\ker(g - \lambda)^{m}$ is an isotropic subspace if $\lambda \neq \pm 1$, on the other hand vectors in $\ker(g - \lambda)^{m_{\lambda}}$ for $\lambda = \pm 1$, can only intersect another vector in the same invariant subspace as them.

The subspaces $\ker(g - \lambda)^{m_{\lambda}}$ can be further decomposed into Jordan blocks. A Jordan block of size $n$ can be spanned by vectors $u_{\lambda,i}$ for $0 \leq i < n$,
\begin{align}
  \label{eq:41}
  g u_{\lambda,0} &= \lambda u_{\lambda,0} ~, \nonumber \\
  g u_{\lambda,i} &= \lambda u_{\lambda,i} + u_{\lambda,i-1} ~.
\end{align}
As a result a Jordan block of size $n$ contains precisely $n$ invariant subspaces $S_{\lambda,i} = \mbox{Span}(\{u_{\lambda,j}\, |~ j \leq i\})$. Furthermore for each Jordan block $J_{\lambda}$ there is a dual Jordan block $J_{\lambda^{-1}}$ such that vectors in $J_{\lambda}$ only intersect those in $J_{\lambda^{-1}}$. Furthermore,
\begin{align}
  \label{eq:42}
  \ev{u_{\lambda,i} , u_{\lambda^{-1},j}} = 0 && \mbox{if } i + j < n - 1 ~.
\end{align}

As a result of this simple structure once the Jordan blocks of $g$ and their degeneracy is known finding invariant lagrangians over $E$ is a matter of combinatorics of various ways of assembling a collection of Jordan (sub)blocks which don't intersect each other.

\subsection{Galois Group}
\label{sec:galois-group}
The above discussion streamlines the discovery of invariant lagrangians over a splitting field of $P(x) = \det(x-g)$. However mostly we will be interested in a field $F$ over which $P(x)$ does not split. Hence we need to figure out which of the lagrangians over $E$ can be thought of as lagrangians over $F$. The ingredient needed to answer this question is the Galois group $\gal(E / F)$ of the field extension $E$ over $F$ i.e. the group of automorphisms of $E$ that acts as identity on $F$. If $\mu \in E$, then $\mathsf{g} \, \mu$ with $\mathsf{g} \in \gal(E / F)$ is called a Galois conjugate of $\mu$.

Since $P(x)$ is a polynomial over $F$, if $\lambda$ is a root of $P(x)$ then $\mathsf{g} \lambda$ must also be a root and with the same multiplicity. Denoting by $C_{\lambda}$ the set of Galois conjugates of $\lambda$, the product
\begin{align}
  \label{eq:43}
  Q_{\lambda}(x) = \prod_{\mu \in C_{\lambda}}(x - \mu)
\end{align}
is invariant under the Galois group action and hence its coefficients must be in $F$. Hence $Q_{\lambda}$ is a factor of $P(x)$ over $F$ and moreover it is irreducible over $F$ i.e. once we know the factorization of $P(x)$ over its splitting field, the irreducible factors over $F$ can obtained by acting on linear factors by Galois group elements. The irreducible factors of $P(x)$ are thus labeled by the orbits of its root under the action of $\gal (E / F)$ i.e.,
\begin{align}
  \label{eq:44}
  P(x) = \prod_{C_{\lambda}} (Q_{\lambda}(x))^{m_{\lambda}} ~.
\end{align}
The $g$-invariant subspaces over $F$ are thus annihilated by $(Q_{\lambda}(g))^{m}$ for some $m \leq m_{\lambda}$. From the discussion in \cref{sec:struct-sympl-matr} it can be deduced that $\ker((Q_{\lambda}(g))^{m})$ is isotropic if $\lambda$ and $\lambda^{-1}$ are not Galois conjugates. If the multiplicity $m_{\lambda}$ is odd the converse also holds i.e.~if $\lambda$ and $\lambda^{-1}$ are conjugates then $\ker(Q_{\lambda}(g))$ is not isotropic. For even multiplicity there are degenerate eigenspaces or (possibly degenerate) Jordan blocks it is always possible to construct invariant lagrangian subspaces by appropriately pairing eigenvectors or Jordan blocks with conjugate eigenvalues which don't intersect each other.

\subsubsection{Example: matrices with distinct eigenvalues}
\label{sec:exampl-matr-with}
Although we will discuss examples of matrices with higher multiplicity, let us start with the simplest case i.e.~over the splitting field $E$, all roots of $\det(x-g)$ are pairwise distinct.\footnote{\ This excludes $\lambda = \pm 1$. For symplectic matrices, these eigenvalues always occur with even algebraic multiplicity.} In that case we can split the vector space $E^{2r}$ into $2r$ eigenspaces spanned by eigenvectors $u_{\lambda}$ with the only non-vanishing intersections being $\ev{u_{\lambda},u_{\lambda^{-1}}}$. Hence we can divide eigenvalues into $r$ pairs of $(\lambda_{i},\lambda_{i}^{-1})$. Each $g$-invariant lagrangian is a span of a collection of $g$-eigenvectors $\{u_{\mu}\}$ such that if $\{u_{\mu}\}$ is among $u_{\mu}$ then $u_{\lambda^{-i}}$ is not. Hence there are precisely $2^{r}$ invariant lagrangians.

Next let us consider the situation over $F$: if there is an eigenvalue $\lambda$ such that $\lambda$ and $\lambda^{-1}$ are Galois conjugates then any invariant subspaces of $g$ over $F$ must contain either both of $u_{\lambda}$ and $u_{\lambda^{-1}}$ or neither of them. Hence no lagrangian can be invariant. On the other hand if no eigenvalue is a Galois conjugate of its inverse then invariant lagrangians exist. In this case the characteristic polynomials splits as in \eqref{eq:43} and vectors in $Q_{\lambda}$ only intersect those in $Q_{\lambda^{-1}}$. Hence the number of factor must be even, let's say $2h$, and there are $2^{h}$ invariant lagrangians.

In fact using the non-vanishing intersections described in \eqref{eq:42} it can be deduced that invariant lagrangians do not exist if there is a pair of eigenvalues $\lambda$ and $\lambda^{-1}$ with odd multiplicity which are Galois conjugates. If no eigenvalue $\lambda$ is Galois conjugate to its inverse $\lambda^{-1}$, invariant lagrangians always exist. This leaves the case of $\lambda$ and $\lambda^{-1}$ which are Galois conjugate but they occur with even multiplicity in the characteristic polynomial. In such a case we need to know the detailed structure of Jordan blocks: invariant lagrangian always exist unless there are Jordan blocks with odd size and odd degeneracy.

\section{Finite order elements of $\mathrm{Sp}(2r,\mathbb{Z})$}
\label{sec:finite-order-elem}

In this section we will combine the discussion of $p$-adic invariant lagrangians in \cref{sec:invar-subsp-polyn} with the application of Galois theory in \cref{sec:invar-lagr-galo} to investigate the existence of invariant lagrangians for elements of finite order of $\mathrm{Sp}(2r,\mathbb{Z})$.

Following \cite{Bashmakov:2022uek} we recall some basic facts. Regarded as a rational matrix, such a matrix $g$ is always diagonalizable over the splitting field of its characteristic polynomial. Its characteristic polynomial can be written as a product of cyclotomic polynomials $\Phi_{n}(x)$, i.e.
\begin{align}
  \label{eq:45}
  \det(x - g) = \prod_{n}(\Phi_{n}(x))^{m_{n}} ~.
\end{align}
The roots of $\Phi_{n}(x)$ are primitive $n$-th roots of unity, i.e.~$\exp(2\pi i \frac{k}{n})$ with $k$ coprime to $n$, each with multiplicity one. If $\lambda$ is a primitive $n$-th root of unity, so is its inverse $\lambda^{-1}$. Therefore the space $\ker((\Phi_{n}(g))^{m_{n}})$ is not only $g$-invariant, but $g$ restricted to this subspace is once again\footnote{\ up to a change of basis} a symplectic transformation w.r.t.~$\Omega$ of its dimension. As a result over a field extension of $\mathbb{Q}$, a $g$-invariant lagrangian consists of a $g$-invariant lagrangian for each of $\ker((\Phi_{n}(g))^{m_{n}})$.

If $p$ is not a divisor of any $n$ with non-zero multiplicity, almost all of this holds verbatim for $\mathbb{F}_{p}$. Primitive $n$-th roots of unity now interpreted as elements of order $n$ in (some extension of) $\mathbb{F}_{p}$ i.e.~to assemble an invariant lagrangian for $g$ over $\mathbb{F}_{p}$ we need a $g$-invariant lagrangian for each of $\ker((\Phi_{n}(g))^{m_{n}})$. If on the other hand $p$ divides $n$, the cyclotomic polynomial $\Phi_{n}$ factorizes over $\mathbb{F}_{p}$ as,
\begin{align}
  \label{eq:46}
  \Phi_{n}(x) = (\Phi_{n^{\prime}}(x))^{\phi(p^{l})} ~,
\end{align}
with $n = p^{l}n^{\prime}$ and $\gcd(p,n^{\prime}) = 1$ and $\phi$ the Euler totient function. This factorization means that the splitting field $E$ of $\Phi_{n}(x)$ over $\mathbb{F}_{p}$ is the splitting field of $\Phi_{n^{\prime}}(x)$. Over $E$, the structure of $\ker((\Phi_{n}(g))^{m_{n}})$ can be described precisely: each primitive $n^{\prime}$-th root of unity contributes $m_{n}$ Jordan blocks of size $p^{l}$ to it.\footnote{\ If there are no other factors of $\Phi_{n^{\prime}}(x)$ present these Jordan blocks span $\ker((\Phi_{n}^{g})^{m_{n}})$, however there can be other factors which come from the $\mod p$ reduction of \eqref{eq:45}. These factors can contribute more Jordan blocks to $\ker((\Phi_{n}(g))^{m_{n}})$.}

These are the only additional relations that arise from reducing \eqref{eq:45} modulo $p$. Taking them into account we can write analogue of \eqref{eq:45} which only contains cyclotomic polynomial $\Phi_{n}(x)$ with $n$ coprime to $p$ and apply the discussion of \cref{sec:invar-lagr-galo}.

\subsection{Galois groups of cyclotomic polynomials}
\label{sec:galo-groups-cyclt}
For this purpose we start by describing Galois groups of cyclotomic polynomials\footnote{\ The Galois group of a polynomial $P(x)$ over a field $F$ is the Galois group $\gal(E / F)$ of the splitting field $E$ of $P(x)$.} $\Phi_{n}(x)$ over $\mathbb{Q},\mathbb{F}_{p}$ and $\mathbb{Q}_{p}$ and as will be important for us the relationship between them.

Over $\mathbb{Q}$, the cyclotomic polynomials are by definition irreducible. The eigenvalues are primitive $n$-th roots of unity $\exp(2\pi i \frac{s}{n})$ with $\gcd(s,n)=1$. Given one such root $\zeta_{n}$ the other roots are $\zeta_{n}^{s}$. As a result the splitting field of cyclotomic polynomial $\Phi_{n}(x)$ (called the $n$-th cyclotomic field) is obtained by adjoining $\zeta_{n}$ to $\mathbb{Q}$. The Galois group thus consists of transformations that map one primitive $n$-th roots of unity of another primitive $n$-th root of unity, i.e.~it consists of maps,
\begin{align}
  \label{eq:47}
  \zeta_{n} \mapsto \zeta_{n}^{s}
\end{align}
with $\gcd(n,s) = 1$. Hence it is isomorphic to $\mathbb{Z}_{n}^{\times}$, i.e.~the group of invertible elements of $\mathbb{Z}_{n}$, and has order $\phi(n)$.

Over $\mathbb{F}_{p}$ we can find solutions to more equations than over $\mathbb{Q}$ and as a result the Galois group is typically smaller. In fact if $n$ and $p$ are coprime it is the cyclic subgroup of $\mathbb{Z}_{n}^{\times}$ generated by $p \in \mathbb{Z}_{n}^{\times}$ and consists of maps \cite{MR1697859}:
\begin{align}
  \label{eq:48}
  \zeta_{n} \mapsto \zeta_{n}^{p^{a}} ~.
\end{align}
with $0 \leq a < \abs{p}_{n}$ where we denote by $\abs{p}_{n}$ the order of $p$ in $\mathbb{Z}_{n}^{\times}$. This means in particular that over $\mathbb{F}_{p}$, $\Phi_{n}(x)$ splits into $\phi(n) / \abs{p}_{n}$ factors, each of degree $\abs{p}_{n}$. The factors of $\Phi_{n}(x)$ are indexed by cosets $[s]$ of the group $\ev{p}_{n}$ generated by $p$ in $\mathbb{Z}_{n}^{\times}$. Explicitly, given one primitive root $\zeta_{n}$, the factor $P_{[s]}(x)$ is given by,
\begin{align}
  \label{eq:49}
  P_{[s]}(x) = \prod_{a = 0}^{\abs{p}_{n} - 1}(x - \zeta_{n}^{p^{a}s}) ~.
\end{align}

Lastly let us describe the case of $\mathbb{Q}_{p}$. When $n$ and $p$ are coprime, the Galois group of $\mathbb{F}_{p}$ and $\mathbb{Q}_{p}$ are isomorphic. Hence regarding $\zeta_{n}$ as the $p$-adic $n$-th root of unity, the Galois group is still generated by \eqref{eq:48}. This is a consequence of Hensel's lemma which ensures that each of the factors in \eqref{eq:49} lift to $p$-adic factors.

When $n$ and $p$ are not coprime i.e.~$n = p^{l}n^{\prime}$ the situation is different because of the factorization over $\mathbb{F}_{p}$ given by \eqref{eq:46}. As a result the Galois group of $\Phi_{n}(x)$ over $\mathbb{F}_{p}$ is the Galois group of $\Phi_{n^{\prime}}(x)$ i.e.~it is the subgroup of $\mathbb{Z}_{n^{\prime}}^{\times}$ generated by $p$. However because of repeated factors, Hensel's lemma does not hold and the factorization in \eqref{eq:46} does not lift to a factorization over $\mathbb{Q}_{p}$. Indeed even when $p$ and $n$ are not coprime, the $p$-adic Galois group of $\Phi_{n}$ is bigger than the Galois group over $\mathbb{F}_{p}$.\footnote{\ More precisely the $p$-series for an $n$-th root of unity takes the form $\zeta_{n^{\prime}} + p R$ i.e.~the difference between $n$-th and $n^{\prime}$-th roots of unity starts are order $p$ and hence is irrelevant over $\mathbb{F}_{p}$.}

More precisely when $n = p^{l}$, the $p$-adic Galois group of $\Phi_{n}(x)$ is same as over rationals i.e.~$\mathbb{Z}_{n}^{\times}$. When $n = p^l n'$, the $p$-adic Galois group has two factors, which can be constructed by noticing that $\zeta_{p^{l} n'}$ can be written as a product $\zeta_{p^l}\zeta_{n'}$. As a result the $p$-adic Galois group is $\mathbb{Z}_{p^{l}}^{\times} \times \ev{p}_{n'}$.

Note that the fact that irreducible factors over $\mathbb{Q}_{p}$ are invariant under the Galois group action means that (monic) factors of $\Phi_{n}(x)$ over $\mathbb{Z}_{p^{k}}$ are invariant under this action for all $k$. By keeping track of this action it is possible to show that for $k > 1$, the irreducible factors\footnote{\ We once again point out that factorization over $\mathbb{Z}_{p^k}$ is not unique for $k > 1$ and irreducibility cannot be defined in the usual way. Here we mean just the factorization into monic polynomials which cannot be further factorized into monic polynomials. Such factorization is unique by virtue of unique factorization in $\mathbb{Q}_{p}(x)$.} of $\Phi_{n}(x)$ over $\mathbb{Z}_{p^{k}}$ are precisely those corresponding to mod $p^{k}$ reduction of the irreducible $p$-adic factors. We leave a proof of this fact to \cref{sec:cyclo_poly_except_primes}.

\subsection{Existence of invariant lagrangians}
\label{sec:exist-invar-lagr}

Having determined the Galois groups of $\Phi_{n}(x)$ over $\mathbb{F}_{p}$ and $\mathbb{Q}_{p}$ we can now use the discussion in \cref{sec:invar-lagr-galo} to give necessary and sufficient conditions for the existence of an invariant lagrangian under the action of an element of $g$ of finite order in $\mathrm{Sp}(2r,\mathbb{Z})$. For such $g$ the condition discussed in \cref{sec:invar-lagr-galo} is particularly simple: over rationals the map in \eqref{eq:47} with $k = -1 \in \mathbb{Z}_{n}^{\times}$ maps a primitive $n$-th root of unity to its inverse. Furthermore no other element of Galois group maps a root to its inverse. Since Galois groups over $\mathbb{Q}_{p}$ and $\mathbb{F}_{p}$ are subgroups of the Galois group over rationals, given by maps in \eqref{eq:48}, the existence of lagrangians depends on whether $-1 \in \mathbb{Z}_{n}^{\times}$ is in the Galois group or not.

Let us now state the consequences for existence of invariant lagrangian over $\mathbb{Q}_{p}$:
\begin{itemize}
  \item If $\det(x-g) = (\Phi_{n}(x))^{m}$, invariant lagrangians always exist over $\mathbb{Q}_{p}$ if $m$ is even. For odd $m$ they exist iff $-1 \notin \ev{p}_{n}$ where $\ev{p}_{n}$ is the group generated by $p$ inside $\mathbb{Z}_{n}^{\times}$.
  \item If $\det(x-g) = \prod_{n} (\Phi_{n}(x))^{m_{n}}$, invariant lagrangians exist iff they exist for $(\Phi_{n}(x))^{m_{n}}$ for all $n$ with non-zero multiplicity $m_{n}$.
\end{itemize}

Over $\mathbb{F}_{p}$, the story goes along much the same lines. The only wrinkle is the additional factorization given by \eqref{eq:46}. If the characteristic polynomials has both $\Phi_{p^{l}n'}(x)$ and $\Phi_{p^{l'}n'}(x)$ as factors with non-zero degeneracy, they will both contribute to the multiplicity of $\Phi_{n'}(x)$ over $\mathbb{F}_{p}$. However since $l \neq l^{\prime}$ these Jordan blocks are of different size and the degeneracy of Jordan blocks is not affected. Hence we need to find invariant lagrangians among these subspaces independently.\footnote{\ This is one of the points where we are using the assumption that $g$ is a finite order element of $\mathrm{Sp}(2r,\mathbb{Z})$. Just knowing the characteristic polynomial over $\mathbb{F}_{p}$ is not enough to determine the size and degeneracy of Jordan blocks.} Taking it into account, we deduce,
\begin{itemize}
  \item If $\det(x-g) = (\Phi_{p^{l}n'}(x))^{m} = (\Phi_{n'}(x))^{\phi(p^l)m}$, invariant lagrangians always over $\mathbb{F}_{p}$ if $\phi(p^l) m$ is even. For odd $\phi(p^l) m$ they exist iff $-1 \notin \ev{p}_{n'}$. Since $\phi(p^l)$ is odd precisely when $l = 0$ or $(p, l) = (2, 1)$, this means in terms of $n = p^l n'$ that they exist for odd $m$ unless $p$ is coprime to $n$ and $-1 \notin \ev{p}_n$, or $p \neq 2$ and $p \mid m$, or $p = 2$ and $4 \mid n$, or $p = 2$, $n = 2 \pmod{4}$ and $-1 \notin \ev{2}_{n/2}$. These are the promised conditions of \Cref{thm:main}.
  \item More generally if $\det(x-g) = \prod_{n,l} (\Phi_{p^{l}n}(x))^{m_{n,l}}$, as an integer polynomial, invariant lagrangians exist over $\mathbb{F}_{p}$ iff they exist for $(\Phi_{p^{l}n}(x))^{m_{n,l}}$ for all $n,l$ with non-zero multiplicity $m_{n}$.
\end{itemize}

\subsection{Invariant lagrangians over $\mathbb{Z}_{p^k}$}
\label{sec:invar-lagr-over}
Having described when invariant lagrangians exist over $\mathbb{Q}_{p}$ and $\mathbb{F}_{p}$ we can now answer the question we started with: the existence of invariant lagrangians over $\mathbb{Z}_{p^k}$. We will divide the task into two steps:

\subsubsection{Free invariant lagrangians}
\label{sec:free-invar-lagr}
The lagrangians which are most directly related to $\mathbb{Q}_{p}$ or equivalently $\hat{\mathbb{Z}}_{p}$ lagrangians are those that are free as a module over $\mathbb{Z}_{p^k}$. As a consequence of Cayley-Hamilton theorem and Hensel lifting, these are just the mod $p^k$ reduction of $\hat{\mathbb{Z}}_{p}$ $g$-invariant lagrangians. These correspond to the boundary conditions with $\mathbb{Z}_{p^k}^{r}$ symmetry.

\paragraph{Single cyclotomic polynomial:}
Let us look at these in more detail starting with the case when
\begin{align}
  \label{eq:50}
  \det(x-g) = \Phi_{n}(x)
\end{align}
is a single cyclotomic polynomial (then $\phi(n) = 2r$). Over $\mathbb{Q}_{p}$ the irreducible factors of $\Phi_{n}(x)$ are $P_{[s]}(x)$ given by \eqref{eq:49}. The indecomposable $g$-invariant subspaces are in one-to-one correspondence with these factors and are non-degenerate i.e.,
\begin{align}
  \label{eq:51}
  \hat{\mathbb{Z}}_{p}^{2r} \cong \bigoplus_{[s]} \mathcal{V}_{[s]} ~, && \mathcal{V}_{[s]} \equiv \ker(P_{[s]}(g)) ~.
\end{align}
Any invariant subspace of $\hat{\mathbb{Z}}_{p}^{2r}$ can be written as a sum of $\mathcal{V}_{[s]}$. If $-1 \notin \ev{p}_{n}$, $\mathcal{V}_{s}$ are isotropic. Furthermore vectors in $\mathcal{V}_{[s]}$ only intersect vectors in $\mathcal{V}_{[-s]}$. The number of $\mathcal{V}_{[s]}$ is even and we denote it by $2\tilde{r} = 2r/\abs{p}_n$. We can pairs $\mathcal{V}_{[s]}$ and $\mathcal{V}_{[-s]}$ to obtain $\tilde{r}$ pairs. An $g$-invariant lagrangian consists of one subspace from each of these $\tilde{r}$ pairs. Hence there are $2^{\tilde{r}}$ invariant lagrangians over $\hat{\mathbb{Z}}_{p}$.

Reducing these mod $p^{2l}$ we obtain $2^{\tilde{r}}$ lagrangians over $\mathbb{Z}_{p^{l}}$. If $p > n$ or if $n$ and $p$ are coprime these are all the invariant lagrangians. However for $n = p^{l}n^{\prime}$ additional lagrangians exists over $\mathbb{F}_{p}$ as a result of \eqref{eq:46}. In the current case it is easy to enumerate the extra invariant subspaces resulting from this factorization. For $n^{\prime} > 2$, these are,
\begin{align}
  \label{eq:52}
  \mathcal{V}_{[s],a} = \ker((P_{[s]}(g))^{a}) ~,
\end{align}
with $a \leq \phi(p^l)$ and $P_{[s]}$ an irreducible factor of $\Phi_{n^{\prime}}(x)$ over $\mathbb{F}_{p}$. The subspace $\mathcal{V}_{[s],a} \oplus V_{[-s], \phi(p^l) -a}$ is invariant and isotropic. An invariant lagrangian consists of such a subspace for each $[s]$. As result there are $p^k 2^{r^{\prime}}$ invariant subspaces where $2r^{\prime}$ is the number of irreducible factors of $\Phi_{n^{\prime}}(x)$ over $\mathbb{F}_{p}$.\footnote{\ Or $\mathbb{Q}_{p}$, since $n^{\prime}$ is coprime to $p$.}

If $n \in {1,2}$, there is a single Jordan block over $\mathbb{F}_{p}$ and it includes a single invariant lagrangian given by $\ker((g - 1)^{\phi(p^l) / 2})$ for $n = 1$ and $\ker((g + 1)^{\phi(p^l) / 2})$ for $n=2$.

\paragraph{Power of a cyclotomic polynomial:}
Next let us consider the case when the characteristic polynomial,
\begin{align}
  \label{eq:53}
  \det(x - g) = (\Phi_{n}(x))^{m} ~.
\end{align}
In this case over the $p$-adic splitting field of $\Phi_{n}(x)$, each primitive $n$-th root of unity has an $m$-dimensional eigenspace. As before let $P_{[s]}(x)$ be an irreducible factor of $\Phi_{n}(x)$ over $\hat{\mathbb{Z}}_{p}$ then the indecomposable $g$-invariant subspaces over $\hat{\mathbb{Z}}_{p}$ are precisely those that have dimension equal to the degree of $P_{[s]}(x)$ i.e.~$\abs{p}_{n}$ and are annihilated by $P_{[s]}(g)$ for some $[s]$. To construct such a subspace we need to pick one eigenvector for each of the Galois conjugates that are factors of $P_{[s]}(x)$. Bigger invariant subspaces also exist and have dimension which are multiples of $\abs{p}_{n}$. To construct a $\tilde{m}\abs{p}_{n}$ dimensional space we need to pick a $\tilde{m}$-dimensional plane from the eigenspace of each of $\abs{p}_{n}$ Galois conjugate. The biggest of these invariant spaces has dimension $\abs{p}_{n} m$ and is $\ker(P_{[s]}(g))$.

If $-1 \notin \ev{p}_{n}$, $g$-invariant lagrangians always exist. To construct one we can choose an invariant subspace $\mathcal{V}_{[s],\tilde{m}} \subset \ker(P_{[s]}(g))$ of dimension $\abs{p}_{n}\tilde{m}$. This subspaces uniquely determines a subspace $\mathcal{V}_{[-s],\tilde{m}}$ of dimension $\abs{p}_{n}(m - \tilde{m})$. It is the subspace consisting of all vectors in $\ker(P_{[-s]}(g))$ that do not intersect with any vector in $\mathcal{V}_{[s],m}$. Any $g$-invariant subspace is then assembled as a sum of such subspaces for pairs $[s]$ and $[-s]$. As a result the space of invariant lagrangians is (at least schematically),
\begin{align}
  \label{eq:54}
  \left(\bigoplus_{\tilde{m} = 0}^{m} \left(\mbox{Gr}(\tilde{m}, \hat{\mathbb{Z}}_{p}^{m}) \right)^{\abs{p}_{n}}\right)^{\frac{\phi(n)}{2\abs{p}_{n}}} ~,
\end{align}
where $\mbox{Gr}(\tilde{m} , M)$ is the space of $\tilde{m}$-dimensional planes in the module $M$.\footnote{\ By plane in this context we mean a submodule $S$ of $M$ which is free and primitive i.e.~$M / S$ is also a free module.}

If $-1 \in \ev{p}_{n}$, $g$-invariant lagrangian only exist for even $m$. In this case $\ker(P_{[s]}(g))$ is not isotropic. To construct an invariant lagrangian we need to choose a maximal isotropic subspace of $\ker(P_{[s]}(g))$ for all $[s]$. To do so we can choose an $\frac{m}{2}$ dimensional eigenspace $\mathcal{V}_{\lambda}$ for an eigenvalue $\lambda$. This uniquely determines an $\frac{m}{2}$ dimensional subspace $\mathcal{V}_{\lambda^{-1}}$ of eigenspace with eigenvalue $\lambda^{-1}$ such that $\mathcal{V}_{\lambda} \oplus \mathcal{V}_{\lambda^{-1}}$ is isotropic. Doing this for all pairs $\lambda,\lambda^{-1}$ of eigenvalues which are roots of $P_{[s]}(x)$ yields a $g-$invariant lagrangian subspace of $\ker(P_{[s]}(g))$. A $g$-invariant lagrangian of $\mathbb{Q}_{p}^{2r}$ then consists of a collection of $g$-invariant lagrangians, one for each $[s]$. Hence in this case, the space of invariant lagrangian can be schematically described as,
\begin{align}
  \label{eq:55}
  \left(\mbox{Gr}\left(\frac{m}{2} , \hat{\mathbb{Z}}_{p}^{m}\right)\right)^{\frac{\phi(n)}{2}} ~.
\end{align}

After this discussion of $p$-adic lagrangians we can construct lagrangians over $\mathbb{Z}_{p^k}$ by simply reducing mod $p^k$. As before for $p$ and $n$ coprime these are all the invariant lagrangians. If on the other hand, $n = p^l n^{\prime}$, additional lagrangians exist over $\mathbb{F}_{p}$. In this case there are degenerate Jordan blocks with size $\phi(p^l)$ and degeneracy $m$. Invariant lagrangian exist whenever $-1 \notin \ev{p}_{n}$ or if $m$ is even. They can be counted using reasoning similar to that used for \eqref{eq:54} and \eqref{eq:55} but the analysis is messier and we will not attempt it here.

\paragraph{General Case:}
The general case of a product of different cyclotomic polynomials just consists of dealing with each factor $\ker((\Phi_{n}(g))^{m_{n}})$ separately i.e.~over $\mathbb{Q}_{p}$ an invariant lagrangian of $g$ consists of an invariant lagrangian of $\mathcal{V}_{n} = \ker((\Phi_{n}(g))^{m_{n}})$ for $g$ restricted to $\mathcal{V}_{n}$. Hence a $g$-invariant lagrangian exists iff it exists for each of $\mathcal{V}_{n}$.

\subsubsection{Torsional invariant lagrangians}
\label{sec:tors-invar-lagr}
Let us now consider lagrangians over $\mathbb{Z}_{p^k}$ with correspond to one-form symmetry other than $\mathbb{Z}_{p^k}^{r}$. To discuss them it is useful to divide them into two classes:
\begin{itemize}
  \item Ones that descend from $p$-adic invariant subspaces i.e.~they can be written as
    \begin{align}
      \label{eq:56}
      \bigoplus_{i} p^{l_i} \mathcal{V}_{i} \mod p^k
    \end{align}
    with $l_i < k$ and $\mathcal{V}_{i}$ is a $p$-adic invariant subspace.
  \item Those that cannot be written in such a form. We will show that these are extremely limited. They appear because of the additional factorization given by \eqref{eq:46} and only when $k = 2\ell + 1$ is odd.
\end{itemize}

\paragraph{Torsional descendants of $p$-adic invariant subspaces}
Let us once again start with the case where $\det(x-g) = \Phi_{n}(x)$ is a single cyclotomic polynomial which factorizes into factors $P_{[s]}(x)$. If $-1 \notin \ev{p}_{n}$ then $\ker(P_{[s]}(g))$ is isotropic. If an invariant lagrangian $L$ is such that $L \cap \ker(P_{[s]}(g)) = p^{l_{[s]}} \ker(P_{[s]}(g))$ then,
\begin{align}
  \label{eq:57}
  L \cap \ker(P_{[-s]}(g)) = p^{k-l_{[s]}} \ker(P_{[-s]}(g)) ~.
\end{align}
As a result a general torsional lagrangian can be written as,
\begin{align}
  \label{eq:58}
  \bigoplus_{([s],[-s])}  \left(p^{l_{[s]}} \ker(P_{[s]}(g)) \oplus p^{k-l_{[s]}} \ker(P_{[-s]}(g))\right) ~.
\end{align}
where the sum is over the distinct $\tilde{r} = \frac{\phi(n)}{2\abs{p}_{n}}$ pairs $([s],[-s])$. The lagrangian above corresponds to a global symmetry,
\begin{align}
  \label{eq:59}
  \bigtimes_{([s],[-s])} \left(\mathbb{Z}_{p^{k - l_{[s]}}} \times \mathbb{Z}_{p^{l_{[s]}}}\right) ~.
\end{align}
If $-1 \in \ev{p}_{n}$ there is no torsional lagrangian when $k$ is odd and only one when $k = 2\ell$ is even. It is given by $p^{\ell} \mathbb{Z}_{p^{2\ell}}^{2r}$.

The case of $\det(x-g) = (\Phi_{n}(x))^{m}$ is not so different. Let's start with considering that $-1 \notin \ev{p}_{n}$. In this case, for any lagrangian $L$,
\begin{align}
  \label{eq:60}
  V_{[s]} \equiv L \cap \ker(P_{[s]}(g)) = p^{l_{1}} V_{[k],1} \oplus p^{l_{2}} V_{[s],2} \oplus \dots \oplus p^{l_{j}}V_{[s],j} ~,
\end{align}
with $0 \leq  l_{i} \leq l$ and $V_{[s],i}$ are all free invariant subspaces of $\ker(P_{[s]}(g))$ and they span $\ker(P_{[s]}(g))$.\footnote{\ This is just makes describing the structure a little simpler. Since we allow $l_{i} = k$, some part of $\ker(P_{[s]}(g))$ can be completely absent from $V_{[s]}$.} To each of $V_{[s],i}$ we can assign a dual subspace $V_{[-s],i} \subseteq \ker(P_{[-s]}(g))$ which is defined as,
\begin{align}
  \label{eq:61}
  V_{[-s],i} = \{v \in \ker(P_{[-s]}(g)) \,|\, \ev{v,\tilde{v}} = 0 ~ \forall\, \tilde{v}\, \in\, V_{[s],i' \neq i}\} ~.
\end{align}
For $L$ to be lagrangian,
\begin{align}
  \label{eq:62}
  L \cap \ker(P_{[-s]}(g)) = p^{k-l_{1}} V_{[-s],1} \oplus p^{k-l_{2}} V_{[-s],2} \oplus \dots \oplus p^{k-l_{j}}V_{[-s],j} ~.
\end{align}
Repeating this process for all independent $[s]$ we can reconstruct all invariant (possibly) torsional $g$-invariant lagrangians. The global symmetry of $L$ can also be obtained from this description. The contribution to the global symmetry from $V_{[s]} \oplus V_{[-s]}$ is,
\begin{align}
  \label{eq:63}
  (\mathbb{Z}_{p^{k-l_{1}}} \times \mathbb{Z}_{p^{k_{1}}})^{\dim(V_{[s],1})} \times (\mathbb{Z}_{p^{k-l_{2}}} \times \mathbb{Z}_{p^{l_{2}}})^{\dim(V_{[s],2})} \times \dots \times (\mathbb{Z}_{p^{k-l_{j}}} \times \mathbb{Z}_{p^{l_{j}}})^{\dim(V_{[s],j})} ~.
\end{align}

Next we deal with the case of $-1 \in \ev{p}_{n}$. For $k = 2\ell$ is even the invariant lagrangians $p^{\ell} \mathbb{Z}_{p^{2\ell}}^{2r}$ always exists. Additional invariant lagrangians only exist when $m = 2\tilde{m}$ is even. We have already described the free lagrangians in this case. These are the only lagrangians that exist when $k$ is odd. When $l=2\ell$ is even there are additional torsional lagrangians. For a lagrangian $L$, $L \cap \ker(P_{[s]}(g))$ is either a free invariant lagrangian or else $p^{\ell}\ker(P_{[s]}(g))$. The invariant lagrangian $L$ is,
\begin{align}
  \label{eq:64}
  L = \bigoplus_{[k]} L \cap \ker(P_{[k]}(g)) ~.
\end{align}

For the most general case i.e.~a product of cyclotomic polynomials, a torsional invariant lagrangian consists of a collection of torsional invariant lagrangians, one for each of $\ker((\Phi_{n}(g))^{m_{n}})$.

\paragraph{Lagrangians without $p$-adic descent}
\label{sec:no-p-adic-descent}
Let us now describe the lagrangians which do not descend from a $p$-adic lagrangian subspace. To explain how these arise let $V$ be a free subspace of $\mathbb{Z}_{p^k}^{2r}$ which is not invariant under $g$-action. Instead for any $v \in V$,
\begin{align}
  \label{eq:65}
  g v = \tilde{v} + p^{\tilde{k}} w ~, && \tilde{v} \in V ~, w \notin V \setminus \{0\} ~.
\end{align}
If this the case with $\tilde{k} < k$, although $V$ is not $g$-invariant the torsional subspace $p^{k-\tilde{k}} V$ is $g$-invariant. Similarly $p^{l} V \cup p^{l+\tilde{k}}\mathbb{Z}_{p^k}^{2r}$ is also an invariant subspace.

To understand when such subspaces exist we can choose an integral lift $\tilde{V}$ of $V$. $\tilde{V} \bmod p$ is then a $g$-invariant subspace over $\mathbb{F}_{p}$ i.e.
\begin{align}
  \label{eq:66}
  \tilde{V} \bmod p = \ker(\tilde{P}(g)) ~.
\end{align}
with $\tilde{P}(x)$ a factor of $\det(x-g) \bmod p$. If $\tilde{P}(x)$ has a $p$-adic lift $P(x)$, then as we have seen $V$ must be $\ker(P(x)) \bmod p^k$ and as a result $V$ is an invariant subspace on its own i.e. $w = 0$ for $v \in V$. Hence $g$-invariant subspaces without $p$-adic descent can exist only when there are factors of $\det(x-g) \bmod p$ for which Hensel lifting fails.

If $g$ is a finite element of $\mathrm{Sp}(2r,\mathbb{Z})$ Hensel lifting fails only if some $\Phi_{p^{l}n'}(x)$ is a factor of $\det(g-x)$. This failure happens due to the degenerate factorization given by \eqref{eq:46}. We have already described the extra lagrangians obtained over $\mathbb{F}_{p}$ in this way. The extra torsional lagrangians are also easy to describe: let $\tilde{L}$ be one of the lagrangians over $\mathbb{F}_{p}$ that don't have a $p$-adic lift. We can then choose any $p$-adic lift $\mathcal{L}$ of it: this lift $\mathcal{L}$ is not invariant instead for $v \in \mathcal{L}$,
\begin{align}
  \label{eq:67}
  g v = \tilde{v} + p w ~, && \tilde{v} \in \mathcal{L} ~, w \notin \mathcal{L} \setminus \{0\} ~.
\end{align}
Furthermore there is always some $v \in \mathcal{L}$ such that $w$ above is a non-zero primitive vector i.e. $w \neq p \tilde{w}$ for any $\tilde{w}$.\footnote{\ This is a consequence of Proposition 7.13 in Chapter 2 of \cite{MR1697859}.} If $k = 2\ell+1$, the subspace
\begin{align}
  \label{eq:68}
  L = \qty(p^{\ell} \mathcal{L} \bmod p^{2\ell+1}) \cup p^{\ell+1} \mathbb{Z}_{p^k}^{2r}
\end{align}
is an invariant lagrangian and any torsional lagrangian which is not of $p$-adic descent takes this form.\footnote{\ To see this, notice that given $L$, reducing to $\mathbb{Z}_p$ gives an $\tilde{L}$ and $\mathcal{L}$ as above; then $L$ must be of the form $(p^l \mathcal{L} \bmod p^k) \cup p^{l+1} \mathbb{Z}_{p^k}^{2r}$ for some $l$, but only $l = \ell = \frac{k-1}{2}$ produces a group of the proper order.}

\section{Non-abelian subgroups of $\mathrm{Sp}(2r,\mathbb{Z})$}
\label{sec:non-abel-subgr}
In the last section we have dealt comprehensively with the case of finite cyclic subgroups of $\mathrm{Sp}(2r,\mathbb{Z})$. We have shown that the salient properties of invariant lagrangians can be deduced entirely from the knowledge of the characteristic polynomial of the generator of the group. The case of a general finite Abelian subgroup of $\mathrm{Sp}(2r,\mathbb{Z})$ is very close to it. However since it tells us nothing about the commutation relations, knowledge of characteristic polynomial of its generators is insufficient in the case of non-abelian subgroups. In this section we show how linear algebra over number fields can be used to efficiently build and classify invariant lagrangians even for non-abelian groups.

\subsection{Invariant subspaces over a field}
\label{sec:invar-subsp-over}
Given a set of generators $\{g_{i}\}$ of a group $G$, we can use the results developed so far to obtain the $g_{i}$ invariant lagrangians for all of $g_{i}$. In particular we can put each of $g_{i}$ in the Jordan normal form. However non-commutativity means that we can't put all $g_{i}$ into Jordan normal form simultaneously. However if one of the generators has a small number of invariant lagrangians the rest of the task in quite simple. We just need to check if one of these lagrangians is also invariant under the action of all the others generators. In the examples we consider there will always be some element $g_{0}$ of $G$ with characteristic polynomial a single cyclotomic polynomial $\Phi_{n}(x)$. As a result the number of indecomposable invariant subspaces over both $\mathbb{F}_{p}$ and $\mathbb{Q}_{p}$ is at most $n$. If the number of indecomposable invariant subspaces is $k$, we can do a basis change such that matrix for $g_{0}$ is block diagonal with $k$-blocks. Choosing a generating set $\{g_{i}\}$ that includes $g_{0}$ we just need to check if this basis change also puts other $g_{i}$ in block diagonal form.

This process is straightforward for a given $p$. However by making some further use of Galois theory and number fields we can reduce the problem of carrying this out for all primes to a finite number of cases.

\subsubsection{Subfields of cyclotomic fields}
\label{sec:subf-cyclt-fields}
The main idea is to use the fact that the factorization of $\Phi_{n}(x)$ over $\mathbb{Q}_{p}$ depends only on $p \mod n$ or more precisely it depends only on the Galois group $\ev{p}_{n}$ generated by $p$ inside $\mathbb{Z}_{n}^{\times}$.\footnote{\ For this section we will exclude primes which divide $n$. They are an additional finite number of cases and can be dealt with separately using the techniques we have already developed.} Hence the different cases we have to deal with should be classified by the cyclic subgroups of $\mathbb{Z}_{n}^{\times}$ and what we are looking for is a field $F$ such that we should be able to map factorization of $\Phi_{n}(x)$ over $F$ to any $\mathbb{Q}_{p}$ such that the Galois group of $\Phi_{n}(x)$ over $\mathbb{Q}_{p}$ is $\ev{p}_{n}$. The existence of such a field is guaranteed by the fundamental theorem of Galois theory. It states that the subfields of a field extension $E$ over a field $F$ are in one to one correspondence with the subgroups of $\gal(E / F)$.

For us the field of interest is the $n$-th cyclotomic field $\mathbb{Q}(\zeta_{n})$ i.e. splitting field of $\Phi_{n}(x)$ over $\mathbb{Q}$. The fundamental theorem tell us that given $\ev{p}_{n}$ there is a field $F_{\ev{p}_{n}}$ such that
\begin{align}
  \label{eq:69}
  \gal(F_{\ev{p}_{n}} ~ /~ \mathbb{Q}) = \mathbb{Z}_{n}^{\times} / \ev{p}_{n} ~.
\end{align}
The cyclotomic field $\mathbb{Q}(\zeta_{n})$ can also be regarded as field extension of $\ev{p}_{n}$ with,
\begin{align}
  \label{eq:70}
  \gal(\mathbb{Q}(\zeta_{n}) ~/~ F_{\ev{p}_{n}}) = \ev{p}_{n} ~.
\end{align}
This means that $F_{\ev{p}_{n}}$ sits in same relation to $\mathbb{Q}(\xi_{n})$ as $\mathbb{Q}_{p}$ sits in relation $\mathbb{Q}_{p}(\xi_{n})$. Since $g \in \mathrm{Sp}(2g,\mathbb{Z})$ we can always map a basis change of $g$ over $F_{\ev{p}_{n}}$ to $\mathbb{Q}_{p^{\prime}}$ for all $p^{\prime}$ such that $\ev{p^{\prime}}_{n} = \ev{p}_{n}$.

Let us unpack this process. Let us recall the irreducible factors of $\Phi_{n}(x)$ over $\mathbb{Q}_{p}$ given by \eqref{eq:49}. Since the polynomial $P_{[k]}(x)$ only involves $\xi_{n}$ and integer coefficients we can also regard it as a polynomial over $\mathbb{Q}(\xi_{n})$. $F_{\ev{p}_{n}}$ is the smallest extension of $\mathbb{Q}$ such that $P_{[k]}(x) \in F_{\ev{p}_{n}}[x]$ i.e. all coefficients $c_{i}$ of $P_{[k]}(x)$ are elements of $F_{\ev{p}_{n}}$. We can also reduce $P_{[k]}(x)$ modulo $p$, this gives us a polynomial over some subfield $F$ of $\mathbb{F}_{p}[\zeta_{n}]$. The matching of Galois groups means that $F$ must be $\mathbb{F}_{p}$. Hensel lifting then ensures that we can lift $P_{[k]}(x)$ to $\mathbb{Q}_{p}$.

So far we have just used a round about way to reproduce our earlier results. What is more important for us is the fact that we can find a basis change over $F_{\ev{p}_{n}}$ to put $g$ with characteristic polynomial $\Phi_{n}(x)$ into block diagonal form. In this basis can reduce $g$ modulo $p$ to get a block diagonal form of $g$ over $\mathbb{F}_{p}$ and we can lift it to $\mathbb{Q}_{p}$. Hence the existence of an invariant lagrangian for any finite subgroup $G$ of $\mathrm{Sp}(2g,\mathbb{Z})$ over $\mathbb{F}_{p}$ and $\mathbb{Q}_{p}$ is equivalent to its existence over $F_{\ev{p}_{n}}$. Since there are only a finite number of $F_{\ev{p}_{n}}$ (the number is bounded by $\phi(n)$) this streamlines the determination of invariant lagrangians for non-abelian groups.

\subsection{Examples}
\label{sec:examples}

In this section, we give detailed examples of finding invariant lagrangians for finite subgroups of $\mathrm{Sp}(2r, \mathbb{Z})$ with $r > 1$. These arise, for example, in theories of class $\mathcal{S}$ defined on a Riemann surface $\Sigma$ where $r$ is the genus of $\Sigma$. The zero-form symmetries $G^{(0)} \subset \mathrm{Sp}(2r, \mathbb{Z})$ are symmetries (conformal automorphisms) of $\Sigma$, combined with appropriate topological manipulations that generically make the symmetry non-invertible. Several examples were analyzed in the paper \cite{Bashmakov:2022uek} in the case when the one-form symmetries are described by the $\mathbb{Z}_p$ BF symmetry TFT, with $p$ prime; that is, a $G^{(0)}$-invariant boundary condition (the existence of which implies that the symmetry is not intrinsically noninvertible) is described by an invariant lagrangian subgroup of $\mathbb{Z}_p \times \mathbb{Z}_p$. Here, we will extend the analysis of two of these examples to lagrangians of $\mathbb{Z}_{p^k} \times \mathbb{Z}_{p^k}$.

\subsubsection{$\mathrm{GL}(2, 3)$: the Bolza surface}
At genus $r = 2$, the surface with the largest (orientation-preserving) conformal automorphism group is the \emph{Bolza surface} \cite{Bolza1887}. The group is of order $48$ and is isomorphic to $\mathrm{GL}(2, 3) = \mathrm{GL}(2, \mathbb{F}_3)$. As a subgroup of $\mathrm{Sp}(4, \mathbb{Z})$, it is generated by two elements (called $M_8$ and $M_7$ respectively in \cite{Bashmakov:2022uek})
\begin{equation}
  A = \mqty(-1 & 1 & 1 & 0 \\ 1 & 0 & 0 & 1 \\ -1 & 0 & 0 & 0 \\ 1 & -1 & 0 & 1)
  \qquad
  B = \mqty(0 & 1 & 0 & 0 \\ 1 & 0 & 0 & 0 \\ 0 & 0 & 0 & 1 \\ 0 & 0 & 1 & 0).
\end{equation}
The characteristic polynomials are $\det(x - A) = x^4 + 1 = \Phi_8(x)$ and $\det(x - B) = (x - 1)^2 (x + 1)^2 = \Phi_1(x)^2 \Phi_2(x)^2$. We summarize the findings of this section in \Cref{tab:GL23-num-lagrs}.

\paragraph{$A$-invariant lagrangians over $\mathbb{Z}_p$}
Let us begin by classifying free lagrangians of $\mathbb{Z}_{p^k}^{4}$ invariant under $A$, as we explain in \cref{sec:free-invar-lagr}, starting with those over $\mathbb{Z}_{p}$. The row $n = 8$ in \cref{tab:thepowerofvodoo} tells us that invariant lagrangians exist iff $p = 2$ or $p = 1, \pm 3 \pmod{8}$.

For $p = 2$, the characteristic polynomial factors as $(x - 1)^4$, and in fact $A$ is similar to a single Jordan block of size 4. Hence there is a unique two-dimensional $A$-invariant subspace, namely $L_2 = \ker((A - 1)^2) = \Span_{\mathbb{Z}_2}((1,1,1,0), (0,0,1,1))$, which is readily checked to be isotropic, hence lagrangian.

For $p \neq 2$, we consider the Galois group $\gal(\mathbb{F}_p(\zeta_8) / \mathbb{F}_p) = \ev{p}_8$, generated by $\zeta_8 \mapsto \zeta_8^p$ with $\zeta_8$ a primitive eighth root of unity, in the different cases for $p \pmod{8}$. Recall from \cref{sec:free-invar-lagr} that the irreducible $A$-invariant subspaces are of the form $\mathcal{V}_{[s]} = \ker P_{[s]}(A)$ with $[s]$ a coset in $\ev{p}_8$, and an invariant lagrangian is a direct sum of these such that for any two terms $\mathcal{V}_{[s]}$ and $\mathcal{V}_{[s']}$, $[s'] \neq [-s]$.

For $p = 1 \pmod{8}$, the Galois group is trivial. Thus $\mathbb{Z}_p$ contains $\zeta_8$, the characteristic polynomial splits into four distinct linear factors $P_{s}(x) = x - \zeta_8^s$ with $s = \pm 1, \pm 3$, and hence there are four indecomposable invariant subspaces $\mathcal{V}_{s} = \ker(A - \zeta_8^s)$. From these, we can construct four $A$-invariant lagrangians: $\mathcal{V}_{\pm 1} \oplus \mathcal{V}_{\pm 3}$.

For $p = \pm 3 \pmod{8}$, the Galois group is $\mathbb{Z}_2\colon \zeta_8 \mapsto \zeta_8^{\pm 3}$, so the characteristic polynomial splits into quadratic factors $P_{[1]}(x) = (x - \zeta_8)(x - \zeta_8^{\pm 3})$ and $P_{[-1]}(x) = (x - \zeta_8^{-1})(x - \zeta_8^{\mp 3})$. The invariant lagrangians are simply the two indecomposable subspaces $\mathcal{V}_{[\pm 1]} = \ker P_{[\pm 1]}(A)$.

\paragraph{$A$-invariant free lagrangians}
We now look for free invariant lagrangians over $\mathbb{Z}_{p^k}$. For $p = 2$, we have found the unique invariant lagrangian over $\mathbb{Z}_2$ above. Meanwhile, $x^4 + 1$ is irreducible over $\mathbb{Z}_4$, so no free invariant lagrangians exist over $\mathbb{Z}_{2^k}$ with $k > 1$.

For $p \neq 2$, the factors of the characteristic polynomial are coprime (the eigenvalues are distinct), so Hensel lifting applies, and the factorization apply also over the $p$-adics. Thus, the free lagrangians over $\mathbb{Z}_{p^k}$ are as above. We write them out explicitly: for $p = 1 \pmod{8}$, the eigenvectors for $\zeta_8$, $\zeta_8^{-1}$, $\zeta_8^3$ and $\zeta_8^{-3}$ respectively are
\begin{equation}
  \label{eq:GL23-1mod8-eigenbasis}
  \mqty(1  \\ 1 + \zeta_8 - \zeta_8^3 \\ \zeta_8^3 \\ \zeta_8 + \zeta_8^2),
  \quad
  \mqty(1  \\ 1 + \zeta_8 - \zeta_8^3 \\ -\zeta_8 \\ -\zeta_8^2 - \zeta_8^3),
  \quad
  \mqty(1  \\ 1 - \zeta_8 + \zeta_8^3 \\ \zeta_8 \\ -\zeta_8^2 + \zeta_8^3),
  \quad
  \mqty(1  \\ 1 - \zeta_8 + \zeta_8^3 \\ -\zeta_8^3 \\ -\zeta_8 + \zeta_8^2),
\end{equation}
and a free invariant lagrangian $\mathcal{V}_{\pm 1} \oplus \mathcal{V}_{\pm 3}$ is spanned by one from the first pair together with one from the second pair.

For $p = 3 \pmod{8}$, the quadratic factors are $P_{[\pm 1]}(x) = x^2 \mp r x - 1$ with $\pm r = \zeta_8^{\pm 1} + \zeta_8^{\pm 3} \in \hat{\mathbb{Z}}_p$ satisfying $r^2 = -2$. The two free invariant lagrangians are
\begin{equation}
  \label{eq:GL23-A-3mod8-free}
  L_{[\pm 1]} = \ker(A^2 \mp rA - 1) = \Span_{\mathbb{Z}_{p^k}}\!\qty(
  \mqty(2 \\ 0 \\ 1 \pm r \\ -1), \mqty(0 \\ 2 \\ -1 \\ 1 \pm r)).
\end{equation}

For $p = -3 \pmod{8}$, the factors are rather $P_{[\pm 1]} = x^2 \mp i$ where $\pm i = \zeta_8^{\pm 1} + \zeta_8^{\pm 5}$ satisfies $i^2 = -1$. The two free invariant lagrangians are
\begin{equation}
  \label{eq:GL23-A-5mod8-free}
  L_{[\pm 1]} = \ker(A^2 \mp i) = \Span_{\mathbb{Z}_{p^k}}\!\qty(
  \mqty(2 \\ 0 \\ 1 \mp i \\ -1 \pm i), \mqty(0 \\ 2 \\ -1 \pm i \\ 1 \pm i)).
\end{equation}

\paragraph{General $A$-invariant lagrangians}
We turn to the torsional $A$-invariant lagrangians of $\mathbb{Z}_{p^k}^{4}$, classified in \cref{sec:tors-invar-lagr}. For $p \neq 2$, any torsional lagrangian (except $p^{k/2} \mathbb{Z}_{p^k}^{4}$ for $k$ even) is of $p$-adic descent. Hence for $p = \pm 3 \pmod{8}$, it is of the form
\begin{equation}
  \label{eq:GL23-A-pm3mod8-general}
  p^{l_+} L_{[+1]} \oplus p^{k - l_+} L_{[-1]}
\end{equation}
as in \eqref{eq:58}. Including the free cases $l_p = 0, k$, there are $k + 1$ $A$-invariant lagrangians.

Similarly, for $p = 1 \pmod{8}$, a general lagrangian is of the form
\begin{equation}
  \qty(p^{l_1} \mathcal{V}_{1} \oplus p^{k - l_1} \mathcal{V}_{-1})
  \oplus
  \qty(p^{l_3} \mathcal{V}_{3} \oplus p^{k - l_3} \mathcal{V}_{-3}),
\end{equation}
making for $(k + 1)^2$ possibilities.

For $p = 7 \pmod{8}$, there are no lagrangians over $\mathbb{Z}_p$; hence the only $A$-invariant lagrangians are $p^{k/2} \mathbb{Z}_{p^k}^{4}$ when $k$ is even.

For $p = 2$, there are no $p$-adic lagrangians. Hence, over $\mathbb{Z}_{2^k}$ with $k$ even, there is one $A$-invariant lagrangian $2^{k/2} \mathbb{Z}_{2^k}^{4}$. When $k = 2\ell + 1$ is odd, there is similarly a single $A$-invariant lagrangian arising from the one over $\mathbb{Z}_2$, as in \eqref{eq:68}: $2^{\ell} L_2 \cup 2^{\ell + 1} \mathbb{Z}_{2^k}^{4}$; written out
\begin{equation}
  \label{eq:GL23-A-2-odd-general}
  L = 2^{\ell} \Span_{\mathbb{Z}_{2^k}}\!\qty\Big(
  (1, 1, 1, 0), (0, 0, 1, 1))
  \cup 2^{\ell + 1} \mathbb{Z}_{2^k}^{4}.
\end{equation}

\paragraph{$\mathrm{GL}(2, 3)$-invariant lagrangians}
We now determine which of the $A$-invariant lagrangians are also invariant under $B$, and hence under the whole group $\mathrm{GL}(2, 3)$.

For $p = 3 \pmod{8}$, we observe that $B$ simply swaps the two basis vectors in \eqref{eq:GL23-A-3mod8-free}. Hence $L_\pm$, and therefore all the $A$-invariant lagrangians \eqref{eq:GL23-A-pm3mod8-general} are in fact invariant under the whole $\mathrm{GL}(2, 3)$.

For $p = -3 \pmod{8}$, on the other hand, $B$ does not leave $L_{[\pm 1]}$ of \eqref{eq:GL23-A-5mod8-free} invariant. Hence the only $A$-invariant lagrangian \eqref{eq:GL23-A-pm3mod8-general} invariant under $B$ is $p^{k/2} \mathbb{Z}_{p^k}^4$ for $k$ even, that is, the one with $l_+ = k - l_+$.

Now we come to $p = 1 \pmod{8}$. In the eigenbasis \eqref{eq:GL23-1mod8-eigenbasis}, $B$ takes the form
\begin{equation}
  B = \mqty(
  -\frac{1}{2} \zeta_{8}^{3} + \frac{1}{2} \zeta_{8} & 0 & \frac{1}{2} \zeta_{8}^{3} - \frac{1}{2} \zeta_{8} + 1 & 0 \\
  0 & -\frac{1}{2} \zeta_{8}^{3} + \frac{1}{2} \zeta_{8} & 0 & \frac{1}{2} \zeta_{8}^{3} - \frac{1}{2} \zeta_{8} + 1 \\
  -\frac{1}{2} \zeta_{8}^{3} + \frac{1}{2} \zeta_{8} + 1 & 0 & \frac{1}{2} \zeta_{8}^{3} - \frac{1}{2} \zeta_{8} & 0 \\
  0 & -\frac{1}{2} \zeta_{8}^{3} + \frac{1}{2} \zeta_{8} + 1 & 0 & \frac{1}{2} \zeta_{8}^{3} - \frac{1}{2} \zeta_{8})
\end{equation}
From the structure of zero entries in this matrix we see that precisely two of the four $A$-invariant free lagrangians, namely $\mathcal{V}_{1} \oplus \mathcal{V}_{3} = \ker\!\qty((A - \zeta_8)(A - \zeta_8^3))$ and $\mathcal{V}_{-1} \oplus \mathcal{V}_{-3} = \ker\!\qty((A - \zeta_8^{-1})(A - \zeta_8^{-3}))$ (which in this basis consist of elements of the form $(a, 0, b, 0)$ and $(0, c, 0, d)$, respectively) are invariant under the whole $\mathrm{GL}(2, 3)$. For a general $\mathrm{GL}(2, 3)$-invariant lagrangian, this forces $l_{[1]} = l_{[3]}$ in \eqref{eq:58}, so that there are $k + 1$ possibilities
\begin{equation}
  p^l \qty(\mathcal{V}_{1} \oplus \mathcal{V}_{3})
  \oplus p^{k-l} \qty(\mathcal{V}_{-1} \oplus \mathcal{V}_{-3}).
\end{equation}

For $p = 7 \pmod{8}$ with $k$ even, $2^{k/2} \mathbb{Z}_{2^k}^{4}$ is automatically invariant.

For $p = 2$, $2^{k/2} \mathbb{Z}_{2^k}^{4}$ is automatically invariant for $k$ even, while for $k$ odd, we consider explicitly the action of $B$ on the lagrangian $L$ of \eqref{eq:GL23-A-2-odd-general}. Since $B$ leaves $(0, 0, 1, 1)$ invariant and $B(1, 1, 1, 0) = (1, 1, 1, 0) + (0, 0, 1, 1)$, $L$ is invariant under the whole $\mathrm{GL}(2, 3)$.

\begin{table}
  \centering
  \begin{tabular}{r|cc}
    $p$ & $A$ & $\mathrm{GL}(2, 3)$ \\
    \hline
    $2$ & $1$ & $1$ \\
    $1 \pmod{8}$ & $(k + 1)^2$ & $k + 1$ \\
    $3 \pmod{8}$ & $k + 1$ & $k + 1$ \\
    $5 \pmod{8}$ & $k + 1$ & ($k$ even) \\
    $7 \pmod{8}$ & ($k$ even) & ($k$ even)
  \end{tabular}
  \caption{Number of lagrangian subgroups of $\mathbb{Z}_{p^k}^{4}$ invariant under $A$ and under the whole of $\mathrm{GL}(2, 3)$, respectively. The notation ($k$ even) means that there is one invariant lagrangian if $k$ is even and none otherwise.}
  \label{tab:GL23-num-lagrs}
\end{table}
In summary, we have arrived at the results of \cref{tab:GL23-num-lagrs} for prime powers $p^k$, which solves the problem for arbitrary $N$. In particular, $\mathrm{GL}(2, 3)$-invariant lagrangians exist over all $\mathbb{Z}_N$ where the prime factorization of $N$ does not contain an odd power of a prime congruent to $5$ or $7 \pmod{8}$; that is
\begin{equation}
  \begin{split}
    2, 3, 4, 6, 8, 9, 11, 12, 16, 17, 18, 19, 22, 24, 25, 27, 32, 33, 34, 36, 38, 41, 43, 44, 48, 49, 50, \\
    51, 54, 57, 59, 64, 66, 67, 68, 72, 73, 75, 76, 81, 82, 83, 86, 88, 89, 96, 97, 98, 99, 100, \dots \\
    &
  \end{split}
\end{equation}
i.e.~numbers of the form $x^2 + 2y^2$. We remark that the characteristic polynomial alone is enough to decide the existence of $A$-invariant lagrangians, but to determine whether there exist lagrangians invariant under the whole group over $\mathbb{Z}_{p^k}$ with $k$ odd, one needs to check this explicitly over $\mathbb{Z}_p$ as we have done.

\subsubsection{$\mathrm{GL}(3, 2)$: the Klein quartic}
At genus $r = 3$, the surface with the largest conformal automorphism group is the \emph{Klein quartic} \cite{Klein1878} (see also \cite{Braden:2009un}). The group is of order $168$ (this in fact saturates the \emph{Hurwitz bound}) and is isomorphic to $\mathrm{GL}(3, 2) = \mathrm{GL}(3, \mathbb{F}_2)$. We can deduce its embedding into $\mathrm{Sp}(6, \mathbb{Z})$ by considering its action on a generating set of homology cycles; see \cref{app:klein-quartic}. We find generators
\begin{equation}
  \label{eq:klein-GL32-generators}
  R = \mqty(
  0 & 0 & -1 & 1 & 0 & 0 \\
  -1 & 1 & 0 & 1 & 1 & 1 \\
  0 & 1 & -1 & 0 & 0 & 0 \\
  -1 & 1 & 0 & 0 & 0 & 0 \\
  0 & -1 & 0 & 0 & 0 & 0 \\
  1 & -1 & 0 & 0 & 0 & -1)
  \qquad
  P = \mqty(
  0 & 0 & 1 & 0 & 0 & 0 \\
  0 & 0 & 1 & -1 & 0 & 0 \\
  0 & 0 & 1 & 0 & 1 & 0 \\
  -1 & 1 & 1 & 0 & 1 & 1 \\
  1 & 0 & -1 & 0 & -1 & 0 \\
  0 & -1 & 0 & 1 & 0 & 0)
\end{equation}
of order $7$ and $3$, respectively. The characteristic polynomials are $\det(x - R) = x^6 + x^5 + x^4 + x^3 + x^2 + x + 1 = \Phi_7(x)$ and $\det(x - P) = (x - 1)^2 (x^2 + x + 1)^2 = \Phi_1(x)^2\Phi_3(x)^2$.\footnote{\ The full matrix form of $R$ is not necessary to determine $\det(x - R)$. Per the discussion in the introduction, the characteristic polynomial of any $6 \times 6$ order $7$ matrix is $\Phi_7(x)$.} The classification of lagrangians invariant under this group is quite analogous to the previous example.

\paragraph{$R$-invariant lagrangians over $\mathbb{Z}_p$}
We begin by classifying the $R$-invariant lagrangian subgroups of $\mathbb{Z}_{p^k}^6$, beginning as above with those over $\mathbb{Z}_p$. \Cref{tab:thepowerofvodoo} with $n = 7$ tells us that these exist iff $p = 7$ or $p = 1, 2, 4 \pmod{7}$.

For $p = 7$, the characteristic polynomial factors as $(x - 1)^6$, and one can check that $R$ is similar to a single Jordan block of size $6$. Hence there is again a single three-dimensional $R$-invariant subspace
\begin{equation}
  \label{eq:GL32-Z7}
  L_7 = \ker((R - 1)^3) = \Span_{\mathbb{Z}_7}\!\qty((1, 0, 0, 6, 3, 3), (0, 1, 0, 3, 6, 2), (0, 0, 1, 3, 2, 6))
\end{equation}
which is in fact isotropic and therefore lagrangian.

For $p = 1 \pmod{7}$, the Galois group $\gal(\mathbb{F}_p(\zeta_7) / \mathbb{F}_p) = \ev{1}_7$ is trivial, so $\zeta_7 \in \mathbb{Z}_p$. The characteristic polynomial factors as $\prod_{k = 1}^6 (x - \zeta_7^k)$ and we have eight $R$-invariant lagrangians $\mathcal{V}_{\pm 1} \oplus \mathcal{V}_{\pm 2} \oplus \mathcal{V}_{\pm 3}$ where $\mathcal{V}_{k} = \ker(R - \zeta_7^k)$.

For $p = 2, 4 \pmod{7}$, the Galois group is\footnote{\ Since $2$ and $4$ are inverses mod $7$, they generate the same subgroup of $\mathbb{Z}_7^\times$.} $\ev{2}_7 = \ev{4}_7 = \mathbb{Z}_3\colon \zeta_7 \overset{2}{\mapsto} \zeta_7^2 \overset{2}{\mapsto} \zeta_7^4 \overset{2}{\mapsto} \zeta_7$. Hence the characteristic polynomial splits into cubic factors $P_{[\pm 1]}(x) = (x - \zeta_7^{\pm 1})(x - \zeta_7^{\pm 2})(x - \zeta_7^{\pm 4})$ and there are two $R$-invariant lagrangians $\mathcal{V}_{[\pm 1]} = \ker(P_{[\pm 1]}(R))$.

\paragraph{$R$-invariant free lagrangians}
For $p = 7$, the characteristic polynomial $\Phi_7(x)$ is irreducible over $\mathbb{Q}_7$ (as the Galois group $\gal(\mathbb{Q}_p(\zeta_7) / \mathbb{Q}_p) \cong \mathbb{Z}_7^\times$ \cite[Chap.~2, Prop.~7.13]{MR1697859} has order $6$), so as explained in \cref{sec:galo-groups-cyclt} it is irreducible even over $\mathbb{Z}_{49}$. Hence there are no free invariant lagrangians over $\mathbb{Z}_{7^k}$ with $k > 1$.

For $p \neq 7$, as always, the lagrangians over $\mathbb{Z}_p$ lift to the free lagrangians over $\mathbb{Z}_{p^k}$, which we write down explicitly. For $p = 1 \pmod{7}$, the eigenvectors for $\zeta_7$, $\zeta_7^{-1}$, $\zeta_7^2$, $\zeta_7^{-2}$, $\zeta_7^4$ and $\zeta_7^{-4}$ are\footnote{\ We choose this ordering of the roots to make the form of \eqref{eq:GL32-P-R-eigenbasis} as clean as possible.}
\begin{equation}
  \label{eq:GL32-1mod7-eigenvectors}
  \footnotesize
  \begin{array}{lccc}
    & \mathbf{1}\colon & \mathbf{2}\colon & \mathbf{4}\colon \\[5pt]
    +\colon & \mqty(
         1 \\
    \zeta_{7}^{3} + \zeta_{7}^{2} + \zeta_{7} + 1 \\
    \zeta_{7}^{2} + 1 \\
    \zeta_{7}^{2} + \zeta_{7} + 1 \\
    \zeta_{7}^{5} + \zeta_{7}^{4} + \zeta_{7}^{3} \\
    -\zeta_{7}^{5} - \zeta_{7}^{3} - \zeta_{7}^{2} - \zeta_{7} - 1),
    & \mqty(
      1 \\
    -\zeta_{7}^{5} - \zeta_{7}^{3} - \zeta_{7} \\
    \zeta_{7}^{4} + 1 \\
    \zeta_{7}^{4} + \zeta_{7}^{2} + 1 \\
    -\zeta_{7}^{5} - \zeta_{7}^{4} - \zeta_{7}^{2} - 1 \\
    \zeta_{7}^{5} + \zeta_{7}),
    & \mqty(
      1 \\
    \zeta_{7}^{5} + \zeta_{7}^{4} + \zeta_{7} + 1 \\
    \zeta_{7} + 1 \\
    \zeta_{7}^{4} + \zeta_{7} + 1 \\
    -\zeta_{7}^{4} - \zeta_{7}^{3} - \zeta_{7} - 1 \\
    \zeta_{7}^{3} + \zeta_{7}^{2}),
    \\[40pt]
    -\colon & \mqty(
         1 \\
    -\zeta_{7}^{3} - \zeta_{7}^{2} - \zeta_{7} \\
    \zeta_{7}^{5} + 1 \\
    -\zeta_{7}^{4} - \zeta_{7}^{3} - \zeta_{7}^{2} - \zeta_{7} \\
    \zeta_{7}^{4} + \zeta_{7}^{3} + \zeta_{7}^{2} \\
    \zeta_{7}^{3} + \zeta_{7}),
    & \mqty(
      1 \\
    \zeta_{7}^{5} + \zeta_{7}^{3} + \zeta_{7} + 1 \\
    \zeta_{7}^{3} + 1 \\
    \zeta_{7}^{5} + \zeta_{7}^{3} + 1 \\
    -\zeta_{7}^{5} - \zeta_{7}^{3} - \zeta_{7}^{2} - 1 \\
    -\zeta_{7}^{5} - \zeta_{7}^{4} - \zeta_{7}^{3} - \zeta_{7} - 1),
    & \mqty(
      1 \\
    -\zeta_{7}^{5} - \zeta_{7}^{4} - \zeta_{7} \\
    -\zeta_{7}^{5} - \zeta_{7}^{4} - \zeta_{7}^{3} - \zeta_{7}^{2} - \zeta_{7} \\
    -\zeta_{7}^{5} - \zeta_{7}^{4} - \zeta_{7}^{2} - \zeta_{7} \\
    \zeta_{7}^{5} + \zeta_{7}^{2} + \zeta_{7} \\
    \zeta_{7}^{5} + \zeta_{7}^{4}),
  \end{array}
\end{equation}
and a free invariant lagrangian $\mathcal{V}_{\pm 1} \oplus \mathcal{V}_{\pm 2} \oplus \mathcal{V}_{\pm 4}$ is spanned by one from each pair.

For $p = 2, 4 \pmod{7}$, the cubic factors of the characteristic polynomial are $P_{[\pm 1]}(x) = x^3 - \Pi^\pm x^2 + \Pi^\mp x - 1$ where $\Pi^\pm = \zeta_7^{\pm 1} + \zeta_7^{\pm 2} + \zeta_7^{\pm 4} \in \hat{\mathbb{Z}}_{p}$ are conjugate \emph{Gaussian periods} satisfying $\Pi^+ + \Pi^- = -1$ and $(\Pi^+ - \Pi^-)^2 = -7$. To express the invariant subspaces, we consider $p = 2$ separately. For $p = 2$, take $\Pi^+ = 0$, $\Pi^- = 1$. Then we calculate
\begin{align}
  \label{eq:GL32-2}
  & \mathcal{V}_{[+1]} = \ker(R^3 + R - 1)
    =\Span_{\mathbb{Z}_{2^k}}\!\!\qty(
    \mqty(1, 0, 1, 0, 1, 0),
    \mqty(0, 1, 1, 0, 1, 1),
    \mqty(0, 0, 0, 1, 1, 1)), \\
  & \mathcal{V}_{[-1]} = \ker(R^3 - R^2 - 1)
    =\Span_{\mathbb{Z}_{2^k}}\!\!\qty(
    \mqty(1, 0, 0, 0, 1, 1),
    \mqty(0, 1, 0, 1, 0, 0),
    \mqty(0, 0, 1, 1, 0, 0)).
\end{align}
For $p \neq 2$, we instead have
\begin{equation}
  \label{eq:GL32-24mod7}
  \mathcal{V}_{[\pm 1]} = \Span_{\mathbb{Z}_{p^k}}\!\qty(
  \mqty(4 \\ 0 \\ 0 \\ 3 \Pi^\pm + 1 \\ -\Pi^\pm + 1 \\ -\Pi^\pm + 1),
  \mqty(0 \\ 4 \\ 0 \\ -\Pi^\pm + 1 \\ 3 \Pi^\pm + 1 \\ -\Pi^\pm - 3),
  \mqty(0 \\ 0 \\ 4 \\ -\Pi^\pm + 1 \\ -\Pi^\pm - 3 \\ 3 \Pi^\pm + 1))
\end{equation}
where $\Pi^\pm = \frac{-1 \pm \sqrt{-7}}{2}$.

\paragraph{General $R$-invariant lagrangians}
As before, we now move to the torsional lagrangians. For $p = 7$, there is a unique invariant lagrangian over $\mathbb{Z}_{7^k}$: $7^{k/2} \mathbb{Z}_{7^k}^{6}$ if $k$ is even and
\begin{equation}
  \label{eq:GL32-0mod7}
  L = 2^{\ell} L_7 \cup 2^{\ell + 1} \mathbb{Z}_{7^k}^6
\end{equation}
with $L_7$ as in \eqref{eq:GL32-Z7} if $k = 2\ell + 1$ is odd.

For $p \neq 7$, every lagrangian is of $p$-adic descent. For $p = 2, 4 \pmod{7}$, there are again $k + 1$ lagrangians
\begin{equation}
  p^{l_+} L_{[+1]} \oplus p^{k-l_+} L_{[-1]}
\end{equation}
including the free ones.

For $p = 1 \pmod{7}$, all invariant lagrangians are likewise of the form
\begin{equation}
  \qty(p^{l_1} \mathcal{V}_1 \oplus p^{k - l_1} \mathcal{V}_{-1})
  \oplus
  \qty(p^{l_2} \mathcal{V}_2 \oplus p^{k - l_2} \mathcal{V}_{-2})
  \oplus
  \qty(p^{l_4} \mathcal{V}_4 \oplus p^{k - l_4} \mathcal{V}_{-4})
\end{equation}
with $\mathcal{V}_i$ defined in \eqref{eq:GL32-1mod7-eigenvectors}; in total $(k + 1)^3$ possibilities.

For $p = 3, 5, 6 \pmod{7}$, there are no lagrangians over $\mathbb{Z}_p$, so the only invariant lagrangians are again $p^{k/2} \mathbb{Z}_{p^k}^{6}$ for $k$ even.

\paragraph{$\mathrm{GL}(3, 2)$-invariant lagrangians}
Finally, we check which of the $R$-invariant lagrangians are also invariant under $P$. For $p = 7$ with $k$ even, $7^{k/2} \mathbb{Z}_{7^k}^{6}$ is automatically invariant, while for $k$ odd, we calculate that $L_7$ from \eqref{eq:GL32-Z7} is $P$-invariant over $\mathbb{Z}_7$; hence $L$ of \eqref{eq:GL32-0mod7} is also invariant.

For $p = 1 \pmod{7}$, in the eigenbasis \eqref{eq:GL32-1mod7-eigenvectors} (ordered $\zeta_7$, $\zeta_7^{-1}$, $\zeta_7^2$, $\zeta_7^{-2}$, $\zeta_7^4$, $\zeta_7^{-4}$), $P$ takes the form
\begin{equation}
  \label{eq:GL32-P-R-eigenbasis}
  P = -\mqty(
  \frac{\zeta^2 + \zeta^{-4}}{(\zeta - \zeta^{-1})^2} & 0 & \frac{\zeta^4}{(\zeta^2 - \zeta^{-2})^2} & 0 & \frac{1}{(\zeta^2 - \zeta^{-2})^2} & 0 & \\
  0 & \frac{\zeta^{-2} + \zeta^4}{(\zeta - \zeta^{-1})^2} & 0 & \frac{\zeta^{-4}}{(\zeta^2 - \zeta^{-2})^2} & 0 & \frac{1}{(\zeta^2 - \zeta^{-2})^2} & \\
  \frac{1}{(\zeta^4 - \zeta^{-4})^2} & 0 & \frac{\zeta^4 + \zeta^{-1}}{(\zeta^2 - \zeta^{-2})^2} & 0 & \frac{\zeta}{(\zeta^4 - \zeta^{-4})^2} & 0 & \\
  0 & \frac{1}{(\zeta^4 - \zeta^{-4})^2} & 0 & \frac{\zeta^{-4} + \zeta}{(\zeta^2 - \zeta^{-2})^2} & 0 & \frac{\zeta^{-1}}{(\zeta^4 - \zeta^{-4})^2} & \\
  \frac{\zeta^2}{(\zeta - \zeta^{-1})^2} & 0 & \frac{1}{(\zeta - \zeta^{-1})^2} & 0 & \frac{\zeta + \zeta^{-2}}{(\zeta^4 - \zeta^{-4})^2} & 0 & \\
  0 & \frac{\zeta^{-2}}{(\zeta - \zeta^{-1})^2} & 0 & \frac{1}{(\zeta - \zeta^{-1})^2} & 0 & \frac{\zeta^{-1} + \zeta^2}{(\zeta^4 - \zeta^{-4})^2})
\end{equation}
where $\zeta = \zeta_7$ for legibility. Again, the structure of zero entries tells us that of the free lagrangians, only $\mathcal{V}_1 \oplus \mathcal{V}_2 \oplus \mathcal{V}_4$ and $\mathcal{V}_{-1} \oplus \mathcal{V}_{-2} \oplus \mathcal{V}_{-4}$ are $R$-invariant. Therefore the general $\mathrm{GL}(3,2)$-invariant lagrangian is
\begin{equation}
  p^l \qty(\mathcal{V}_1 \oplus \mathcal{V}_2 \oplus \mathcal{V}_4)
  \oplus p^{k - l}
  \qty(\mathcal{V}_{-1} \oplus \mathcal{V}_{-2} \oplus \mathcal{V}_{-4}).
\end{equation}

For $p = 2, 4 \pmod{7}$, a direct computation verifies that both free lagrangians $L_{[\pm 1]}$ of \cref{eq:GL32-2,eq:GL32-24mod7} are $P$-invariant; therefore also all torsional lagrangians are.

For $p = 3, 5, 6 \pmod{7}$, $p^{k/2} \mathbb{Z}_{p^k}^{6}$ for $k$ even is automatically $\mathrm{GL}(3, 2)$-invariant.

\begin{table}
  \centering
  \begin{tabular}{r|cc}
    $p$ & $R$ & $\mathrm{GL}(3, 2)$ \\
    \hline
    $7$ & 1 & 1 \\
    $1 \pmod{7}$ & $(l + 1)^3$ & $l + 1$ \\
    $2, 4 \pmod{7}$ & $l + 1$ & $l + 1$ \\
    $3, 5, 6 \pmod{7}$ & ($l$ even) & ($l$ even)
  \end{tabular}
  \caption{Number of lagrangian subgroups of $\mathbb{Z}_{p^l}^{6}$ invariant under $R$ and under the whole of $\mathrm{GL}(3, 2)$, respectively. For $p = 3, 5, 6 \pmod{7}$, there is one invariant lagrangian if $l$ is even and none otherwise.}
  \label{tab:GL32-num-lagrs}
\end{table}
In summary, we have the results of \cref{tab:GL32-num-lagrs} for prime powers, and hence for all $N$. $\mathrm{GL}(3, 2)$-invariant lagrangians exist over $\mathbb{Z}_N$ unless the prime factorization of $N$ contains an odd power of a prime congruent to $3, 5, 6 \pmod{7}$. The admissible $N$ are
\begin{equation}
  \begin{split}
    & 2, 4, 7, 8, 9, 11, 14, 16, 18, 22, 23, 25, 28, 29, 32, 36, 37, 43, 44, 46, 49, \\
    & 50, 53, 56, 58, 63, 64, 67, 71, 72, 74, 77, 79, 81, 86, 88, 92, 98, 99, 100, \dots
  \end{split}
\end{equation}
i.e.~numbers of the form $x^2+xy+2y^2$ with $x$ and $y$ non-zero integers.

\section{Repeated roots of quadratic polynomials}
\label{sec:rep-roots}
Here we give more details on the roots of the $r = 1$ characteristic polynomial \eqref{eq:2x2-char-poly} when ordinary Hensel lifting fails. This happens when $\pm 1$ is a repeated root over $\mathbb{Z}_p$. Then the polynomial takes the form
\begin{equation}
  f(x) \equiv \det(x - g) = x^2 - (c p^h \pm 2) x + 1
\end{equation}
over the integers, for some $c$ coprime to $p$ and $h > 0$ (with the $p$-adic metric, so $h \to \infty$ corresponds to $x^2 \mp 2x + 1$). We search for roots of $f(x)$ over $\mathbb{Z}_{p^k}$. We will show:
\begin{itemize}
  \item When $k \leq h$: There are $p^{\floor{\frac{k}{2}}}$ roots $\pm 1 + p^{\ceil{\frac{k}{2}}} \mathbb{Z}_{p^k}$,
  \item When $k > h$ and $h$ is even:
    \begin{itemize}
      \item If $p \neq 2$ and $\pm c$ is a quadratic residue mod $p$:
        There are $2 p^{h/2}$ roots $\hat{y} p^{h/2} \pm 1 + p^{k - h/2} \mathbb{Z}_{p^k}$ and $-\hat{y} p^{h/2} \pm 1 + p^{k - h/2} \mathbb{Z}_{p^k}$, where $\hat{y}$ is a $p$-adic square root of $\pm c$.
      \item If $p = 2$ roots exist iff $y^2 \mp c - 2^{h/2} c y$ has a root $y$ mod $2^{k - h}$. If $k - h \ge 3$ this is equivalent to existence of roots mod $8$, which in turn is equivalent to $\pm c = 1 \pmod{8}$ if $h \ge 6$.
    \end{itemize}
\end{itemize}

Now we prove this. The roots must be of the form $\lambda = y p^\ell \pm 1$ with $y$ coprime to $p$ and $\ell > 0$. A root satisfies
\begin{equation}
  0 = f(\lambda) = y^2 p^{2\ell} - y c p^{\ell + h} \mp c p^h \pmod{p^k}.
\end{equation}
If $k \le h$, this reduces to $2\ell \geq k$, which is satisfied by the $p^{\floor{\frac{k}{2}}}$ roots $\lambda = \tilde{y} p^{\ceil{\frac{k}{2}}} \pm 1$ for $\tilde{y} \in \mathbb{Z}_{p^{\floor{\frac{k}{2}}}}$. If $k > h$, reducing mod $p^{\ell + h}$ shows that $2\ell = h$. The equation now becomes
\begin{equation}
  p^{2\ell} \qty(y^2 \mp c - c p^\ell y) = 0 \pmod{p^k},
\end{equation}
or
\begin{equation}
  \label{eq:q-of-y}
  0 = q(y) \equiv y^2 \mp c - c p^\ell y \pmod{p^{k - 2\ell}}.
\end{equation}
Since $y$ is coprime to $p$, so is $q'(y) = 2 y - c p^\ell$ as long as $p \neq 2$. Hence any root $y$ of \eqref{eq:q-of-y} is the reduction of a unique $p$-adic root $\hat{y} \in \hat{\mathbb{Z}}_p$. These are the lifts of the roots of $q(y)$ over $\mathbb{Z}_p$, which exist iff $c$ is a quadratic residue. The roots of $f(\lambda)$ are thus the values $y p^\ell \pm 1 \pmod{p^k}$ with $y = \hat{y} \pmod{p^{k - 2\ell}}$, that is, the $p^\ell$ values in $\mathbb{Z}_{p^k}$ given by $\lambda = \hat{y} p^\ell \pm 1 \pmod{p^{k - \ell}}$.

To reason about solutions of \eqref{eq:q-of-y} in the case of $p = 2$, we note that by a strengthened version of Hensel's lemma (see e.g.~\cite[theorem~4.1]{Conrad}), if $k - 2\ell \ge 3$ then any root mod $2^{k - 2\ell}$ is congruent to a $2$-adic root mod $2^{k - 2\ell - 1}$. In particular, there exist $2$-adic roots iff there exist roots mod $8$. If in addition $\ell \ge k - 2\ell$, the equation becomes $y^2 = \pm c$, which has solutions iff $\pm c = 1 \pmod{8}$.

\section{Cyclotomic polynomials and exceptional primes}
\label{sec:cyclo_poly_except_primes}
In this section we discuss the factorization of $\Phi_{n}(x)$ over $\mathbb{Z}_{p^k}(x)$ when $p$ divides $n$. This case is an exception since the case Galois group over $\mathbb{F}_{p}$ is a subgroup of the Galois group over $\mathbb{Q}_{p}$. As a result $\Phi_{n}(x)$ over $\mathbb{F}_{p}$ has more factors than $\Phi_{n}(x)$ over $\mathbb{Q}_{p}$. In fact, as we have seen, if $n = p^l n'$, then,
\begin{align}
  \label{eq:81}
  \Phi_{n=p^l n'}(x) = (\Phi_{n'}(x))^{\phi(p^l)} ~.
\end{align}
Since $\Phi_{n}(x)$ has repeated factors, Hensel lifting fails. We now show that for $k > 1$, the monic irreducible factors of $\Phi_{n}(x)$ are precisely those that are $\mod p^k$ reduction of irreducible factors of $\Phi_{n}(x)$ over $\mathbb{Q}_{p}$. The idea here is to use the Galois group over $\mathbb{Q}_{p}$ and $p$-adic roots of unity to constrain the factorization over $\mathbb{Z}_{p^k}$.

Let us suppose $\mathcal{P}(x)$ is an irreducible factor of $\Phi_{n}(x)$ over $\mathbb{Q}_{p}$ that splits over $\mathbb{Z}_{p^k}$, i.e.,
\begin{align}
  \label{eq:82}
  \mathcal{P}(x) = Q_{1}(x) Q_{2}(x) \mod p^k ~.
\end{align}
If $Q_{1}(x)$ and $Q_{2}(x)$ are coprime we can use Hensel lifting to find a pair polynomials $\mathcal{Q}_{1}(x), \mathcal{Q}_{2}(x) \in \hat{\mathbb{Z}}_{p}[x]$ such that $\mathcal{P}(x) = \mathcal{Q}_{1}(x)\mathcal{Q}_{2}(x)$ and,
\begin{align}
  \label{eq:83}
  Q_{1}(x) = \mathcal{Q}_{1}(x) \mod p^k &&,&& Q_{2}(x) = \mathcal{Q}_{2}(x) \mod p^k
\end{align}
which contradicts the assumptions that $\mathcal{P}(x)$ is irreducible. Hence $Q_{1}(x)$ and $Q_{2}(x)$ must have common factors, in fact iterating this argument we conclude that any factorization of $\mathcal{P}(x) \bmod p^k$ must take the form,
\begin{align}
  \label{eq:84}
  \mathcal{P}(x) = (P(x))^{a} \mod p^k ~,
\end{align}
where the polynomial $P(x)$ cannot be written as a product of monic polynomials. Hence $P(x) \bmod p$  must be an irreducible factor of $\Phi_{n}(x)$ over $\mathbb{F}_{p}$. Let us recall that irreducible factors of $\Phi_{n}(x)$ over $\mathbb{Q}_{p}$ are given by $\mathcal{P}_{[s]}$,
\begin{align}
  \label{eq:85}
  \mathcal{P}_{[s]}(x) = \prod_{s \in [s]}(x - \zeta_{n}^{s}) ~,
\end{align}
with $[s] \in \mathbb{Z}_{n}^{*} / \mathsf{G}$ and $\mathsf{G}$ the $p$-adic Galois group. Without loss of generality we consider $\mathcal{P}_{[1]}$. For $\mathcal{P}_{[1]}$ to factorize as in \eqref{eq:84}, some roots of $\mathcal{P}_{[1]}(x)$ must coincide modulo $p$ i.e. there is some $t \in [1]$ such that
\begin{align}
  \label{eq:86}
  \zeta_{p^l} = \zeta_{p^l}^{t} \mod p^k \implies \zeta_{p^l}^{t-1} = 1 \mod p^k ~.
\end{align}
i.e.~over $\mathbb{Z}_{p^k}$, $\zeta_{n}$ has order less than than $n$, indeed if the order is $\tilde{n}$, $\zeta_{p^l}$ becomes a $\tilde{n}$-th primitive root of unity which occurs in $\Phi_{n}(x)$ with some multiplicity $a > 1$. As a result,
\begin{align}
  \label{eq:87}
  \Phi_{n}(x) = (\Phi_{\tilde{n}}(x))^{a} \mod p^k ~.
\end{align}

Let us now consider the case of $n = p^l$. Since $x-1$ is the only irreducible factor over $\mathbb{F}_{p}$, the arguments above shows that any monic factor of $\Phi_{p^l}(x)$ over $\mathbb{Z}_{p^k}$ must take the form,
\begin{align}
  \label{eq:88}
  (x - 1)^{\phi(p^l)},
\end{align}
As a result,
\begin{align}
  \label{eq:89}
  \prod_{i< p^l , \gcd(i,p)=1} (\zeta_{p^l}^{i}-1) = 0 \mod p^k ~.
\end{align}
However from Proposition 7.13 in Chapter 2 of \cite{MR1697859},
\begin{align}
  \label{eq:90}
  \prod_{i < p^l, \gcd(i,p)=1} (\zeta_{p^l}^{i}-1) = p ~.
\end{align}
Comparing with \eqref{eq:89}, we see that factorization can only happen when $k=1$.

The more general case of $n = p^l n'$ follows from this. As explained above the order of $\zeta_{n}$ must be reduced modulo $p^k$ for any factorization over $\mathbb{Z}_{p^k}$ to be possible. Since its order modulo $p$ is $\varphi(m)$, modulo any $p^k$ its order can only be a multiple of $\varphi(m)$. As a result $\zeta_{n}^{n'}$ (which is a primitive $p^l$-th root of unity) must have order less $\varphi(p^l)$ modulo $p^k$. We have shown above that this is only possible when $k = 1$.

\section{Conformal automorphisms of the Klein quartic}
\label{app:klein-quartic}

\begin{figure}[ht]
  \centering
  \includegraphics[width=.9\textwidth]{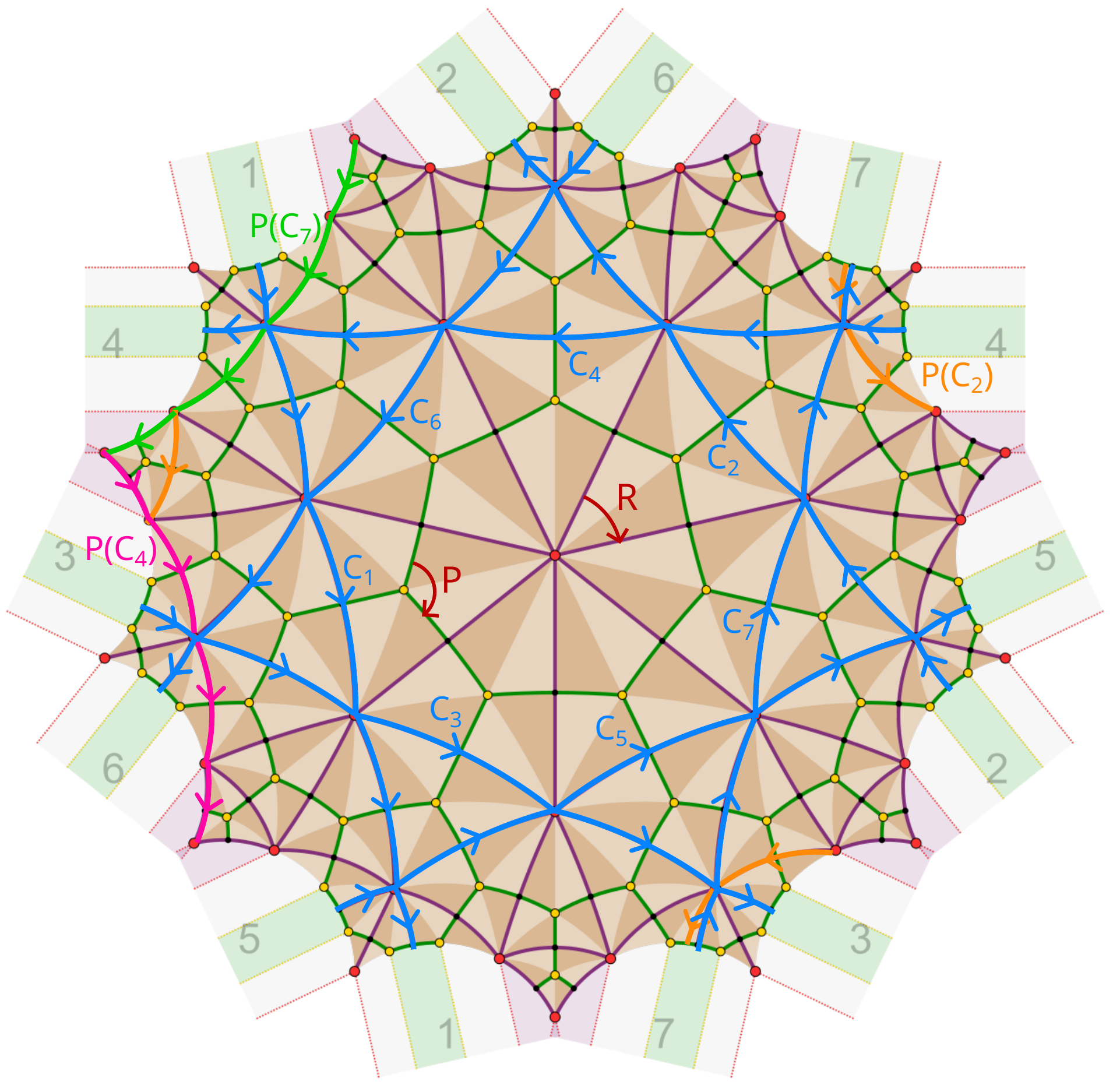}
  \caption{The Klein quartic with the homology cycles $C_1, \dots C_7$, the order $7$ rotation $R$ and the order $3$ rotation $P$. The images $R(C_i)$ and $P(C_i)$ can be computed from the figure; some less obvious cases $P(C_{2,4,7})$ are shown.}
  \label{fig:klein-quartic}
\end{figure}
To find the $\mathrm{Sp}(6, \mathbb{Z})$ action of the automorphisms of the Klein quartic on the 1-form symmetry of the class $\mathcal{S}$ theory, we realize the surface as a quotient of the hyperbolic plane with fundamental domain a regular 14-gon; see \cref{fig:klein-quartic}.\footnote{\ The figure is adapted from work by T.~Piesk; \url{https://commons.wikimedia.org/wiki/File:Klein_quartic_with_dual_graphs.svg}} The sides of the 14-gon are identified in a nontrivial way. A symmetry is specified by mapping a fixed shaded triangle to any of the 168 triangles of the same color. As we will check momentarily, the displayed order 7 rotation $R$ and the order $3$ rotation $P$ generate the symmetry group.

There is a natural set of homology generators $C_1, \dots, C_7$ (\cref{fig:klein-quartic}), subject to a single relation
\begin{equation}
  C_1 + C_2 + C_3 + C_4 + C_5 + C_6 + C_7 = 0.
\end{equation}
By carefully tracking where these end up after a rotation, we find expressions
\begin{equation}
  R =
  \begin{pmatrix}
    0 & 0 & 0 & 0 & -1 & 1 \\
    0 & 0 & 0 & 0 & -1 & 0 \\
    1 & 0 & 0 & 0 & -1 & 0 \\
    0 & 1 & 0 & 0 & -1 & 0 \\
    0 & 0 & 1 & 0 & -1 & 0 \\
    0 & 0 & 0 & 1 & -1 & 0
  \end{pmatrix} \qquad
  P =
  \begin{pmatrix}
    0 & 1 & -1 & 0 & 0 & 0 \\
    0 & 1 & 0 & 0 & 0 & 0 \\
    0 & 0 & 0 & 0 & 0 & 1 \\
    0 & 0 & 0 & -1 & 1 & 0 \\
    0 & 1 & 0 & -1 & 0 & 0 \\
    -1 & 1 & 0 & 0 & 0 & 0
  \end{pmatrix}
\end{equation}
in the basis $\{C_1, \dots, C_6\}$. From here we can verify that they generate a
group of order 168, i.e.~the full symmetry group. In this basis, the intersection form is
\begin{equation}
  \Omega' =
  \begin{pmatrix}
    0 & 0 & 1 & -1 & 1 & -1 \\
    0 & 0 & 0 & 1 & -1 & 1 \\
    -1 & 0 & 0 & 0 & 1 & -1 \\
    1 & -1 & 0 & 0 & 0 & 1 \\
    -1 & 1 & -1 & 0 & 0 & 0 \\
    1 & -1 & 1 & -1 & 0 & 0
  \end{pmatrix}.
\end{equation}
We can bring it to the standard symplectic form \eqref{eq:5} by a change of basis $B$ such that $B^T \Omega' B = \Omega$. One such $B$ is
\begin{equation}
  B =
  \begin{pmatrix}
    0 & 0 & 0 & 1 & 0 & 1 \\
    0 & 0 & 1 & 1 & 1 & 1 \\
    1 & 0 & 0 & 0 & 0 & 0 \\
    0 & 1 & 0 & 0 & 0 & 1 \\
    0 & 0 & 1 & 0 & 0 & 1 \\
    0 & 0 & 1 & 0 & 0 & 0
  \end{pmatrix}.
\end{equation}
The expressions for $R$ and $P$ in the new basis are those presented in \eqref{eq:klein-GL32-generators}.

\bibliography{dualitydefects2}
\bibliographystyle{ytphys}
\end{document}